\documentclass[onecollarge,runningheads]{svjour2}

\smartqed  

\usepackage[pdftex]{graphicx}
\usepackage{amsmath,bm,mathrsfs}
\usepackage{amssymb}
\usepackage{framed}

\begin{document}

\title{Geometric analysis on the unidirectionality of the pulmonary veins for atrial reentry}

\titlerunning{Geometric analysis on the PVs}   

\author{ Sehun Chun \thanks{This work was initiated at the Institute for Mathematical Sciences, Imperial College London, London, United Kingdom, SW7 2PE}}

\institute{African Institute for Mathematical Sciences and Stellenbosch University, 5 Melrose road, Muizenberg, Cape Town, 7945, South Africa \\
Tel.: +27 (0)21 787 9325 \\
\email{schun@aims.ac.za}}

\date{\today}

\maketitle

\begin{abstract}
It is widely believed that the pulmonary veins (PVs) of the atrium play the central role in the generation of atrial reentry leading to atrial fibrillation, but its mechanism has not been analytically explained. In order to improve the current clinical procedures for atrial reentry by understanding its mechanism, geometrical analysis is proposed on the conditions of conduction failure at the PVs and is validated by various computational modeling. To achieve this, a new analytic approach is proposed by adapting the geometric relative acceleration analysis from spacetime physics on the hypothesis that a large relative acceleration can translate to a dramatic increase in the curvature of the wavefront and subsequently to conduction failure. This analytic method is applied to a simplified model of the PV to reveal the strong dependency of the propagational direction and the magnitude of anisotropy for conduction failure. The unidirectionality of the PVs follows directly and is validated by computational tests in a T-shaped domain, computational simulations for three-dimensional atrial reentry and previous in-silico reports for atrial reentry.
\end{abstract}

\section{Introduction}

Cardiac electric propagation starts from a small ellipsoid strip, known as the sinoatrial node (SAN), located in the right atrium right below and slightly lateral to the opening of the superior vena cava \cite{Guyton}. In cardiac electrophysiology, the impulse propagation from other sources than the SAN is a serious problem and often regarded as the cause of many cardiac electrophysiological diseases such as atrial fibrillation. In the core mechanism of atrial fibrillation, there lies atrial reentry by a single re-entry circuit \cite{Nattel}. For activation propagation in the atria, the pathway of atrial reentry is difficult to understand, especially with the extremely thin wall of the atrium that restricts any diversional pathway such as  a three-dimensional route in the depth of cardiac tissues. Even in the two-dimensional plane, only a limited design of reentry, consisting of a unidirectional tapered pathway and schematically placed anisotropy, could simulate atrial reentrance, even without consideration of its realistic scale \cite{Kogan}. In three-dimensional geometry, the presumed complexity of geometry for atrial reentry seems to cast doubts on the discovery and reconstruction of realistic geometry to generate atrial reentry, rather leaving the causes of atrial reentry on the refractory period (RP), or a period of time during which cardiac tissues are not excitable, or conduction property of myocardial tissue.

Nevertheless, there have been many observations and scientific studies suggesting that the pulmonary veins (PVs)--the large veins carrying oxygenated blood from the lung into the left atrium and are covered with cardiac tissues a few inches from the atrium--play the most significant role in atrial reentry. Numerous experimental and clinical studies have been reported to support this phenomenon \cite{Aora} \cite{Haissaguerre} \cite{Nattel} \cite{Wu} and have been used to develop a series of successful surgical procedures by isolating the PVs with ablation lesions \cite{Rostock} \cite{Sueda}. In view of the clinical observations, the anatomical pattern \cite{Roux}, myocardial sleeves \cite{Ho}, and pathology \cite{Hocini} of the PVs correlate the PVs with the cause of atrial fibrillation. Some computational modeling with biologically detailed data also successfully simulated the generation of atrial reentry by the PVs. Cherry et. al. computationally demonstrated the strong correlation between the reentry and the geometrical size or conducting properties of the PVs \cite{Cherry} and Aslanidi et. al. also computationally reconstructed the reentry from the realistic 3D geometry and fibre orientation from micro-CT \cite{Aslanidi2} \cite{Aslanidi}. Despite a host of \textit{in vivo} or \textit{in situ} reports to conjecture the role of the PV for the reentry and \textit{in silico} studies to regenerate the reentry around the PV, the reentry mechanism remains largely unexplained and this lack of understanding has been a major obstacle in improving current surgical procedures to effectively prevent or terminate atrial reentry.

\begin{figure}[ht]
\centering
\vbox{
  \includegraphics[height=4cm,width=8cm]{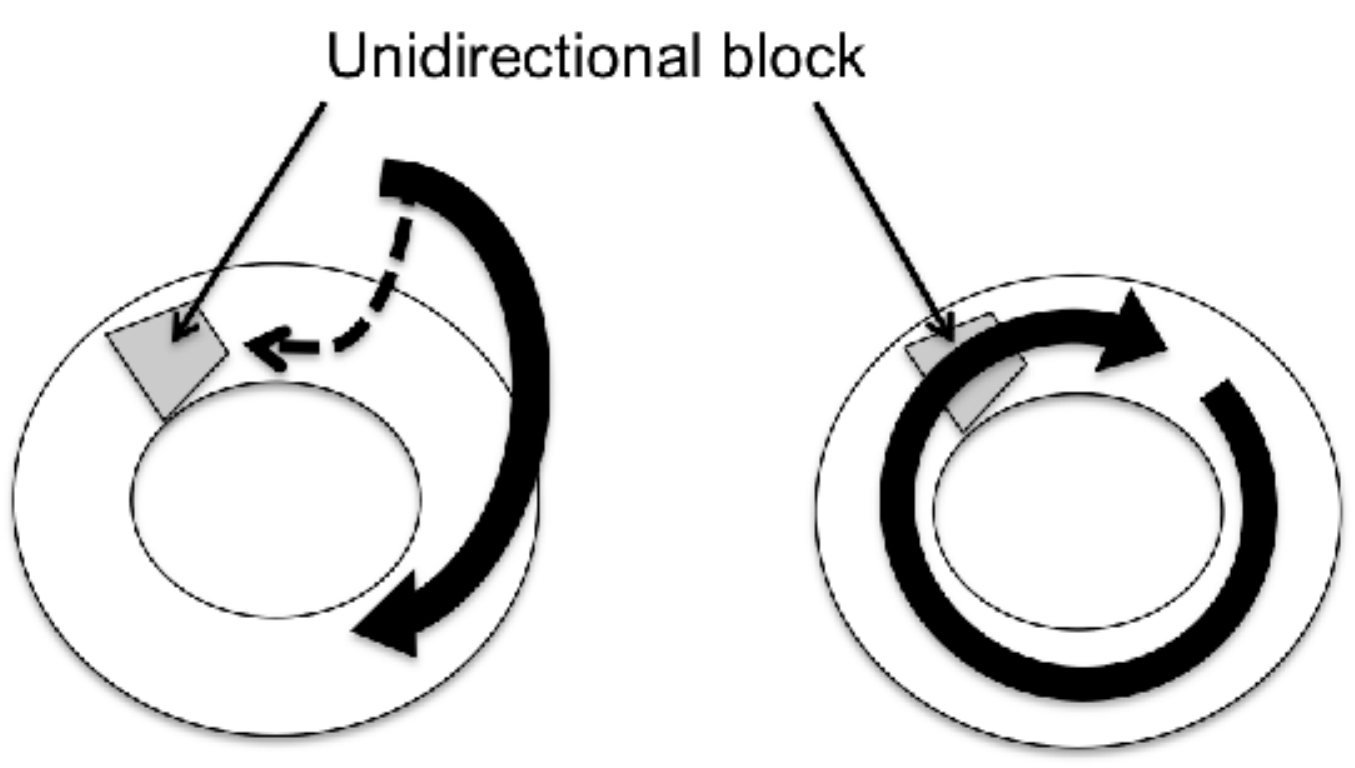} }
\caption{Illustration of unidirectional block. Modified from \cite{Nattel}}
\label {Unidirblock}
\end{figure}

In order to explain the mechanism of reentry, many theories have been proposed, but two of them are accepted most broadly: the first is the theory of \textit{refractory block} caused by periodic but non-symmetric stimulations which was explained briefly in the previous paragraph (see chapter 35 in \cite{Zipes}). The second is the theory of geometrical \textit{unidirectional block} \cite {Nattel} which we pay our attention to in the paper. But, two theories can be explained coherently. As the deviation of propagational direction from a converging point to cause self-excitation, atrial reentry can only occur if some myocardial tissues are not excited along the shortened path from the SAN, but excited along a meandering route (Figure \ref{Unidirblock}). A sufficient arrival time difference between the short path and the meandering path allows the neighboring myocardial cells to be in an excitable state after the RP. This may lead to a phenomenon of general automaticity, especially for atrial reentry. The crucial property of unidirectional block is that excitability strongly depends on the \textit{direction} of propagation toward the block, called directional sensitivity for excitation of myocardial tissue. Even in the plane of homogeneous media, some tissues in the RP may also work as a unidirectional block to become a refractory block, but here we only consider the unidirectionality generated by geometry or anisotropy, not by recovering regions from the RP.

The goal of this paper is to provide mathematical analysis and computational modeling to support the proposition that the peculiar geometry of the PV and specially-aligned anisotropy could be built as a unidirectional block. This conjecture has been widely acknowledged in the cardiology community \cite {Hocini}  \cite {Perez-Lugones} \cite {Jean} \cite {Verheule}, but has never been supported mathematically nor has been computationally modeled. To achieve this goal, we use the new analytic tool for geometrical analysis, called the \textit{relative acceleration approach} in the perspective of \textit{trajectory}, as well as the proven computational schemes for computational simulations of atrial reentry, known as the \textit{method of moving frames} (MMF), to solve the FitzHugh-Nagumo equations on curved surfaces \cite{MMF1} \cite{MMF2}. The motivations and analysis of the MMF are displayed in the corresponding reference papers, thus in the remainder of this section, we mention only the motivations of the relative acceleration approach.

\subsection{Trajectory instead of wavefront}

The mathematical analysis of the action potential, or in general the wave propagation in excitable media, was first introduced by means of the kinematics of the wavefront, known as the \textit{kinematics approach}, to analyze the effects of the geometrical shape or anisotropy on the behavior of excitation propagation. As first appeared in ref. \cite{Davydov1991K} in English, the kinematic approach has been used to study the propagation of excitation in inhomogeneous and anisotropic media. The kinematic equation defines the relationship between the curvature of the wavefront and its arc length on curved surfaces. Kinematic analysis focuses on deriving the conditions of the \textit{critical curvature}, or the critical amount of line curvature of the wavefront which results in breaking up the propagation. The kinematics approach has been successfully used to describe the geometrical effects on surfaces such as non-uniformly curved surfaces \cite{Davydov1991A}, periodically modulated curved surfaces \cite{Davydov2000P}, and torus \cite{Davydov2003} as well as ring-shaped propagation on curved surfaces \cite{Davydov2000R} and propagation on moving excitable media \cite{Davydov2004}. In addition, this approach was also used to analyze the role of anisotropy in the plane: the breakup of the propagation \cite {Mikhailov} and the propagation of curved fronts in anisotropic excitable media \cite {Morozov1999}. Recently, the kinematic study using the curvature of the wavefront has been adapted in various ways for cusp waves \cite{Bernus}, scroll wave filaments \cite{Verschelde}, and accurate eikonal-curvature relation \cite{Dierckx}.

However, when considering both anisotropy and curved surfaces as the actual cardiac structure, which will be denoted as \textit{anisotropic curved surfaces}, the kinematics analysis becomes very complicated even on the simplest curved surface. Consequently, the higher dimensional anisotropic space, which is a more realistic approximation of the atrium and ventricle, seems to remain beyond the scope of the kinematic analysis. Consider the following kinematic equation \cite{Davydov1991K} to describe the relationship between the curvature of the wavefront $(K)$ and arc length $(\ell)$ on curved surfaces
\begin{equation*}
\frac{\partial K}{\partial \ell} \left ( \int_0^{\ell} K V d \xi + C    \right )   + \frac{\partial K}{\partial t} + K^2 V + \frac{\partial^2 V}{\partial \ell^2} = - \Gamma V,
\end{equation*}
where $V$ is the normal propagation velocity, $\Gamma$ is the local Gaussian curvature of the surface, and $C$ is the tangential growth velocity of free ends of the front. Even disregarding the anisotropy and the shape of geometry affecting the multiple variables in the above equation, the most difficult complexity arises due to the local Gaussian curvature ($\Gamma$) which depends on both anisotropy and geometry. This complexity remains the same even for the time-independent case such as the spiral wave.

To overcome this problem, the new approach uses the concept of the \textit{trajectory} in the cardiac action potential propagation, instead of the \textit{wavefront} in the kinematic approach. This implies that the propagation delivering the electric signal to cardiac cells will be regarded as a \textit{wave} such as physical waves. Briefly stated, the trajectory is the path of a particle of the wave which is best described in the context of classical mechanics. If we put a particle on the wavefront of the cardiac action potential at time $t_0$, the continuous observation of the particle in time $t>t_0$ generates a trajectory. Consequently, in homogeneous media, the direction of the trajectory is orthogonal to the direction of the wavefront.

The mathematical definition of the trajectory will be given more rigorously in Section 3B, but the \textit{physical} meaning of the trajectory can be ambiguous in the cardiac action potential. The concept of the trajectory requires the presence of traveling particles, but the existence of any particle propagating through myocardial tissues on macroscopic scale has been unknown. There are complex microscopic movements of ions, such as $Na^+, ~K^+,~C\ell^-, Ca^{2+}$, through ion channels and gap junctions, but it has not been accepted that certain kinds of ions can be identified as \textit{continuously} traveling particles. However, in the neighborhood of every point on a sufficiently smooth surface, a trajectory can be uniquely constructed by a smooth wavefront, which is well defined clinically and mathematically. This is another reason we stick to sufficiently smooth surfaces so that the physical concept of the trajectory should remain the same as any other physical wave involving particle motion.

The relative acceleration approach is therefore based on the study of the trajectory in the cardiac action potential propagation; the relative divergence and convergence of each trajectory. Note that the kinematic approach focuses on the shape of the wavefront, while the relative acceleration approach focuses on the distance between the trajectories. From this perspective, the relative acceleration approach can be regarded as a complementary method to the kinematics approach and vice versa, since the distribution of the trajectories reflects the shape of the wavefront. Nevertheless, we will show that the relative acceleration approach can be expressed much more concisely for anisotropic curved surfaces and a higher-dimensional anisotropic space. Maybe the relative acceleration approach is only reasonably useful for higher dimensional space as we glimpse from many successful applications in spacetime physics. The convenience of the relative acceleration approach becomes clear because the trajectory can be easily written in terms of the metric tensor which incorporates anisotropy and curved space into the same geometric term. The comparison between the relative acceleration approach and the kinematic approach is not discussed in this paper, but will be shown in the future publications.

\subsection{Justification on the use of curved surfaces and its properties}
In the rest of this paper, the relative acceleration analysis and the method of moving frames will be given only for curved surfaces. The justification of the use of the surface for the domain is clear from the anatomical property of the PV and the thin layer of the atrium which we are mostly interested in. Contrary to the ventricles which have an average thickness from $12.0$ to $15.0~mm$ for an adult heart \cite {Ho2009}, the left atrium is a thin-wall structure with an average thickness of $1.89 \pm 0.48 ~mm$ ranging from $0.5$ to $3.5~mm$ \cite{Beinart}. The question therefore remains whether the atrium can be uniformly approximated as a curved surface. On the other hand, the PV is well modeled as a curved surface. Hocini et. al \cite{Hocini} established that the cardiac tissues of the PV are thickest close to the atrium and thinner toward the lung, ranging approximately from $0.5$ to $0.8~mm$ that is relatively constant and  less than half the average thickness of the left atrium. The aim of the following computational model is to simulate and observe the behavior of the propagation in the area of the PV, thus it is justifiable to model the atrium with the PV by a smooth curved surface. But this does not mean that the relative acceleration analysis and the method of moving frames are only valid for two-dimensional space. The extension to a higher dimensional anisotropic space can easily be derived and will be reported in later publications.

For the sake of simplicity of analysis, we suppose that the curved surface that will be used in the remaining of this paper is \textit{two-dimensional manifold} and \textit{locally-Euclidean}. Roughly stated, a curved surface is a two-dimensional manifold in the sense that, for the neighborhood of each point on the curved surface, there exists a one-to-one representation of each curved element onto a simply connected region of the Euclidean plane \cite {Cartan}. Moreover, the curved surface is locally-Euclidean in the sense that the curved element can be represented as a small domain of the Euclidean space in a small neighborhood of any point \cite {Cartan2}. Rigorous mathematical definitions will not be repeated here, but can be found the cited references. For example, the surface of revolution, or any surface which is isometric to the surface of revolution, is always manifold and locally Euclidean. This supposition yields the direct consequences of the following properties of the surface that we frequently use in this paper. (1) The first is  \textit{the existence of a pair of orthogonal curved axis at every point}, which is referred as a \textit{surface of an orthogonal net}, and (2) the second is that \textit{scalars, vectors, and tensors are locally continuous and differentiable}.

Since all the derivations and definitions are achieved in a curved element of arbitrary size from a tessellation of the curved surface, the above properties should be well-defined \textit{locally}, but not necessarily \textit{globally}. For example, the existence of the orthogonal curved axis, the existence of orthogonal diffusivity tensor, and the existence of trajectory for smooth wavefront can be easily justified \textit{point-wisely} due to the fact that the surface is assumed to be locally-Euclidean. Accordingly, the quantities and equations in the following sections are given locally within a small $\Delta x$ from each point. Since the use of the surface of an orthogonal net has been standard in studying waves on curved surfaces \cite{Grindrod1991} \cite {Mulhalland1996}, we do not need further mathematical justification for this supposition. Biologically, this is also acceptable because the shape of the cardiac tissue of the atrium and the PV is normally regarded smooth and because the action potential propagates when the tissue is expanded, thus the tissue is likely to be smooth everywhere. Moreover, we will make the problem more generally applicable by using a \textit{locally} orthogonal curved surface rather than a curved surface with globally continuous orthogonal curved axes. Moreover, in the rest of the paper, we suppose that there exists a unique trajectory in the neighborhood of a point $p$ on the sufficiently smooth wavefront on the curved manifold. A rigorous definition of trajectory in the context of classical mechanics and the proposition of the existence and uniqueness of trajectory for each point are shown in Appendix A.

\subsection{Goal, notations, and order of the paper}
The goal of this paper can be summarized as follows: (i) The eikonal equation for the action potential propagation is derived on anisotropic curved surfaces. (ii) The curvature flow of the action potential propagation is expressed and its component is identified for possible conduction failure. (iii) The relative acceleration analysis is proposed and validated from computational simulations or the known results from the kinematic analysis. (iv) The proposition that the PV can be a unidirectional block is analytically proved from the proposed analysis and its computational validations are provided. (v) The reentry mechanism predicted by the proposition is simulated in a simplified PV and atrium structure.

The \textit{wavefront} refers to the isochrone of the depolarization phase in cardiac action potential, while the \textit{propagational direction} refers to the direction of the excitation propagation in line with the velocity vector and is orthogonal to the direction of the wavefront in isotropic media. The propagational direction at a point $p$ at the wavefront represents the infinitesimal change of $p$ during an infinitesimal time interval. Anisotropy has one direction reflecting the continuous alignment of the myocardial fibre and will be denoted as \textit{$x^m$-anisotropy} to represent the anisotropy which is aligned along the $x^m$-axis and only one component of the diffusivity is different from $1.0$. But, anisotropy can be also represented as \textit{both directions} if both components of the diffusivity tensor are different from $1.0$, to possibly represent the discontinuous alignment of the fibre that can be approximated as an isotropic material with a different conductivity. The domain and the propagation behavior is on macroscopic scale if not mentioned otherwise. The following notations are used in this paper. The (double) subscript and (double) superscript index mean that the corresponding quantity is a covariant and contravariant tensor, respectively. In principle, scalars are represented with the regular font without any index and vectors are represented with the bold font, but a scalar variable with the upper index, as the component of a vector, is also used to indicate that the corresponding quantity is a vector. The lists of frequently used notations and definitions are displayed in Table \ref{table1} and \ref{table2}.

The remaining sections are organized as follows: Section 2 describes the FitzHugh-Nagumo equations on anisotropic curved surfaces and by using the properties of the cardiac excitation propagation as the traveling wave solution, the eikonal equation for anisotropic curved surfaces is derived. In Section 3, the hypothesis is proposed on the stooping condition of the propagation and consequently the relative acceleration equation is derived. Section 4 provides the mathematical analysis and validations of the proposed relative acceleration equation. In Section 5 the geometric characteristics of the PV is described and the relative acceleration equation is applied to a model of the PV. Section 6 illustrates the generation of three-dimensional atrial reentry based on this analysis and predictions and discussions follow in Section 7. In addition, the appendix sections are organized as follows: Appendix A describes the meaning of trajectory and required proposition. Appendix B explains the motivation of the relative accelerate analysis from a discrete model and provides an intuitive reasoning of the proposed hypothesis. In Appendix C, the relative acceleration equations are validated on anisotropic plane, anisotropic sphere, and anisotropic torus. Appendix D proves a lemma on the differentiation of $\sqrt{g} g^{kk}$ and Appendix E displays the geometric factors of the surface of revolution of the PV.

 \begin{table}[htdp]
\caption{Definitions and Notations I}
\begin{center}
\begin{tabular}{c | l}
&  \\
\hline
 $P_i, P_i (j) $ & $i$th trajectory, $j$th point in the $i$th trajectory  \\
  $S_j, S_j (i) $ & $j$th wavefront, $i$th point in the $j$th wavefront  \\
  $\lambda$ & Affine parameter to parameterize the trajectory \\
  $n$  &  Selector parameter to parameterize the wavefront \\
  $y^1$ & Trajectory with sufficiently small curvature \\
   $y^2$ & Wavefront with sufficiently small curvature \\	
 $x^k$ & Orthogonal curved axis \\
 $ z^i$ & Cartesian coordinate axis  \\
 $\Pi$ & curved surface \\
 $\pi^e$ & curved element indexed $e$ such that $\Pi = \cup_e \pi^e$
 \end{tabular}
\end{center}
\label{table1}
\end{table}

\begin{figure}[ht]
\centering
\vbox{
\includegraphics[height=4cm, width=6cm] {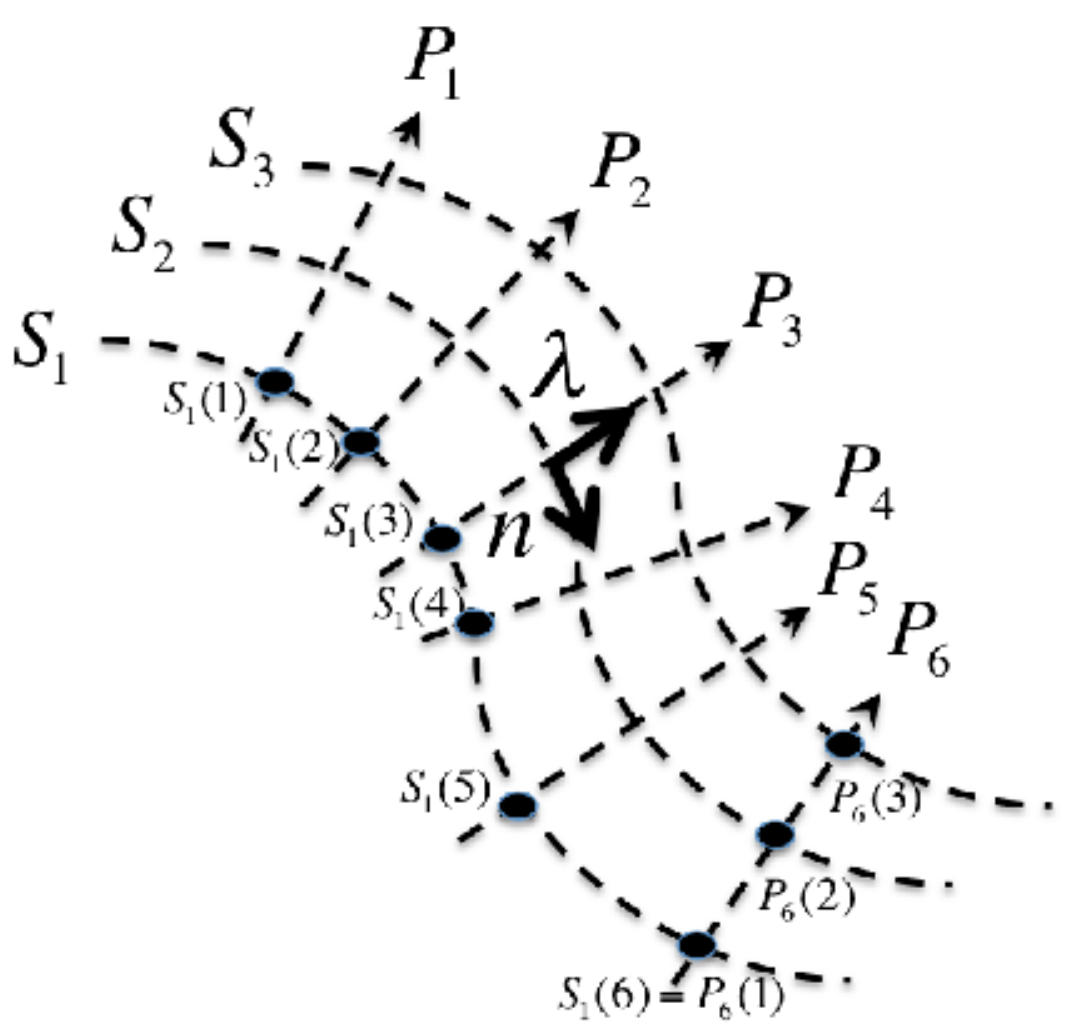} \includegraphics[height=4cm, width=6cm] {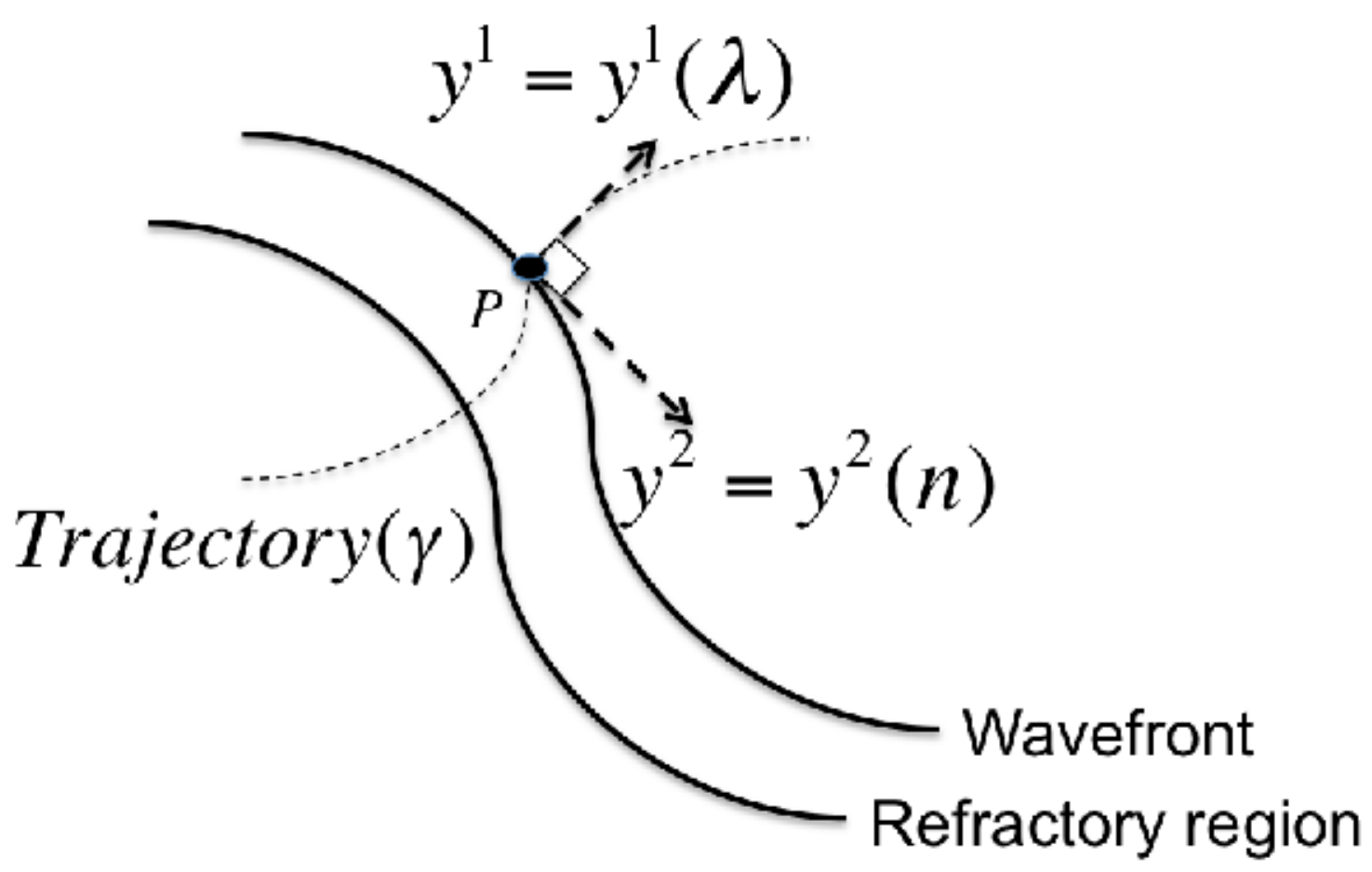}}
\caption{The smooth curve $P_i$ is the trajectory with an affine parameter $\lambda$ and the smooth curve $S_j$ is the wavefront with the selector parameter $n$ (left). $y^1$: direction of the propagation, $y^2$: direction of the wavefront (right).}
\label{fig1}
\end{figure}

\section{Deriving the eikonal equation}

For a smooth curved surface $\Pi$ equipped with anisotropy, suppose that $\Pi$ is a \textit{two-dimensional manifold} which can be divided into a finite number of regions for a one-to-one map from a simply connected region of the Euclidean plane. Moreover, we suppose $\Pi$ is a surface of an orthogonal net with two orthogonal axes $(x^1, x^2)$ and the corresponding metric tensor $g_{ij}$. Consider piecewise orthogonal surfaces $\pi^e$ from the tessellation of the surface $\Pi$ such as,
\begin{equation}
\Pi= \left \{ \cup_e \pi^e | \pi^i \cap \pi^j = \delta^i_j     \right \}, \label{surface}
\end{equation}
where $i,~j$ is the index for $\pi^e$ and $\delta^i_j$ is the Kronecker delta. Note that each $\pi^e$ being equipped with the orthogonal axis $x^k$ constitutes the surface $\Pi$ with two orthogonal axes everywhere, but $x^k$ can be discontinuous across the interfaces of $\pi^e$. The cardiac excitation propagation on anisotropic curved surfaces can be modeled by the following Fitzhugh-Nagumo (FHN) equations \cite{Davydov2000R} \cite{Fitzhugh} \cite{Rosenberg}: in $\mathbf{x} \in \pi^e$,
\begin{align}
\frac{\partial u (\mathbf{x}) }{\partial t} &= \frac{1}{\sqrt{g}} \sum_{\alpha, \beta =1}^2  \frac{ \partial }{\partial x^{\beta} } \left ( \sqrt{g} {d}^{\beta \alpha}  \frac{\partial u (\mathbf{x})}{\partial x^{\alpha} }  \right ) + F(u,v),  \label{FHNu0}  \\
\frac{\partial v (\mathbf{x}) }{\partial t} &= G(u,v),  \label{FHNv}
\end{align}
where $g$ is the determinant of the metric tensor $g_{ij}$. The variable $u$ is an activator and represents the membrane potential, while the variable $v$ is an inhibitor and represents the ion channel openness or the refractoriness. The following analysis is independent of choice of the function $F(u,v)$ and $G(u,v)$ for the shape of the cardiac action potential. For example, we may use the following functions \cite {Rogers} 
\begin{align}
F(u(\mathbf{x}) ,v(\mathbf{x})) &= c_1 u ( u - a) ( 1 - u ) - c_2 v ,     \label{reactionF}\\
G(u(\mathbf{x}),v(\mathbf{x})) &= b_1 ( u- b_2 v ).          \label{reactionG}
\end{align}
The constants $a,~b_1,~b_2,~c_1,~c_2$ determine the excitability of myocardial tissues, but in the rest of this paper, we assume that these variables remain constant in every $\pi^e$ and consequently in the global domain $\Pi$. Thus, inhomogeneity is not considered in this paper. As for anisotropy, we only consider the simple anisotropy that is aligned along one of two axes so that the diffusivity tensor $d^{\beta \alpha}$, written in the orthogonal curved axis $x^{\alpha}$, is expressed as
\begin{equation}
d^{\beta \alpha} = d^{\beta \alpha} \delta^{\beta}_{\alpha} = \left \{
\begin{array}{l}
  d^{\alpha \alpha},~~ \mbox{if}~~ \alpha=\beta  \\
  0,~~~~ \mbox{otherwise}~
  \end{array}
\right .  \label{difftensoroth}
\end{equation}
The justification of this diffusivity tensor is obvious because the domain $\Pi$ is a locally-Euclidean manifold and the cardiac fibre always has a certain orientation in the neighborhood of every point of $\pi^e$. Moreover, the existence of this diffusivity tensor is only enforced in $\pi^e$, not globally in $\Pi$, thus mathematically this is not a strong restriction if two curved axes are chosen arbitrarily in $\pi^e$. The above equality is made possible in the orthogonal curved axis since the diffusivity tensor $d^{\beta \alpha}$ is regarded as a covariant tensor of rank 2, the same as the conjugate metric tensor $g^{\beta \alpha}$. This is justifiable in tensor calculus \cite{Spivak2} \cite{Weatherburn}  and in the context of Riemannian geometry of the cardiac tissue \cite{Panf}. With this diffusivity tensor, equation \eqref{FHNu0} is simplified as
\begin{equation}
\frac{\partial u}{\partial t} =  \frac{1}{\sqrt{g}} \sum_{\alpha =1}^2  \frac{ \partial }{\partial x^{\alpha} } \left ( \sqrt{g} {d}^{\alpha \alpha}  \frac{\partial u}{\partial x^{\alpha} }  \right ) + F(u,v).  \label{FHNu}
\end{equation}
The FHN equations derived in equation \eqref{FHNu} and equation \eqref{FHNv} model the cardiac excitation propagation on anisotropic curved surfaces. To derive the \textit{eikonal equation}, or to extract the dynamics involving motion in the direction of propagation for the PDE, we use the unique property of cardiac excitation propagation as a \textit{traveling wave solution}. The traveling wave property of the membrane potential can be \textit{in vitro} observable from the almost constant shape of the traveling membrane potential in a homogeneous media, it has therefore been widely accepted and used for various analysis for the cardiac action potential propagation \cite {Guerri} \cite {Keener1991}. With this traveling wave of function $\psi$ and a wave-speed $c=c(y^1,y^2)$, the membrane potential $u$ can be expressed as
\begin{equation}
u(y^1, y^2, \tau) = \psi(y^1 - c(y^1,y^2) \tau, y^2), \label{travelingwave}
\end{equation}
where $y^1$ is the path of the propagation and $y^2$ is the isochrone where the membrane potential $u$ is constant with respect to $y^2$ such as ${\partial u} / {\partial y^2} = 0$ as shown in the right plot of Figure \ref{fig1}. Also, $\tau$ indicates the elapsed time from the wavefront with respect to the axis $y^1$ and $y^2$. Consequently, $\tau=0$ indicates the transition layer or just the wavefront for each $\lambda$. The definition of the trajectory for $y^1$ and the wavefront for $y^2$ are also confirmed by differentiating equation \eqref{travelingwave} with respect to $y^2$ to yield $  - ( {\partial \psi} / {\partial y^1} )( {\partial c } / {\partial y^2} ) \tau +  {\partial \psi} / {\partial y^2} = 0 $, which shows that ${\partial \psi} / {\partial y^2}  = 0$ at the moving frame where $\tau=0$. In order to align the trajectory and the wavefront along each axis in $\pi^e$, the following supposition will be required: \\
\\
\textbf{Supposition}: For an orthogonal curved axis $x^{\beta}$ at a point $p \in \pi^e$, the curvature of the trajectory $y^1$ and the wavefront $y^2$ with respect to $x^{\beta}$ is sufficiently small to satisfy, for $\alpha,~\beta = 1,2$
\begin{equation}
\left \| \frac{\partial}{\partial y^{\alpha}} \left ( \frac{\partial y^{\alpha} }{ \partial x^{\beta}}  \right )  \right \| < \varepsilon \ll 0.  \label{supposition}
\end{equation}
This supposition implies that the following analysis is only valid when the wavefront is \textit{slightly curved}. If we consider the procedure of the break-ups of the wavefront as the increase of the curvature up to a critical value (see Appendix B for more details), the use of this supposition means that we are only concerned with the initial deformation of the wavefront which will eventually lead to the break-ups. This causality is supported by the acceleration term in the relative acceleration approach as shown in the later part of this paper. In fact, this supposition has been frequently used in the kinematics approach, as shown in refs. \cite{Davydov1991K}  \cite{Davydov1991A}  \cite{Davydov2000R}, to omit the high order correction terms. The normal propagation velocity $V$ of a curved wavefront is represented by a linear function $V = V_0 - D k$, where $D$ is the diffusivity constant and $k$ is the curvature of the wavefront.

In equation  \eqref{FHNu}, the use of the traveling wave assumption \eqref{travelingwave} yields the following equality at the transition layer where $\tau=0$:
\begin{equation}
\label{Eik2}
  {d}^{11}_y \frac{\partial^2 \psi}{\partial {y^1}^2} + \left [ \frac{1}{\sqrt{g}} \sum_{\alpha=1}^2  \frac{ \partial }{\partial y^{\alpha} } \left ( \sqrt{ g_y} d^{\alpha 1}_y  \frac{\partial \psi}{\partial y^1 }  \right )  +  c \right ]  \frac{\partial \psi}{\partial y^1 } + F = 0 ,
\end{equation}
where the subscript $y$ is used to indicate that the corresponding quantity is expressed with respect to the generally non-orthogonal $y^j$ axis. This implies that $d^{21}_y$ may not be zero in spite of our assumption of the orthogonal diffusivity tensor \eqref{difftensoroth}. Since the axis $y^j$ depends on the shape of the wavefront, the above eikonal equation is written in a time-dependent \textit{moving axis} $y^j = y^j(t)$ for $\tau=0$ on the curved surface. Remember that this eikonal equation is similar to the three dimensional eikonal curvature equation by Keener \cite{Keener} with one main difference. Due to the difference between the orthogonal curved axis and the Euclidean axis, the component $ ( 1 /  \sqrt{ g_y } ) ( \partial \sqrt{ g_y } / {\partial y^{\alpha} }  ) {d}^{\alpha 1}_y$ in the above equation replaces the following term in the three dimensional eikonal curvature equation, $ ( \partial x^k /  \partial y^1) (\partial / \partial y^{\alpha} )$ $\left ( \partial y^1 / \partial x^k \right ) {d}^{\alpha 1}_y $. Equation \eqref{Eik2} is well defined in $\Pi$, but is not convenient for the further analysis because the axis $y^{j}$ depends on the behavior of the propagation such as the initialization of the propagation and subsequently the shape of the wavefront and the time variable, thus is neither orthogonal or time-independent. Because of this, it is inconvenient to express all the tensors with the axis $y^{\alpha}$. Instead, we express every tensors on a fixed and orthogonal curved coordinate axis $x^{\alpha}$ which is independent of the direction of propagation or the direction of the wavefront.

Consider that the diffusivity tensor ${d}^{\alpha \beta}_y$ is expanded with respect to the curved axis $x^k$ such as \cite {Spivak2} \cite {Weatherburn} 
\begin{equation}
{d}^{\alpha 1}_y  \left ( \equiv \sum_{k,\ell=1}^2  d^{k \ell} \frac{\partial y^{\alpha} }{\partial x^k} \frac{\partial y^{1}} {\partial x^{\ell} } \right ) = \sum_{k=1}^2 d^{k k} \frac{\partial y^{\alpha} }{\partial x^k} \frac{\partial y^{1}} {\partial x^{k} } ,   \label{difftensor}
\end{equation}
where the diffusivity coefficient $d^{kk}$ is expressed with respect to $x^j$ and the second equality is obtained from our choice of anisotropy that is only aligned along one of the orthogonal curved axes (equation \eqref{difftensoroth}). As for $\sqrt{g_y}$, the transformation rule by multiplying the determinant of the Jacobian $J \equiv \left[ { \partial x^j} / {\partial y^i }\right ]$ applies as $\sqrt{g_y} = J \sqrt{g}$ \cite {Lipschultz}. As a result, by substituting the equations \eqref{difftensor} into equation \eqref{Eik2}, we obtain
\begin{equation}
\label{Eik2f}
\left ( \sum_{k=1}^2 E^k \right )  \frac{\partial^2 \psi}{\partial {y^1}^2} + \left \{ \sum_{k=1}^2 \frac{1}{\sqrt{g}} \frac{ \partial {U}^k }{\partial x^k } +  c_r \right \} \frac{\partial \psi}{\partial y^1 } + F = 0 ,
\end{equation}
where we introduced the new variables $\Lambda^k \equiv  {\partial y^1} / { \partial x^k}$, ${U}^k \equiv \Lambda^k \sqrt{g} d^{kk}$, and $E^k = \Lambda^k U^k / \sqrt{g}$ with the speed variable $c_r$ that is defined as
\begin{align*}
 c_r \equiv c + \sum_{k=1}^2  \frac{{U}^k}{\sqrt{g}} \left \{  \frac{1}{J} \frac{\partial J}{\partial x^k} + \sum_{\alpha=1}^2 \frac{\partial} { \partial y^{\alpha} } \left ( \frac{ \partial y^{\alpha}}{ \partial x^k } \right )  \right \}.
\end{align*}
Moreover, according to the supposition \eqref{supposition}, for a sufficiently small positive constant $\varepsilon_1$ and $\varepsilon_2$ and for $k,~\alpha = 1,~2$, the following quantities are sufficiently small;
\begin{equation*}
\left \| \frac{1}{J} \frac{\partial J}{\partial x^k} \right \| < \varepsilon_1 \ll 0 ,~~~\mbox{and}~~~ \left \| \frac{\partial} { \partial y^{\alpha} } \left ( \frac{ \partial y^{\alpha} }{ \partial x^k } \right ) \right \| < \varepsilon_2 \ll 0 .
\end{equation*}
Consequently, the speed function ${c}_r$ is approximately the same as $c$, i.e., ${c}_r \approx c $, independent of the time variable and the geometrical factors such as $d^{kk}$, $\Lambda^k$ and $g_{kk}$.

 \begin{table}[htdp]
\caption{Definitions and Notations II}
\begin{center}
\begin{tabular}{c | l}
&  \\
\hline
$d^{kk}$ & Diffusivity coefficient in curved axis \\
$ \varsigma^{kk}$ & Diffusivity coefficient in the Cartesian coordinate  \\
 $c_r$ & Speed function of the action potential propagation \\
  $J$ & Jacobian from the $y^j$-axis to the $x^k$-axis \\
  $\Lambda^k$ & $\partial y^1 / \partial x^k$ \\
  ${U}^k$ & Tensor defined as $\Lambda^k \sqrt{g} d^{kk}$  \\
  ${E}^k$ & Tensor defined as $  \Lambda^k {U}^k / \sqrt{g} $ \\
  $\mathcal{S}_{kk}$ & Non-symmetric Ricci-type tensor in equation \eqref{tensordef1} \\
  $\mathcal{G}_{kk}$ & Gravitational tensor due to anisotropy in equation \eqref{tensordef2} \\
 \end{tabular}
\end{center}
\label{table2}
\end{table}

\section{The relative acceleration equation}

\subsection{Hypothesis on the stopping condition}

Before proceeding further to the relative acceleration equation, we will briefly describe the meaning of relative acceleration in the bundle of trajectories and explain how this can be translated into the conduction failure of the cardiac action potential propagation. Consequently, this section aims to provide the background and justification of the proposed hypothesis, which plays a critical role for the rest of the analysis, from discrete models similar to cellular automata \cite{Chopard}. Based on the crucial characteristics of the action potential propagation, the validity of the hypothesis is obviously independent of the scale of the system, but the proof remains challenging and beyond the scope of this paper. Thus, we just leave it as a hypothesis for later rigorous validation. 

\begin{figure}[h]
\centering
\vbox{
\includegraphics[height=4.0cm, width=4.0cm] {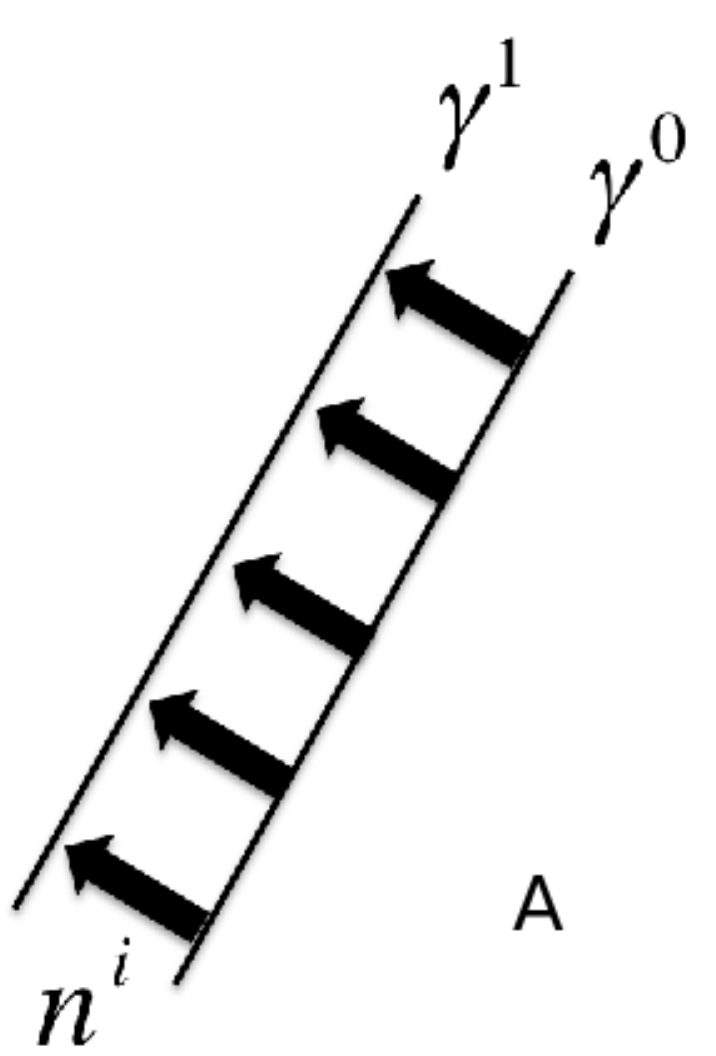} \includegraphics[height=4.0cm, width=4.0cm] {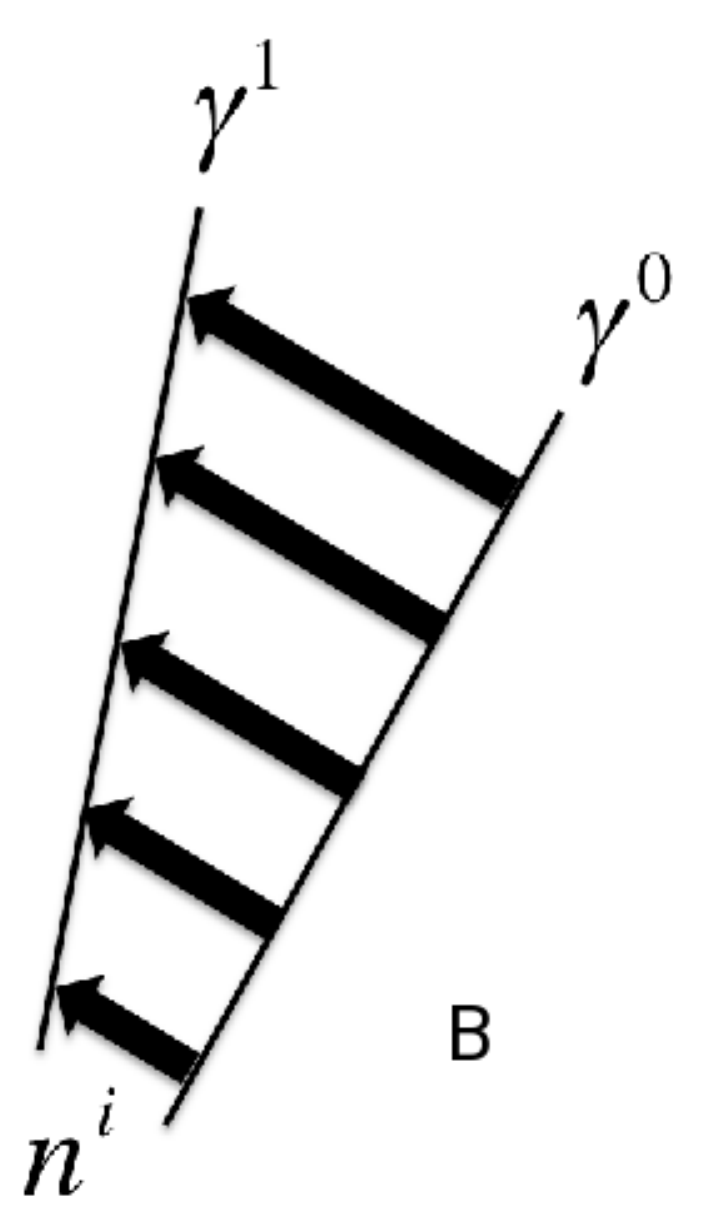} \includegraphics[height=4.0cm, width=4.0cm] {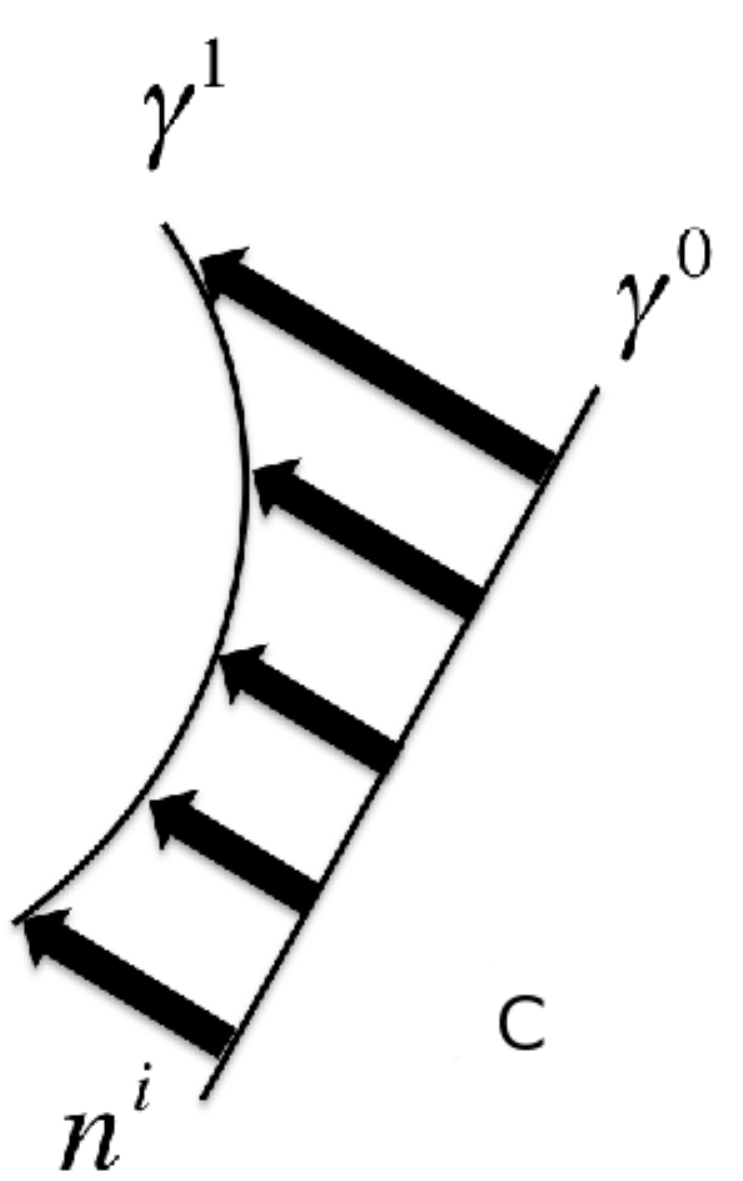}  }
\caption{Three different cases of the relative acceleration in the normal direction. $n^i$ is the separation vector connecting the same $\lambda$ for the trajectories.}
\label{relaccel1}
\end{figure}

Let $P_0$ be a trajectory which is assigned to be fiducial. In the neighborhood of $P_0$, consider the other trajectories $P_k$ for $k = \pm 1, ~\pm 2, ~\pm 3, ~...$ (Figure \ref{relaccel1}). All the trajectories are given the same affine parameter $\lambda$ so that $P_0 (\lambda)$ and $P_k (\lambda)$ lie on the same wavefront $S_{\lambda}$. In other words, the \textit{relative acceleration} of $P_0$ measures how $P_0 (\lambda)$ advances with respect to other $P_k (\lambda)$. Two kinds of relative accelerations can be considered: the first relative acceleration is in the \textit{propagational direction} and the second relative acceleration is in the direction of wavefront, or in the \textit{normal direction}. Relative acceleration in the propagational direction is obvious. Suppose that the trajectories are all parallel. If the rate of changes of $\| P_0(\lambda) - P_0(\lambda-1)\|$ is more than the first order compared to the other trajectories, then $P_0$ is said to be \textit{relatively accelerated in the propagational direction}.

On the other hand, the relative acceleration in the normal direction should be considered for three different cases: let's suppose that the rate of change $\| P_k(\lambda) - P_k(\lambda-1)\|$ is the same for all the trajectories. First, when all trajectories are parallel as shown in Figure \ref{relaccel1}A, the change of the separation vectors (as defined below) along each trajectory is zero, thus there is no relative acceleration in the normal direction. Secondly, if the trajectory $P_0$ linearly diverges from the other trajectories as shown in Figure \ref{relaccel1}B, then the first derivative with respect to $\lambda$ is nonzero, but the second derivative is still zero. Consequently, there is no relative acceleration in the normal direction, either. However, if the trajectory $P_0$ diverges quadratically from the other trajectories as shown in Figure \ref{relaccel1}C, neither of the first nor the second derivative is nontrivial, thus the trajectory $P_0$ is said to be \textit{relatively accelerated in the normal direction}.

More rigorously in mathematics, we use the following definition of the relative acceleration \cite{Misner} which is equivalent to what has been explained:\\
\\
\textbf{Definition}: Let $\mathbf{n}_i \in \Pi$ be the \textit{separation vector} to indicate how one fiducial trajectory is separated from another for the same affine parameter $\lambda$. If the separation vector $\mathbf{n}= \{ n^i \} \in \Pi$ is defined such as
\begin{equation}
\mathbf{n} (i,k) \equiv \lim_{\Delta n \rightarrow 0 } S_{k} ( i + \Delta n) - S_{k} ( i ),  \label{normalvec}
\end{equation}
then the \textit{relative acceleration} can be expressed as
\begin{equation}
\mbox{Relative acceleration} \equiv  \frac{\partial^2 \mathbf{n} }{\partial \lambda^2}  = \left \{ \frac{\partial^2 n^i }{\partial \lambda^2} \right \}.  \label{relaccdef}
\end{equation}

Note that the term \textit{relative} is only relevant to trajectory, not to the propagational velocity. The relative acceleration of trajectory indicates the changes of the propagational velocity along the wavefront, thus the relative acceleration is a point-wise value at the wavefront and does not indicate the relative velocity compared to the propagational velocity of the neighboring points.

A brief intuitive justification of the use of the relative acceleration is provided in Appendix B. We use the fact that the stopping conditions of the cardiac action potential propagation in the discrete models, known as impedance mismatch or sink-source mismatch \cite{Rohr} \cite{Xie}, can be translated as a large relative acceleration in the continuous wave model. In Figure \ref{relaccel2}E, a high ratio between excited cells and excitable cells is interpreted as  the abrupt expansion of the distances between the trajectories which can described as a large relative acceleration in the normal direction, that is, a non-uniform rate of change of speed along the wavefront. In homogeneous media, this means that the distance between particles continues to diverge along the wavefront as the wave propagates. On the other hand, Figure \ref{relaccel2}G represents the wavefronts and trajectories of the second stopping condition, where we observe a relative acceleration in the propagational direction instead of the spreading-out of the trajectories. If the acceleration occurs in the direction of the wavefront, then a large \textit{relative acceleration of trajectory} seems to result in the \textit{deceleration of the propagational velocity}. If the accelerate occurs in the propagational direction, then a large relative acceleration of trajectory seems to result in the acceleration of the propagational velocity, but if the magnitude is sufficiently large, it leads to conduction failure. Without regarding the directions of relative acceleration, conduction failure (stopping condition) for the cardiac action potential propagation can be expressed as the following hypothesis:\\
\\
\textbf{Hypothesis}: A sufficiently large relative acceleration of the cardiac action potential propagation causes conduction failure in the homogeneous and anisotropic myocardial tissues. \\
\\
This hypothesis may not hold in inhomogeneous media such as damaged or malfunctioning myocardial tissue, but has no restriction on the dimension of space or the changes of the conducting velocity. In spite of clear mechanism relating to the relative acceleration, the complexity of the rigorous mathematical proof of this hypothesis forces us to choose whether this hypothesis is valid or even preferred only by the results which it is able to yield, coincident with the previously known conditions or the computational simulations.

\subsection{Deriving the relative acceleration equation}

With the proposed hypothesis, what remains is to obtain the critical factors for the relative acceleration in the action potential propagation. In the ambient space of $\pi^e$, i.e. in $\mathbb{R}^3$, we regard $y^1$-axis as the trajectory and project it into the ambient space of a higher dimension \cite{Cartan2} \cite{Misner}. Let us mathematically define trajectory as follows: \\
\\
\textbf{Definition}: Let $\mathbf{v} = (v^1,v^2,v^3)$ be the velocity of the propagation at the point $p \in \Pi$ which is embedded in $\mathbb{R}^3 = \mathbb{R}^3(z^1,z^2,z^3)$. Then, define the \textit{trajectory} of the vector field by the following differential equations 
\begin{equation*}
\frac{dz^1}{v^1} = \frac{dz^2}{v^2} =\frac{dz^3}{v^3} .
\end{equation*}
Consequently, the trajectory of the cardiac action potential propagation is naturally defined by deriving the tangent vector in $\pi^e$. In the eikonal equation \eqref{Eik2} being originally derived in $\pi^e$, we extend the curved surface $\pi^e$ in the neighborhood of each point in the following way: consider a curved surface as a two dimensional submanifold embedded in three dimensional Euclidean space $z^i$. The function $\psi(y^1, y^2)$ can be extended to a function $\psi(z^1, z^2, z^3)$ in a tubular neighborhood of the surface where the directional derivative of $\psi$ along the surface normal is zero. Define $\psi$ on an open subset around $\pi^e$. Let $\mathbf{Y}_1:\mathbb R^2 \to \mathbb{R}^3$ be the coordinate map for the $y^1$-axis in the ambient space $\mathbb{R}^3$. With the affine parameter $\lambda$ and the selector parameter $n$, we introduce the following notation,
\begin{equation*}
P_n (\lambda) = \mathbf{Y}_1(\lambda, n) \equiv ( \gamma^1_n,~ \gamma^2_n, ~ \gamma^3_n) \in \mathbb R^3,
\end{equation*}
where $\gamma^i_n$ is the component of $\mathbf{Y}_1$ for the Cartesian coordinate axis $z^i$. Moreover, since $\psi(x^1,x^2)$ was extended to $\psi(z^1, z^2, z^3)$, we can express the differentiation with respect to $y^1$ by the chain rule as follows
\begin{equation}
\frac{\partial \psi}{\partial y^1} = \sum_{i=1}^3 \frac{\partial \psi}{\partial z^i} \frac{\partial z^i }{\partial y^1} =  \sum_{i=1}^3 \frac{\partial \psi}{\partial z^i} \frac{\partial \gamma^i_n }{\partial \lambda},  \label{expandR1}
\end{equation}
where the second equality is obtained by using the fact that $\partial z^i / \partial y^1$ is the $i$th component of the tangential vector of the path $\mathbf{Y}_1$ that is the same as $\partial \gamma^i_n / \partial \lambda$. Note that the above equality it nothing but the expression of the derivative in terms of the moving frames as $\nabla \psi \cdot \mathbf{v}^1$ where $\mathbf{v}^1$ is the tangent vector of the $y^1$-axis \cite{Cartan}. Moreover, differentiating again with respect to $y^1$ yields
\begin{equation}
\frac{\partial^2 \psi}{ {\partial y^1}^2} = \sum_{i=1}^3 \left ( \frac{\partial \psi}{\partial z^i} \frac{\partial^2 \gamma^i_n}{\partial \lambda^2} +   \sum_{j=1}^3 \frac{\partial^2 \psi}{\partial z^i \partial z^j } \frac{\partial \gamma^j_n}{\partial \lambda} \frac{\partial \gamma^i_n}{\partial \lambda}  \right ) .  \label{expandR2}
\end{equation}
By substituting the equalities \eqref{expandR1} and \eqref{expandR2} into equation \eqref{Eik2f}, we obtain
\begin{equation}
\frac{\partial \psi}{\partial z^i} \left\{ E^k  \frac{\partial^2 \gamma^i_n}{\partial \lambda^2} +  \left ( \frac{1}{\sqrt{g}} \frac{ \partial U^k }{\partial x^k } + {c}_r \right )  \frac{\partial \gamma^i_n}{\partial \lambda} \right \}  + E^k  \frac{\partial^2 \psi}{\partial z^i \partial z^j } \frac{\partial \gamma^j_n}{\partial \lambda} \frac{\partial \gamma^i_n}{\partial \lambda}  +  F = 0,  \label{Eik2b}
\end{equation}
where the summation notation is used for easier reading and the index $i$ and $j$ is summed up to three and the index $k$ is summed up to two. To obtain the relative acceleration equation, we differentiate the above equation with respect to the selector parameter $n$ along the wavefront. Considering that the reaction function $F$ is constant along the wavefront, leading to $\partial F / \partial n = 0$, we obtain
\begin{equation}
 \frac{\partial {\psi}}{\partial z^i} \left \{ E^k  \frac{\partial^2 {n}^i}{\partial \lambda^2}  + \frac{\partial E^k }{\partial n} \left ( \frac{\partial v^i}{\partial \lambda} + \frac{\partial^2 \psi}{\partial z^i \partial z^j } v^j v^i  \right ) + \frac{\partial }{\partial n} \left [  \left ( \frac{1}{\sqrt{g}} \frac{ \partial U^k }{\partial x^k } + {c}_r \right ) v^i \right ] \right \} = 0  , \label{Eik5}
 \end{equation}
where we used the variable $n^i$ for the \textit{separation vector} as defined in equation \eqref{normalvec} and $v^i$ for the \textit{tangent vector} of the trajectory $\mathbf{Y}_1$, which are defined as $ n^i \equiv {\partial \gamma_n^i } / {\partial n}$ and $v^i \equiv {\partial \gamma_n^i } / {\partial \lambda}$ for $1 \le i \le 3$. Each upper index indicates the component of vector in the Cartesian coordinate $z^i$, for example, $\mathbf{n} = (n^1,~n^2,~n^3)$. In addition, we used the interchangeability of the differentiation \cite{Misner} in the direction of $n$ and $\lambda$ for the above equation such that
\begin{equation*}
\frac{\partial }{\partial n} \left ( \frac{\partial^2 \gamma_n^i}{\partial \lambda^2}  \right ) = \frac{\partial }{\partial n} \left ( \frac{\partial }{\partial \lambda}   \right ) \left ( \frac{\partial }{\partial \lambda}   \right ) \gamma_n^i  =\left ( \frac{\partial }{\partial \lambda}   \right ) \left ( \frac{\partial }{\partial \lambda}   \right )  \frac{\partial \gamma_n^i }{\partial n} = \frac{\partial^2 {n}^i}{\partial \lambda^2}  .
\end{equation*}
Without loss of generality, we may consider the case only when the quantity inside the bracket of equation \eqref{Eik5} is zero, independent of ${\partial {\psi}} / {\partial z^i}$. Then we obtain
\begin{equation}
- E^k \frac{\partial^2 {n}^i}{\partial \lambda^2}  = \frac{\partial }{\partial n} \left ( \frac{1}{ \sqrt{g}} \frac{\partial U^k}{ \partial x^k}  v^i \right ) + \frac{\partial {E}^k }{\partial n} \left ( \frac{\partial v^i}{\partial \lambda} + \frac{\partial^2 \psi}{\partial z^i \partial z^j } v^j v^i \right ) +  \frac {\partial ( {c}_r  v^i ) }{\partial n} .  \label{relacceqn}
 \end{equation}
Let us refer to this equation the \textit{relative acceleration equation} of the FHN equation. Note that each component on the right hand side is only dependent on the coordinate axis, not on the initiation of the propagation or on the time variable. As a consequence, the following proposition being based on the hypothesis is just proved. \\
\\
\textbf{Proposition 1}: Suppose that the curved surface $\Pi$ is a locally Euclidean manifold. If the magnitude of equation \eqref{relacceqn} is sufficiently large at a point $p \in \Pi$, then the action potential propagation stops and conduction failure occurs in the neighborhood of the point $p$. \\
\\
A large magnitude of the right hand side in equation \eqref{relacceqn} may also mean a large relative deceleration of trajectory, but the use of an anisotropy larger than $1.0$ meaning a faster conducting velocity than normal tissue, rules out the case of a large deceleration of trajectory in the propagational direction. Remind that a large deceleration of trajectory is related to a large acceleration of the propagation velocity if the acceleration occurs in the direction of the wavefront. Thus, independent of the direction of deceleration, the propagation does not stop. Note that the relative acceleration does not have to occur in a large area simultaneously for conduction failure, but it suffices for it to occur only at one point since the propagational failure can subsequently diffuse from one point to other points along the wavefront. This can be easily pictured by the \textit{game} explained in Appendix B.

\section{Analysis and validations}

\subsection{Analysis of the relative acceleration equation}
In this section we study each component to verify which component is the major contributor to the magnitude of the relative acceleration.\\
\\
\textbf{(1) Velocity along the wavefront}: Taking into account the fact that the propagation is slowly varying (as frequently used in the biological propagation as well as in cardiac electrophysiology \cite{Keener}), the supposition \eqref{supposition} on the small changes of deformation with respect to the curved axis also implies the relatively small variations of the traveling velocity $c_r$ along the wave front. Moreover, this supposition is another expression of the constant magnitude of the tangent vector, so we notice that the last term of equation \eqref{relacceqn} is also sufficiently small, i.e.
\begin{equation}
 \frac {\partial ( {c}_r  v^i ) }{\partial n}  < \varepsilon \ll 0  .  \label{relcomp1}
 \end{equation}
\textbf{(2) Static electric force and acceleration by media}: The component $( \partial^2 \psi / {\partial z^i \partial z^j } )$, the twice differentiation of $\psi$ with respect to the spacial coordinates, indicates the magnitude of the \textit{static electric force} produced by the shape of the cardiac action potential. Considering the smooth and uniform shape of the cardiac action potential and the relatively weak amount of the electric voltage, the static force is relatively small and uniform along the wavefront compared to other dynamic forces. However, the acceleration term is not trivial because of the presence of anisotropy. For the same reason, $\partial E^k / \partial n$ is not trivial either. Thus, again resorting to the small deformation of the wavefront, the second term can be approximated as
\begin{equation}
\frac{1}{E^k} \frac{\partial {E}^k }{\partial n} \left ( \frac{\partial v^i}{\partial \lambda} + \frac{\partial^2 \psi}{\partial z^i \partial z^j } v^j v^i \right ) \approx \Lambda^k \frac{\partial ( \log {d^{kk} ) } }{\partial n}  \frac{\partial v^i}{\partial \lambda} .  \label{relcomp2}
\end{equation}
\textbf{(3) Curvature flow}: For simplicity, first consider the case $d^{kk} = g^{kk}$. Let the index $m$ indicate another index different from $k$ such that $k \neq m$. With the same property of the propagation as used in equation \eqref{relcomp1}, the first term in equation \eqref{relacceqn}, after being divided by $E^k$, is derived as 
\begin{equation}
\frac{1}{E^k}  \frac{\partial }{\partial n}  \left ( \frac{1}{ \sqrt{g}} \frac{\partial U^k}{ \partial x^k}  v^i \right ) = \frac{g_{kk}}{\Lambda^k}  \frac{\partial}{\partial x^m } \left [ \frac{1}{\sqrt{g}} \frac{\partial }{\partial x^k} \left ( \frac{g_{mm}}{\sqrt{g}} \right ) \right ] v^i.  \label{curvatureflow1}
\end{equation}
Note that the above tensor is very similar to the \textit{Ricci curvature tensor} which is obtains from the Riemannian curvature tensor $\mathcal{R}_{\gamma \alpha \beta \delta}$ on curved surfaces \cite{Eisenhart}
\begin{equation*}
\mathcal{R}_{\alpha \beta} = - (g_{\alpha \beta} / g ) \mathcal{R}_{1212} =  \frac{ g_{\alpha \beta } }{2 \sqrt{g} } \sum_{k=1}^2  \frac{\partial}{\partial x^k } \left [ \frac{1}{\sqrt{g}} \frac{\partial g_{mm} }{\partial x^k} \right ].
\end{equation*}
Since the only significant difference is the differentiation of $x^m$ for the bracket, not $x^k$, this means that if we differentiate equation \eqref{Eik2b} with respect to the affine parameter $\lambda$, we obtain the Ricci curvature tensor. Due to this similarity, we may define \textit{non-symmetric Ricci-type tensor} $\mathcal{S}_{kk}$, by considering the original Ricci tensor \textit{symmetric}, such that 
\begin{equation}
\mathcal{S}_{kk}  \equiv  g_{kk }  \frac{\partial}{\partial x^{m} } \left [ \frac{1}{\sqrt{g}} \frac{\partial  }{\partial x^{k}} \left ( \frac{g_{m m }}{\sqrt{g}} \right )   \right ]. \label{tensordef1}
\end{equation}
In this case, equation \eqref{curvatureflow1} becomes
\begin{equation}
\frac{1}{E^k}  \frac{\partial }{\partial n}   \left ( \frac{1}{ \sqrt{g}} \frac{\partial U^k}{ \partial x^k}  v^i \right )=  ( \Lambda^k )^{-1}  \mathcal{S}_{kk} v^i.   \label{curvatureflow2}
\end{equation}
For general diffusivity tensors, we consider  $d^{kk} = \varsigma^{kk} g^{kk}$ where $\varsigma^{kk}$ is the diffusivity coefficient in the plane and its Cartesian coordinate axis is mapped from each $x^k$. This construction of the diffusivity tensor can be easily verified for a surface of an orthogonal net. The reason we separate the diffusivity coefficient from the conjugate metric tensor is to construct the same strength of anisotropy independent of curvature. As a consequence, $\varsigma^{kk}$ is constant for all the anisotropy, while $d^{kk}$ is not, and this is also coincidence with the clinical observations. Now, equation \eqref{curvatureflow1} is expressed as, 
\begin{equation}
\frac{1}{E^k}  \frac{\partial }{\partial n}   \left ( \frac{1}{ \sqrt{g}} \frac{\partial U^k}{ \partial x^k}  v^i \right ) = ( \Lambda^k )^{-1} \left [   \mathcal{S}_{kk} + \mathcal{G}_{kk} \right ] v^i,  \label{curvatureflow3}
\end{equation}
where we introduce the new tensor $\mathcal{G}_{k k}$ that is defined as
\begin{equation}
\mathcal{G}_{kk}  \equiv   \frac{1}{\varsigma^{kk}}  \frac{\partial^2 \varsigma^{kk}}{\partial x^k \partial x^m}  +  \left ( \frac{\partial \log{g^{kk}} }{\partial x^m}   +  \frac{\partial \log{g^{mm}} }{\partial x^k} + \frac{1}{2}  \frac{\partial {g}}{\partial x^k}  \right )  \frac{\partial \log{\varsigma^{kk}} }{\partial x^k}  .   \label{tensordef2}
\end{equation}
The first component is the second order differentiation of the diffusivity coefficient with respect to the axis and consequently means the \textit{gravitational force} induced by anisotropy. The second component is the first order differentiation of $\log{\varsigma^{kk}}$ with respect to the axis multiplied by the change of the metric tensor in the neighborhood. Thus, we may refer to $\mathcal{G}_{kk}$ a \textit{gravitational tensor}, similarly used in the context of spacetime physics \cite{Misner}.

By combining equations \eqref{relcomp1}, \eqref{relcomp2}, and \eqref{curvatureflow3}, we obtain the new approximation of the relative acceleration equation \eqref{relacceqn}.
\begin{equation}
 - \Lambda^k \frac{\partial^2 {n}^i}{\partial \lambda^2}  =  \left [ \mathcal{S}_{kk}  + \mathcal{G}_{kk}  \right ] v^i + \frac{\partial ( \log {d^{kk} ) } }{\partial n}  \frac{\partial v^i}{\partial \lambda}.  \label{relacceqntensor}
 \end{equation}
The first component represents the relative acceleration induced by the velocity $v^i$, while the second component represents the relative acceleration induced by the acceleration $\partial v^i/ \partial \lambda$. Therefore, we have the following proposition:\\
\\
\textbf{Proposition 2}: In an isotropic locally Euclidean manifold, the conduction failure of the action potential propagation \textit{only} happens when the Ricci-type tensor $\mathcal{S}_{kk}$ for the curvature flow on the surface is sufficiently large. In an anisotropic locally Euclidean manifold, the conduction failure can happen, regardless of $\mathcal{S}_{kk}$, when the acceleration caused by anisotropy is sufficiently large or when the gravitational tensor $\mathcal{G}_{kk}$ induced by anisotropy is sufficiently large.\\
\\
Nevertheless, for simple curved surfaces which are frequently observed in cardiac structure, conduction failure by $\mathcal{S}_{kk}$ seems to be very rare or even impossible according to numerous observations and analyses leading to the following conjecture. \\
\\
\textbf{Conjecture}: If cardiac tissue lies on a locally Euclidean manifold, then conduction failure can be only caused by anisotropy, not by the curvature flow $\mathcal{S}_{kk} $. \\
\\
This means that conduction failure cannot be solely caused by the curvature of the surface. A rigorous mathematical proof of this conjecture is not necessary for the remaining analysis and is beyond the scope of this paper. Instead, this conjecture can be exemplified by several models in the following sections and in Appendix C.

\subsection{Validations of \eqref{relacceqntensor} on simple anisotropic surfaces}

To demonstrate the validity of the proposed relative acceleration analysis we choose several simple anisotropic curved surfaces which can be validated by previously known results from the kinematics approach and/or by simple computational simulations to be compared with the relative acceleration analysis from equation \eqref{relacceqn}.

For the computational simulations we used Nektar++ \cite{Nektar}, which is a C++-object-oriented partial differential equation solver in the context of spectral/hp element methods, to implement the method of moving frames \cite{Cartan2} to solve partial differential equations on anisotropic curved surfaces \cite{MMF1} \cite{MMF2}. Consider the mixed formulation of the FHN equations such as for $\mathbf{q},~\mathbf{d} \in \Pi$
\begin{equation*}
\frac{\partial u}{\partial t} =  \nabla \cdot  \mathbf{q}  + F(u,v),~~~\mathbf{q}  = \mathbf{d} \nabla u, \\
\end{equation*}
where $F(u,v) = c_1 u ( u - a)(1- u) - c_2 v,~G(u,v)=b_1 (u - b_2 v),~a = 0.13,~b_1=0.013, ~b_2=1.0, ~c_1=0.26,~c_2=0.1$. For the above FHN equations, 1 time unit is equivalent to 0.63 $ms$ and 1 space unit to $0.99~mm$ \cite{Rogers}. By adapting the orthogonal moving frames $\mathbf{e}^m$ \cite{Cartan} of the unit length, we integrate the above equation with respect to test functions $\varphi \mathbf{e}^m$ which are also defined in $\Pi$, thus we obtain a Galerkin formation of the above diffusion-reaction equation so that, for $\mathbf{x},~\mathbf{e}^m \in \pi^e$,
\begin{align*}
& \int \frac{\partial u (\mathbf{x}) }{\partial t} \varphi dx = \sum_{m=1}^2 \int  q_m(\mathbf{x}) ( \mathbf{e}^m \cdot \nabla \varphi ) dx  + \sum_{m=1}^2 \int_{\partial \pi^e} \tilde{q}_m (\mathbf{x}) ( \mathbf{e}^m \cdot \mathbf{n} ) \varphi ds  + \int  F(u(\mathbf{x}),v(\mathbf{x})) dx, \nonumber \\
& \int q_m(\mathbf{x}) ({d}^m)^{-1} \varphi dx = - \int (\nabla \cdot (\varphi \mathbf{e}^m ) ) u(\mathbf{x}) dx + \int_{\partial \pi^e} (\mathbf{e}^m \cdot \mathbf{n} ) \varphi \tilde{u} ds, ~~~m=1,~2,
\end{align*}
where $d^m = \mathbf{d} \cdot \mathbf{e}^m$ and the \textit{tilde} sign represents the approximated value at the interfaces of edges, known as \textit{flux} in the context of the discontinuous Galerkin methods. The validation and analysis of this scheme is provided in \cite{MMF1} \cite{MMF2} in the context of numerical analysis and is beyond the scope of this paper, so we will leave interested readers to refer to the numerical papers.

As shown in Appendix C, we observe that the relative acceleration equation \eqref{relacceqntensor} provides quantitatively effective predictions on the behavior of the action potential propagation. On various anisotropic curved surfaces such as plane, sphere, and torus, the analyses by equation \eqref{relacceqntensor} coincide with the results by the kinematic approaches. Even for more complex anisotropic surfaces for which the kinematic approach becomes too complicated to analyze, the relative acceleration analysis also explains the conditions of conduction failures of the computational simulations. We may not determine the exact magnitude of anisotropy or the exact radius of the curved lines to cause conduction failure, but we can understand the effect of anisotropy whether it increases the curvature of the surface leading to possible conduction failure or whether it compromises the critical curvature to guarantee the propagation. This analysis could in fact be an effective tool in the consideration of a biological dynamical system with large resiliency and safety factors. The next step is to apply the relative acceleration equation \eqref{relacceqntensor} to a simplified PV geometry with anisotropy and to predict the behavior of the action potential propagation on it.

\section{Geometric analysis on the PV with anisotropy}

\begin{figure}[ht]
\centering
\vbox{
 \includegraphics[height=4cm, width=4cm] {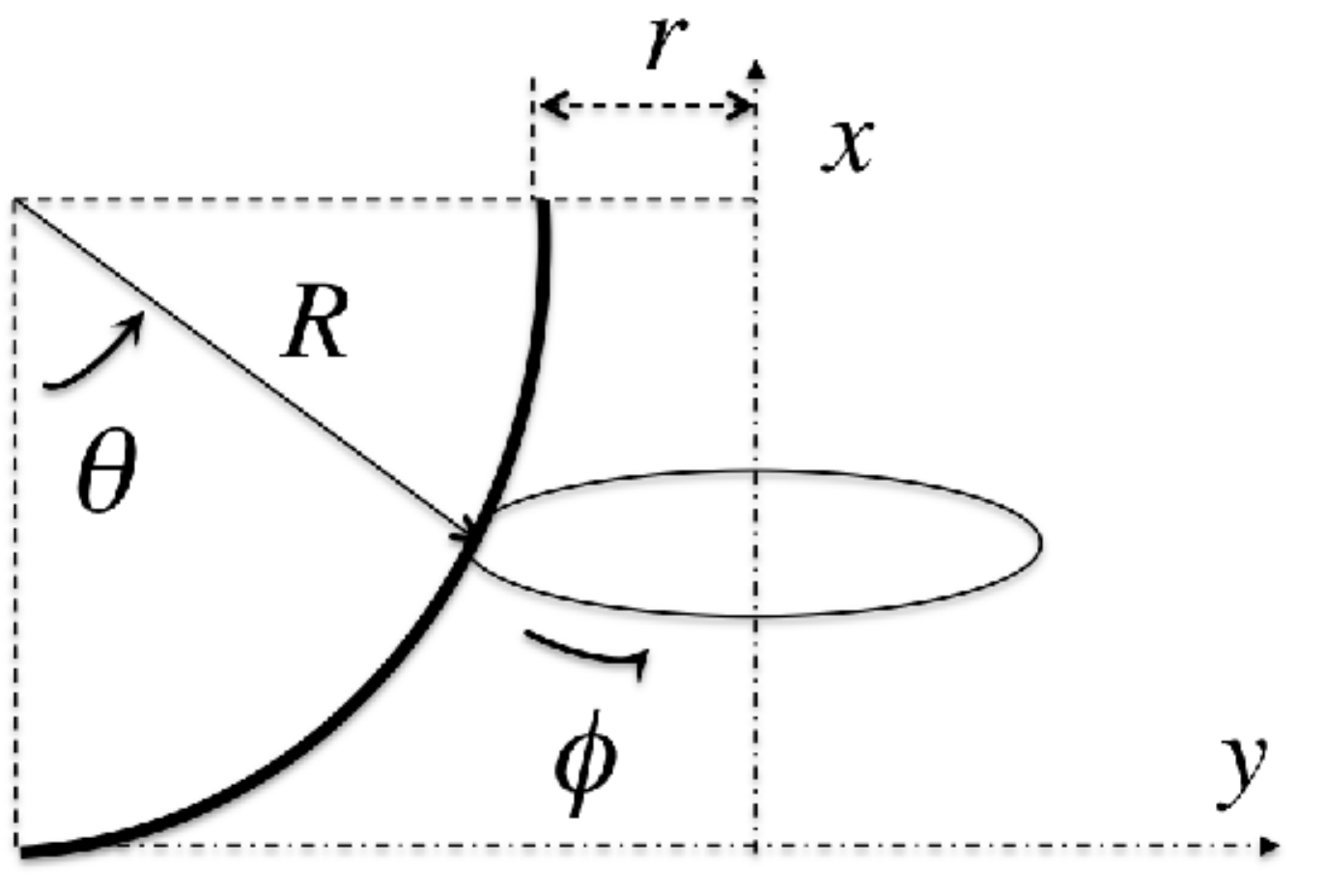} \hspace{1cm} \includegraphics[height=4cm, width=4cm] {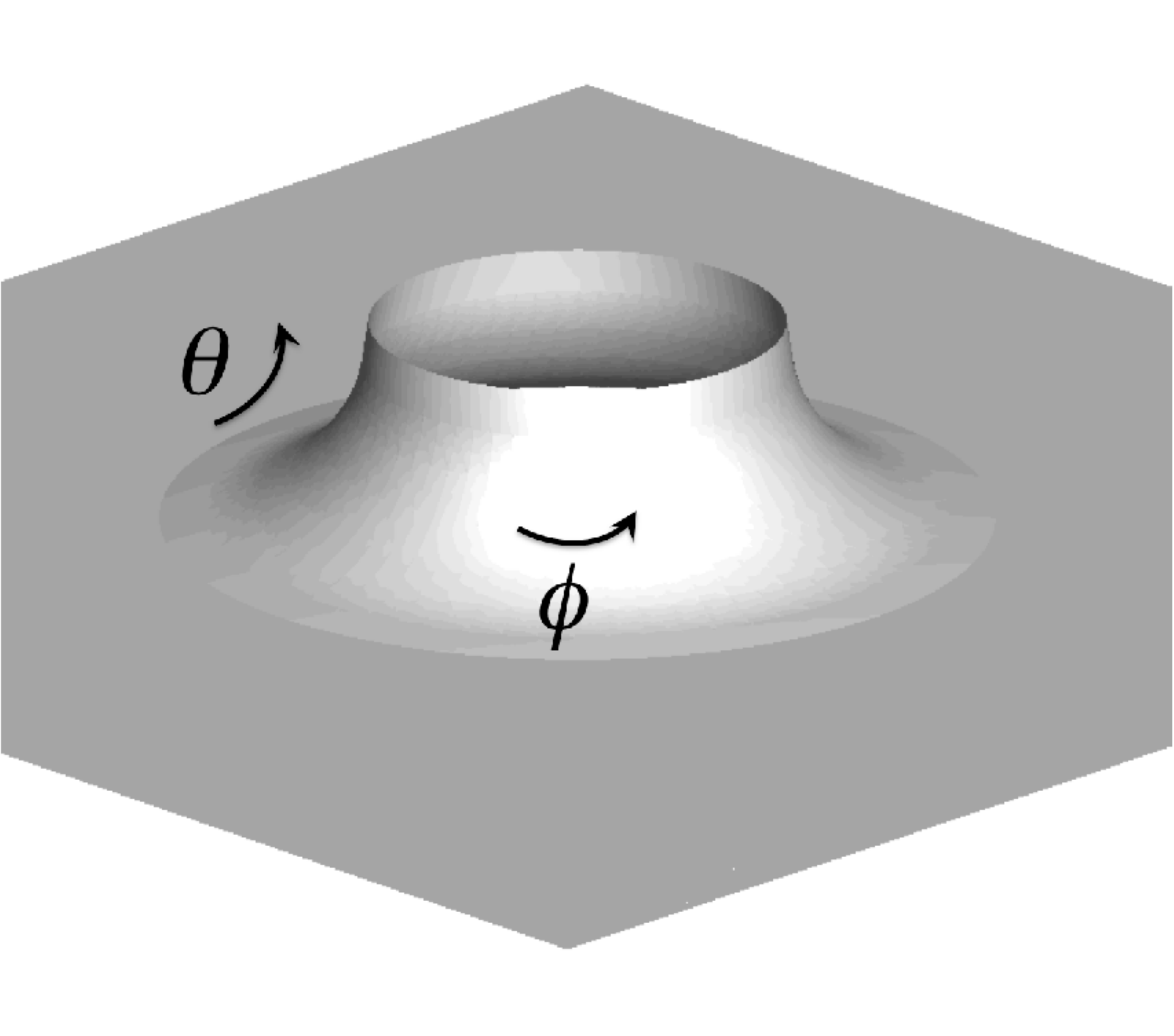} }
\caption{Modeling of the PV by a surface of revolution. The $\phi$-axis is in the circumferential direction of the column and the $\theta$-axis is orthogonal to the $\phi$-axis.}
\label{figgeom2}
\end{figure}

In the following analysis we suppose that the PV and some part of the atrium in the neighborhood of the PV are a locally Euclidean manifold. Also, we suppose for some cases that anisotropy in these area is continuously aligned. This supposition may be viewed as a convenient simplification from the real cardiac anatomy, but this geometry seems to provide sufficient analytical explanations of the reported behavior of the action potential propagation on the PV. Thus in the remainder of this paper, we do not consider other factors such as the variable depth of the tissue, the complex alignment of anisotropy, irregular curvature or the location of the PV which may or may not contribute the critical behavior of the propagation.

\subsection{Geometric characteristics of the PV}

As explained in the introduction, the PV can be well approximated as a surface, but the three-dimensional shape of the PV is very peculiar in the following ways: \\
\\
\textbf{[G-i]} \textit{The PV has a different exterior curvature compared to that of the atrium}. \\
\\
Even when the shape of the atrium is approximated as a simple smooth surface, the PV should be constructed as a separate surface and should be planted on the atrium as a smooth column on a sphere without stretching it. This means that the \textit{intrinsic curvature of the PV} can be similar to that of the atrium, while the \textit{exterior curvature} of it is very different. This is possible because the shape of the PV can be constructed by cutting, pasting and/or turning inside out some part of the atrium without stretching. Mathematically speaking \cite{Lee}, the Gaussian curvature of the PV can be approximately the same as that of the atrium, but the mean curvature can be strikingly different from that of atrium. \\
\\
\textbf{[G-ii]} \textit{The anisotropy of the PV contains at least one discontinuity}. \\
\\
The real atrium has multiple pulmonary veins (four for the human atrium and five or six for the canine atrium), not even mentioning the PV-like geometries such as the inferior vena cava, the superior vena cava, the coronary sinus and possibly the valve annuli. Consequently, analogous to the Hairy-Ball theorem \cite{Hairyball}, it is geometrically impossible to construct a continuous distribution of anisotropy at all interfaces between the PVs and the atrium. This problem is similar to the mathematical construction of a continuous and differentiable vector field on a connected curved surface with $n$ holes, for example, eight holes for the human atrium. As a consequence, the discontinuities of anisotropy inevitably happen at any point of the atrium, but according to anatomical observations they normally appear close to the root of each PV. Hocini et. al. \cite{Hocini} reported that the anisotropy of the canine PVs abruptly changes its direction or is aligned in the mixed orientations at the root of the PVs, but it is aligned along the longitudinal direction as the vein gets closer to the lung. Also, Verheule et. al. \cite{Verheule} reported the circumferential orientation of anisotropy in a vein of the canine atrium being sectioned transversely at the various tips of the PV sleeves. Note that the circumferential orientation of anisotropy in the PV also presupposes the discontinuity of anisotropy at some roots of the multiple PVs. It still remains a question of how to construct or approximate the mixed orientation of anisotropy, but for the sake of simplicity we only consider the longitudinal anisotropy and/or the circumferential anisotropy which has discontinuities at the root of the PV.\\
\\
\textbf{[G-iii]} \textit{Almost all the circumferential lines of the PV are not geodesic and consequently the propagation is not always parallel to the circumferential direction}.\\
\\
Let the circumferential line be aligned along the $\theta$-axis in Figure \ref{figgeom2}. If the propagation only follows the circumferential line, the behavior of the propagation is only affected by the $\theta$-anisotropy. However, the propagation is not likely to follow the circumferential line all the time because the line is not geodesic in isotropic media and not likely to be geodesic even in anisotropic media. The circumferential path at $\theta=\pi/2$ is actually the only geodesic path in the PV because the tangent vector of the other generating curves is not parallel to the axis of revolution \cite{Carmo}. For example consider the root of the PV when $\theta = 0$. When the propagation first reaches the root of the PV without any further acceleration due to anisotropy or curvature, the propagation does not follow the root of the PV because the propagation induced by diffusion naturally follows the geodesic path. As a consequence, the propagation climes up to the middle or upper part of the PV through various $\theta$ and reaches the other end of the PV. In view of propagational angle, this means that the propagation should be at an oblique angle with the orthogonal axis at least at some points of the PV.

\subsection{Surface of revolution for a modeling of the PV}

For the modeling of the PV without compromising its above mentioned geometric properties, a simple surface of revolution will be used as the geometric model of the PV in the plane as shown in Figure \ref{figgeom2}. Consider a circular arc with radius $R$ which is in the distance of $r$ from the center of the column. The $\theta$-axis is aligned along the circular arc and the $\phi$-axis is aligned along the revolving direction. Note that the $\theta$-axis lies in the longitudinal direction and the $\phi$-axis lies in the circumferential direction. Revolve the circular arc around the $x$-axis, i.e. perpendicular to the plane, to generate a smooth column on the plane which has the following parameterization of the surface $\mathbf{x}$: For a $R,~r \in \mathbb{R}^+$ and $0 \le \theta \le \frac{\pi}{2},~ -\pi \le \phi \le \pi$, 
\begin{equation*}
\mathbf{x} = ( R (1- \cos \theta), ( R (1- \sin \theta) + r ) \sin \phi, ( R (1- \sin \theta) + r ) \cos \phi   ). 
\end{equation*}
The corresponding metric tensors and the Christoffel symbols of this surface are displayed in Appendix E. As for anisotropy, a specific orientation of anisotropy can be chosen for some analysis, but we may also consider anisotropy aligned in both directions such that the diffusivity tensor $\mathbf{d}$ is expressed as $\mathbf{d} = \varsigma^{\theta \theta} \boldsymbol{\theta} + \varsigma^{\phi \phi} \boldsymbol{\phi}$. Biologically, anisotropy aligned in both directions may be regarded the media with complex conducting properties. For example, if $\varsigma^{\theta \theta} = \varsigma^{\phi \phi}$, then it is just an isotropic media with conductivity coefficient of $\varsigma^{\theta \theta}$. Moreover, the discontinuously aligned anisotropy can be also expressed as anisotropy of both directions which can be generally approximated as an isotropic media of a different conductivity velocity. From now one, the PV means \textit{this surface of revolution} unless it is indicated specifically as the real PV.

\subsection{Case I: Propagation towards the $\theta$-axis}

\textbf{(1) PV with $\phi$-anisotropy}: Consider a simple case where the propagation follows only the $\theta$-axis of the PV. This type of propagation happens only when the propagation with the planar or convex wavefront reaches the facet of the PV. Let all the anisotropy of the PV be aligned in the circumferential direction such that $\mathbf{d} = \varsigma^{\phi \phi} \boldsymbol{\phi},~~\varsigma^{\phi \phi}>1.0$ in the block of anisotropy and $\varsigma^{\theta \theta}=1.0$ everywhere. 

\vspace{0.3cm}

\begin{figure}[ht]
\centering
\vbox{
 \includegraphics[height=4.0cm, width=4.0cm] {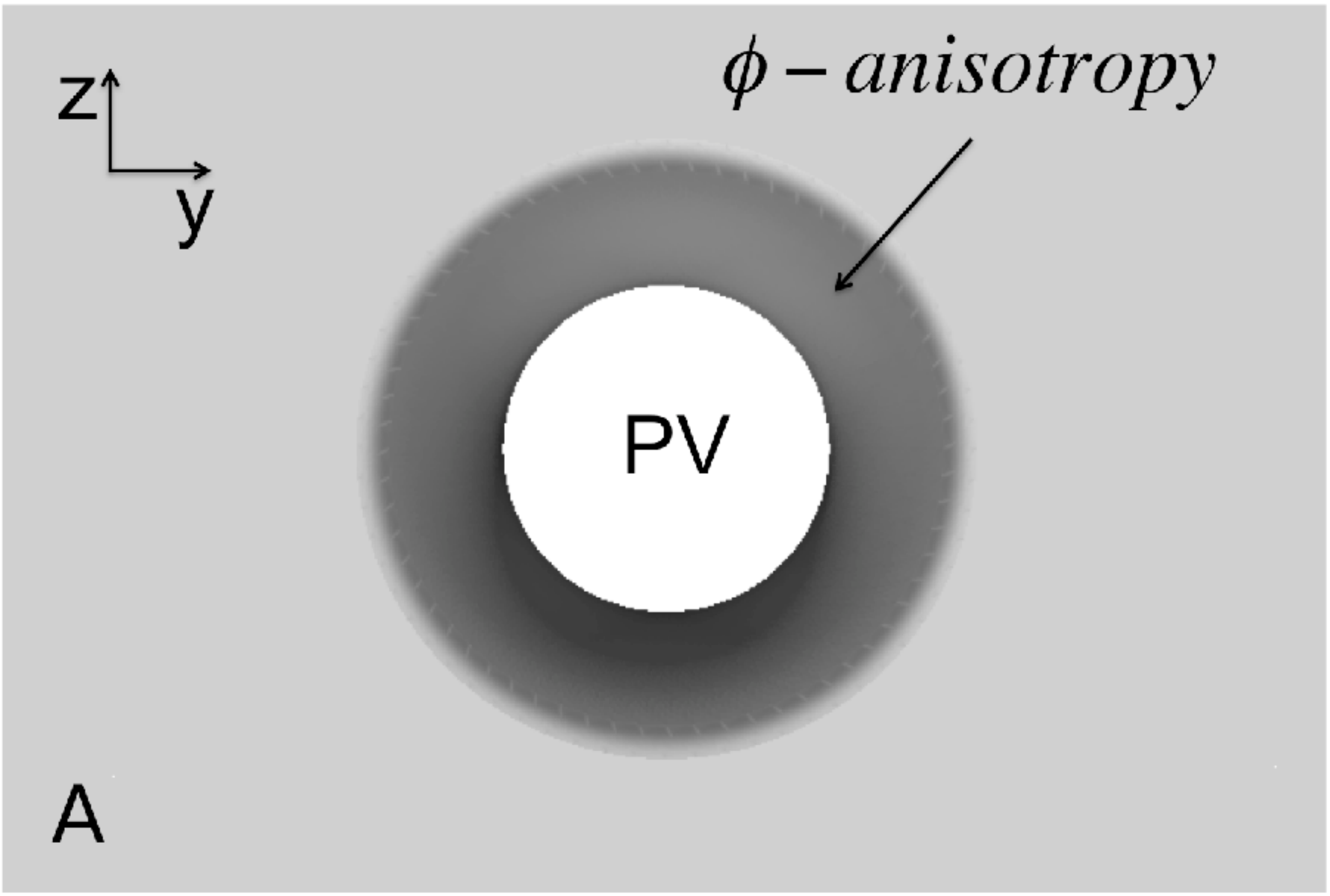}  \includegraphics[height=4.0cm, width=4.0cm] {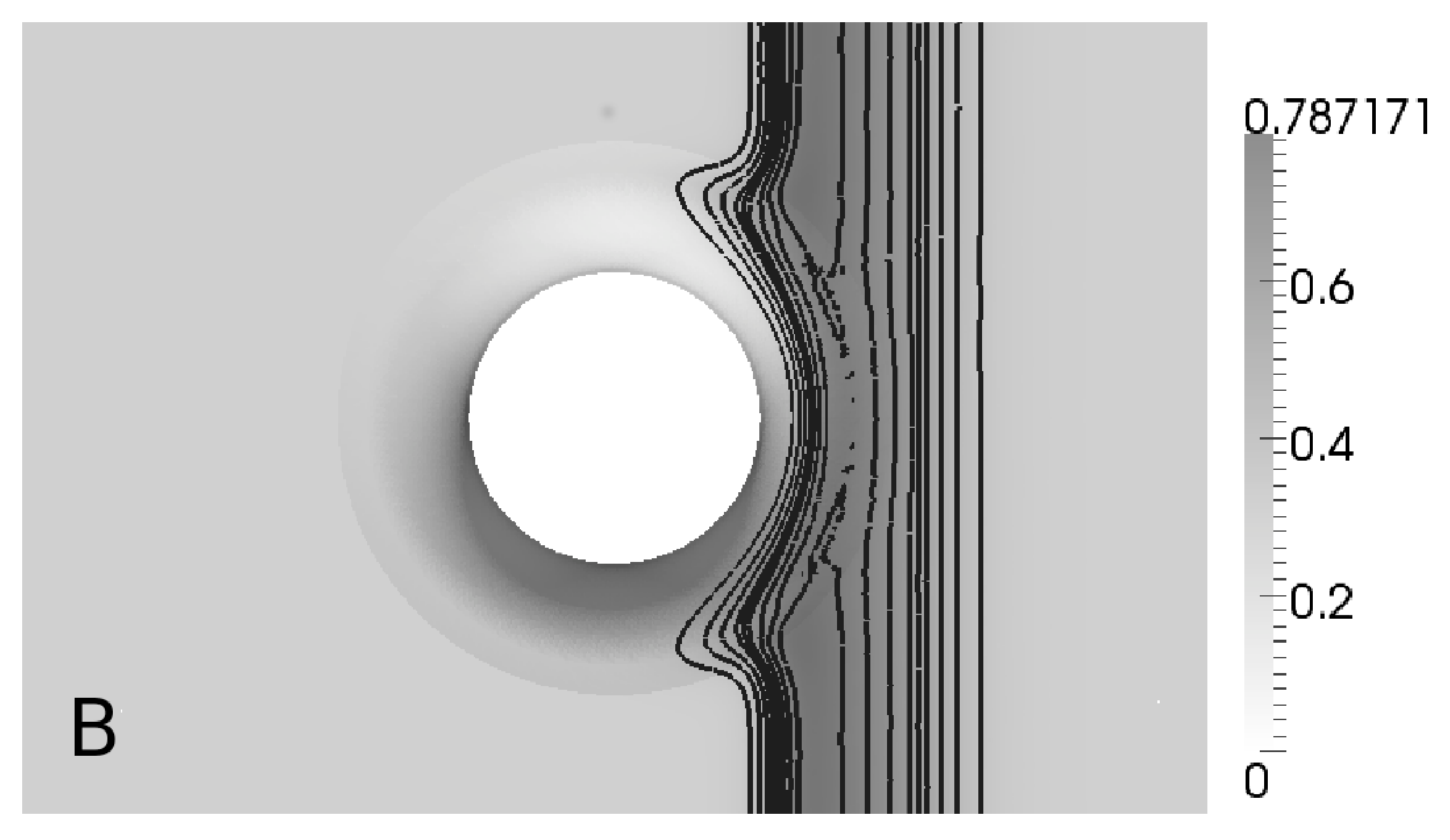}  \includegraphics[height=4.0cm, width=4.0cm] {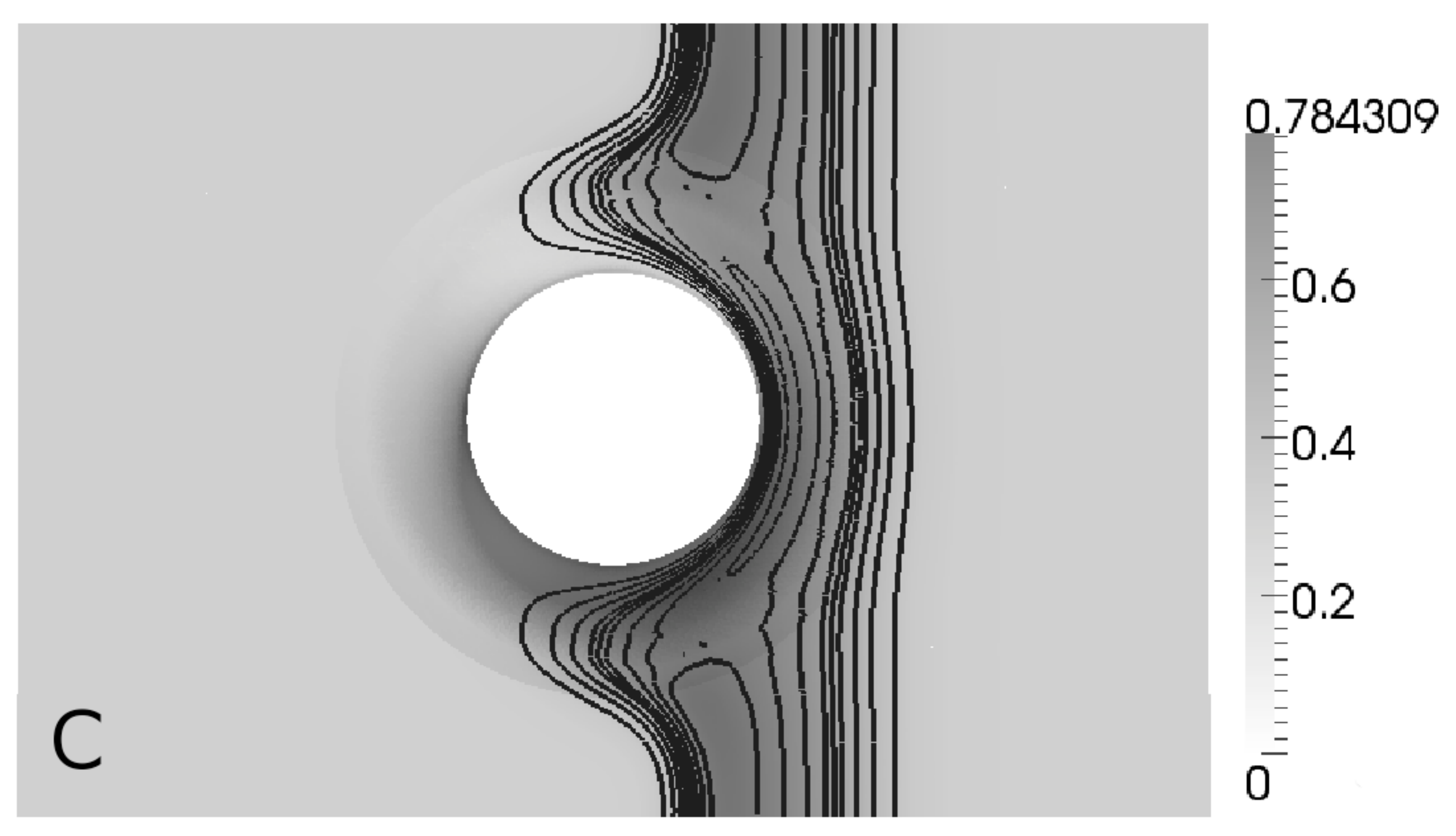} }
\caption{The PV with the $\phi$-anisotropy $\varsigma^{\phi \phi}=8.0$ in the region of $0 \le \theta \le \pi/2$ (A). Initiated from the right wall, the action potential ($u$) at $T = 750.0$ (B), $T=850.0$ (C) represents the slow-down of the propagation in the PV.}
\label{relpvsim1}
\end{figure}

\begin{figure}[ht]
\centering
\vbox{
 \includegraphics[height=4.0cm, width=4.0cm] {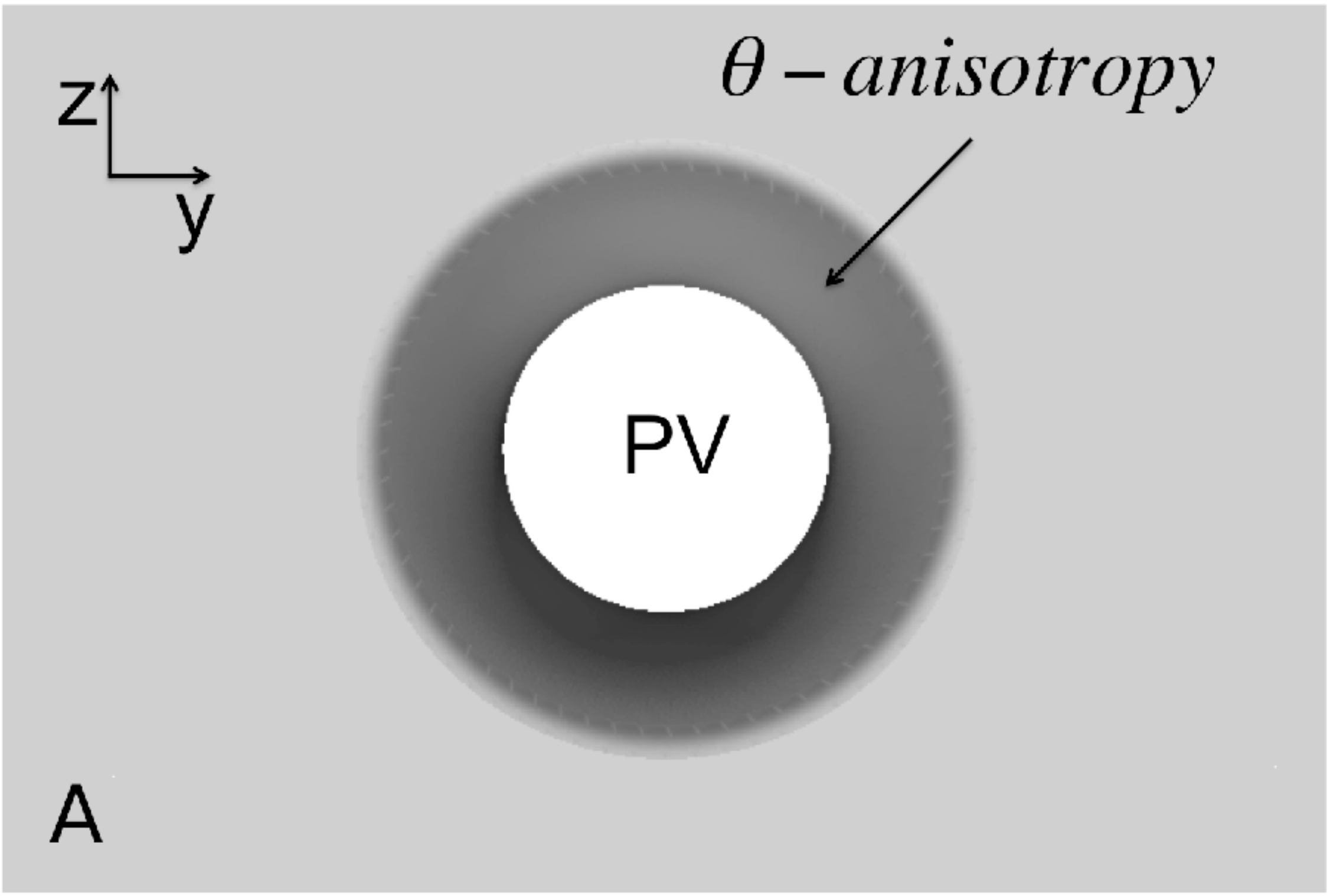}  \includegraphics[height=4.0cm, width=4.0cm] {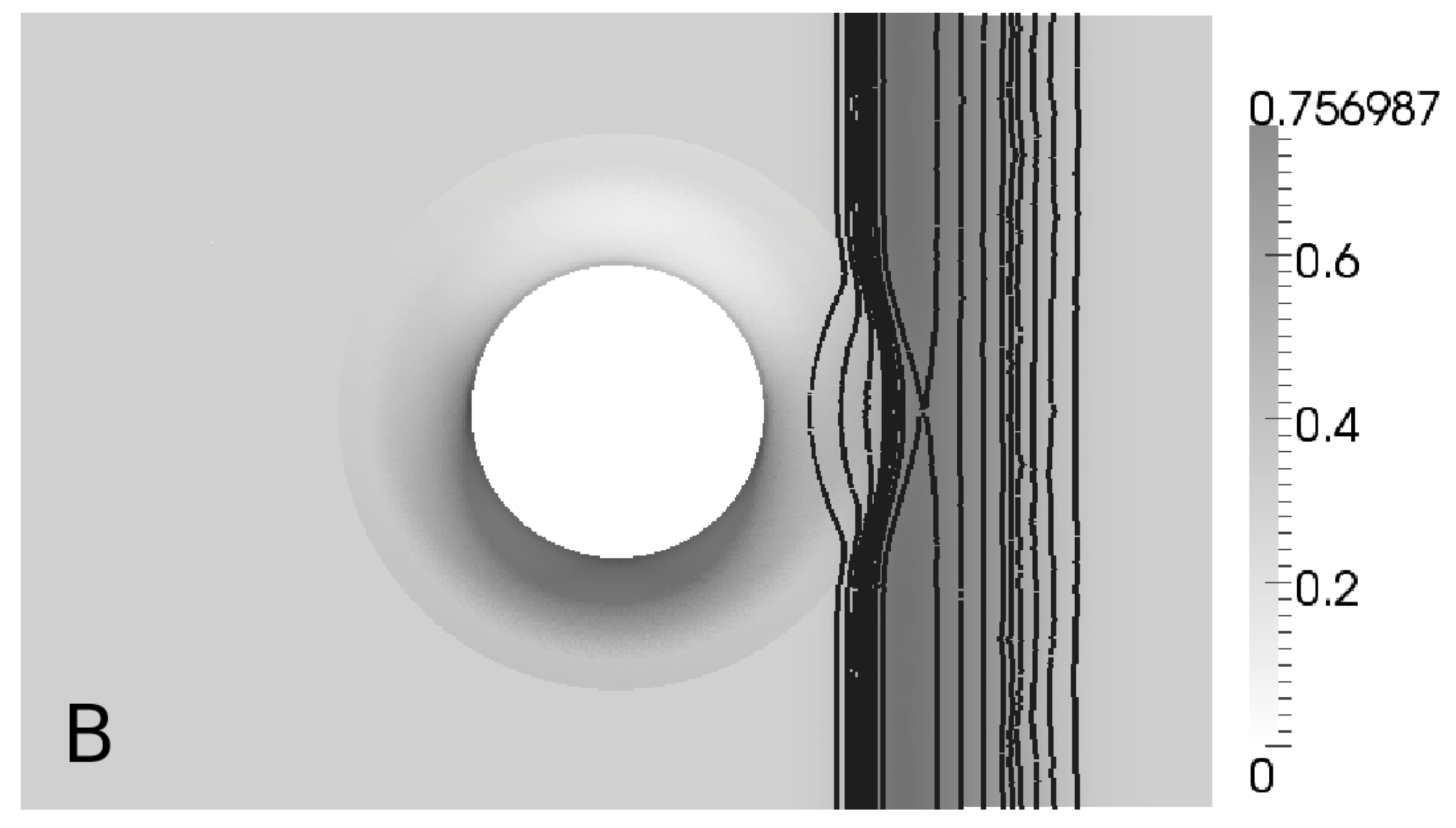}  \includegraphics[height=4.0cm, width=4.0cm] {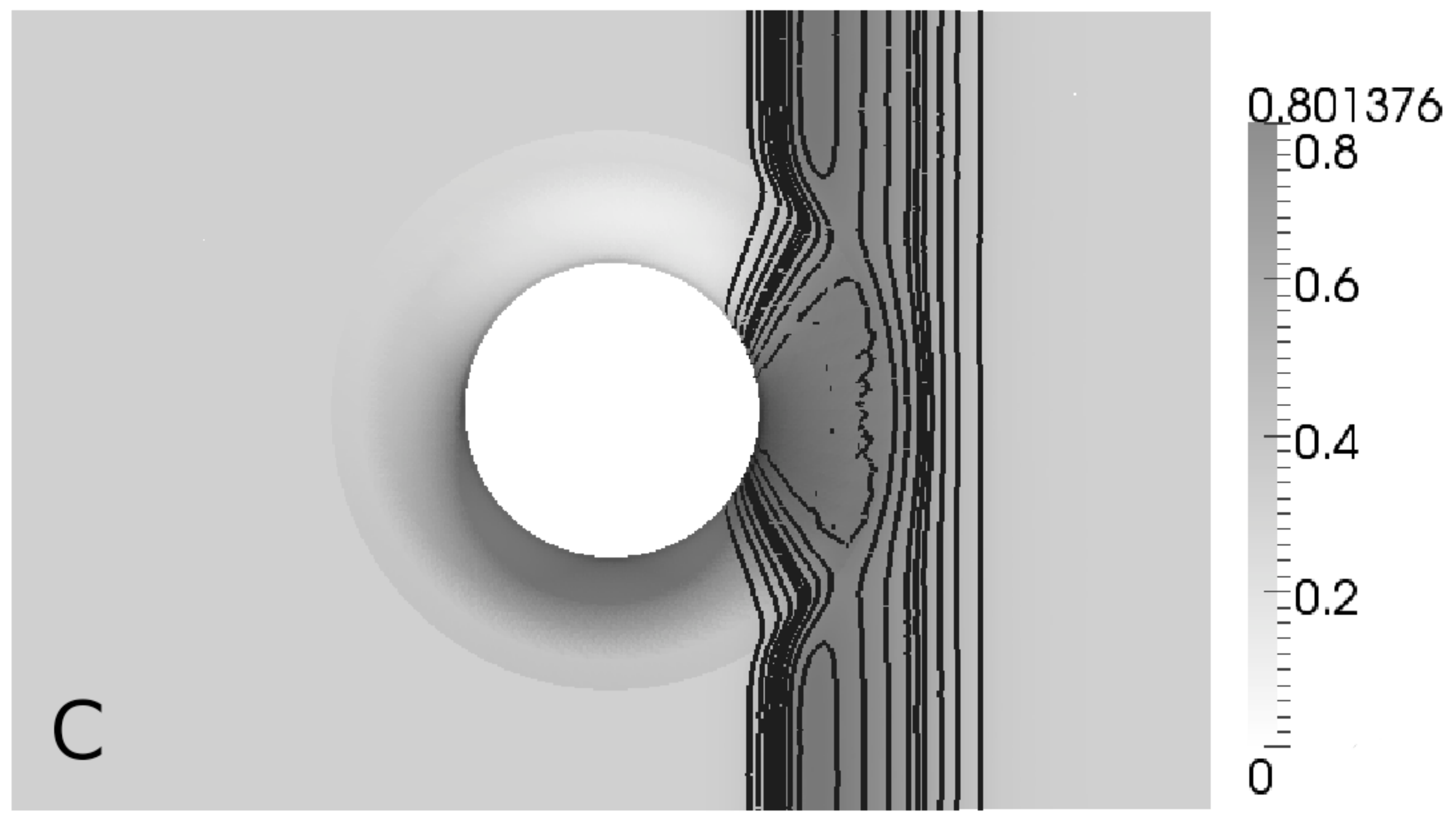} }
\caption{The PV with the $\theta$-anisotropy $\varsigma^{\theta \theta}=8.0$ in the region of $0 \le \theta \le \pi/2$ (A). Initiated from the right wall, the action potential ($u$) at $T = 650.0$ (B), $T=750.0$ (C) displays the acceleration due to the anisotropy, but the general behavior is not much different from the propagation in the anisotropic plane.}
\label{relpvsim2}
\end{figure}

\textbf{RA-analysis}: The above conditions imply that $\Lambda^{\theta}=1.0,~\Lambda^{\phi}=0.0$, and $\varsigma^{\theta \theta}=1.0$ near the root of the PV, i.e., at $\theta \approx 0$. Then, we obtain the relative acceleration equation as
\begin{equation}
- \frac{\partial^2 n^i}{\partial \lambda^2} = - \frac{1 - \sin \theta ( 1 + r / R ) }{(1 - \sin \theta + r / R)^2 }  \left ( \frac{\partial \theta}{\partial n}   \right ) v^i   - \frac{\cos \theta}{ 1 - \sin \theta + r/ R} \frac{\partial v^i}{\partial n}.  \label{relacc1}
\end{equation}
Being derived from the Ricci-type tensor $\mathcal{S}_{kk}$, the magnitude of the first component is relatively small and bounded, thus we also consider the second component which we normally neglect. Since $\partial v^i / \partial n$ is approximately aligned along the direction of the wave front, we can say that the first component is in the propagational direction and the second component is in the direction of the wavefront. Consider the above equation near the root of the PV at $\theta \approx 0$ implying $\partial \theta / \partial n \approx 0$ from $n \approx \phi$. Then, the first component becomes zero and the second component reduces to ${\partial^2 n^i} / {\partial \lambda^2} = {R} /(R+r) {\partial v^i} /{\partial n}$. This indicates that there is no relative acceleration in the propagational direction when it is aligned along the $\theta$-axis. However, a certain magnitude of the relative acceleration can be generated in the direction of the wavefront caused by the geometry of the PV. Because the sign of the second component is positive, this actually means a decrease in the propagational velocity. \\
\\
\textbf{Lemma 1}: The PV with $\phi$-anisotropy decreases the speed of the action potential propagating along the $\theta$-axis.\\
\\
\textbf{Computational modeling}: Figure \ref{relpvsim1} validates this analysis. For the planar propagation approaching the PV from the right wall, the propagational speed decreases in the PV due to the relative acceleration in the direction of the wavefront caused by the geometry of the PV. But, conduction failure does not occur possibly due to the absence of anisotropy. Since the propagation follows the direction of anisotropy in the atrium with increased conduction velocity, the decrease of the conduction speed can be more dramatic in the actual atrium as \textit{in vivo} observed by Arora et. al. \cite{Aora}. This phenomenon seems to confirm the dominance of the circumferential anisotropy in the actual PV as reported in ref. \cite{Verheule}.\\
\\
\textbf{(2) PV with $\theta$-anisotropy, RA-analysis and computational modeling}: We also consider the anisotropy that is aligned along the $\theta$-axis such as $\mathbf{d} = \varsigma^{\theta \theta} \boldsymbol{\theta},~~\varsigma^{\theta \theta}>1.0$ in the block of anisotropy and $\varsigma^{\phi \phi}=1.0$ everywhere. But, for the sake of simplicity, we let $\varsigma^{\theta \theta}$ be constant along the $\phi$-axis. Since these conditions can be expressed as
\begin{align*}
& \frac{\partial \varsigma^{\theta \theta}}{\partial \theta} \gg 0 ~~~~ \mbox{and} ~~~~ \frac{\partial \varsigma^{\theta \theta}}{\partial \phi} = 0 \\
& \frac{\partial \varsigma^{\theta \theta}}{\partial n} = \frac{\partial \varsigma^{\theta \theta}}{\partial \phi} \frac {\partial \phi}{\partial n} +  \frac{\partial \varsigma^{\theta \theta}}{\partial \theta} \frac {\partial \theta}{\partial n} =   \frac{\partial \varsigma^{\theta \theta}}{\partial \theta} \frac {\partial \theta}{\partial n} ,
\end{align*}
with $\partial \theta / \partial n \approx 0$, we obtain the relative acceleration equation at the PV junction where $\theta=0$ such that
\begin{equation*}
 - \frac{\partial^2 n^i}{\partial \lambda^2} = \frac{\partial (\log \varsigma^{\theta \theta})}{\partial n} \frac{\partial v^i}{\partial \lambda}  + \frac{1}{\varsigma^{\theta \theta} } \frac{\partial^2 \varsigma^{\theta \theta}}{\partial \theta^2} v^i  + \left [ \frac{\partial (\log \varsigma^{\theta \theta})}{\partial \theta} - \frac{R}{R+r}  \right ]  \frac{\partial v^i}{\partial n}  .
\end{equation*}
Observe that the first two components are the same as those in the relative acceleration equation \eqref {relplane1} for anisotropy in the plane as shown in Appendix C. The only additional factor is the last component which represents the relative acceleration caused by the unique shape and the anisotropy of the PV. In the last component, $\partial (\log \varsigma^{\theta \theta}) / \partial \theta $ has a nontrivial magnitude as shown in Figure \ref{planeani1}A of Appendix C, but is a negative value which is not summed up with the positive $R/(R+r)$. Intuitively, this component indicates that the relative acceleration in the direction of the wavefront is compromised by the $\theta$-anisotropy. As a consequence, the relative acceleration does not significantly increase. Figure \ref{relpvsim2} displays that the geometry or the $\theta$-anisotropy of the PV does not add a significant amount of additional relative acceleration to yield conduction failure. \\
\\
\textbf{Lemma 2}: The PV with $\theta$-anisotropy does not significantly change the behavior of the action potential propagating along the $\theta$-axis, compared to the propagation of the anisotropic plane.

\subsection{Case II: Propagation with an oblique angle}
\textbf{(3) PV with $\phi$-anisotropy and $\theta$-anisotropy}: Because of the geometric characteristics of the PV (G-iii) that the excitation always propagates at oblique angle with the $\theta$-axis, we must take into account both the $\phi$-direction and the $\theta$-direction to better understand the mechanism of the propagation in the PV. But the consideration of both directions without any simplification may bring incomprehensible complexity into the analysis, thus, for the sake of simplicity, we suppose that the propagational direction and the direction of the wavefront are \textit{locally uniform} along the wavefront. In addition, although we consider both directions of anisotropy due to the second geometric characteristics of the PV (G-ii), we also suppose that the diffusivity tensor $\varsigma^{\theta \theta}$ and $\varsigma^{\phi \phi}$ remain constant along the wavefront in the PV. These simplifications can be regarded as reasonable considering the relatively small width of the myocardial tissue along the $\theta$-axis of the PV and the periodicity of $\varsigma^{\theta \theta}$ and $\varsigma^{\phi \phi}$ along the $\phi$-axis. \\
\\
\textbf{RA-analysis}: The above conditions lead to the following mathematical expressions for $\Lambda^{\theta}$, $\Lambda^{\phi}$, and $\mathbf{d}$ such that ${\partial \Lambda^{\theta}} / {\partial n} = {\partial \Lambda^{\phi}} / {\partial n} = 0$, ${\partial \varsigma^{\theta \theta}}/{\partial n} = {\partial \varsigma^{\phi \phi}}/{\partial n} =0$ at the wavefront in the PV. Moreover, if we consider that the propagational direction is at an angle $\mathcal{A}$ with the $\theta$-axis such as $\Lambda^{\theta} = \cos \mathcal{A}$ and $\Lambda^{\phi} = \sin \mathcal{A}$ for $0 \le  \mathcal{A} \le \pi$, and consequently $\partial \theta / \partial n = - \sin \mathcal{A}$. Then, the above equations can expressed as the function of $\mathcal{A}$ as
\begin{align*}
 \frac{\partial^2 n^i}{\partial \lambda^2} &= \frac{  R^2 \sin \mathcal{A} }{ \varsigma^{\theta \theta } Q^2  {\cos \mathcal{A} }^2  + \varsigma^{\phi \phi} R^2 {\sin \mathcal{A}}^2  }   \left [ \varsigma^{\theta \theta} \cos \mathcal{A}   \left [ R ( 1 - \sin \theta ) - r \sin \theta \right ] v^i - 2 \varsigma^{\phi \phi} { \sin \mathcal{A} }^2 \frac{ R \cos \theta}{Q} \frac{\partial v^i}{\partial \lambda}  \right ] ,   
\end{align*}
where we use the new variable $Q \equiv R ( 1 - \sin \theta ) + r$ ignoring the $\partial v^i / \partial n$ component.

\begin{figure}[ht]
\centering
\vbox{
 \includegraphics[height=4cm, width=4cm] {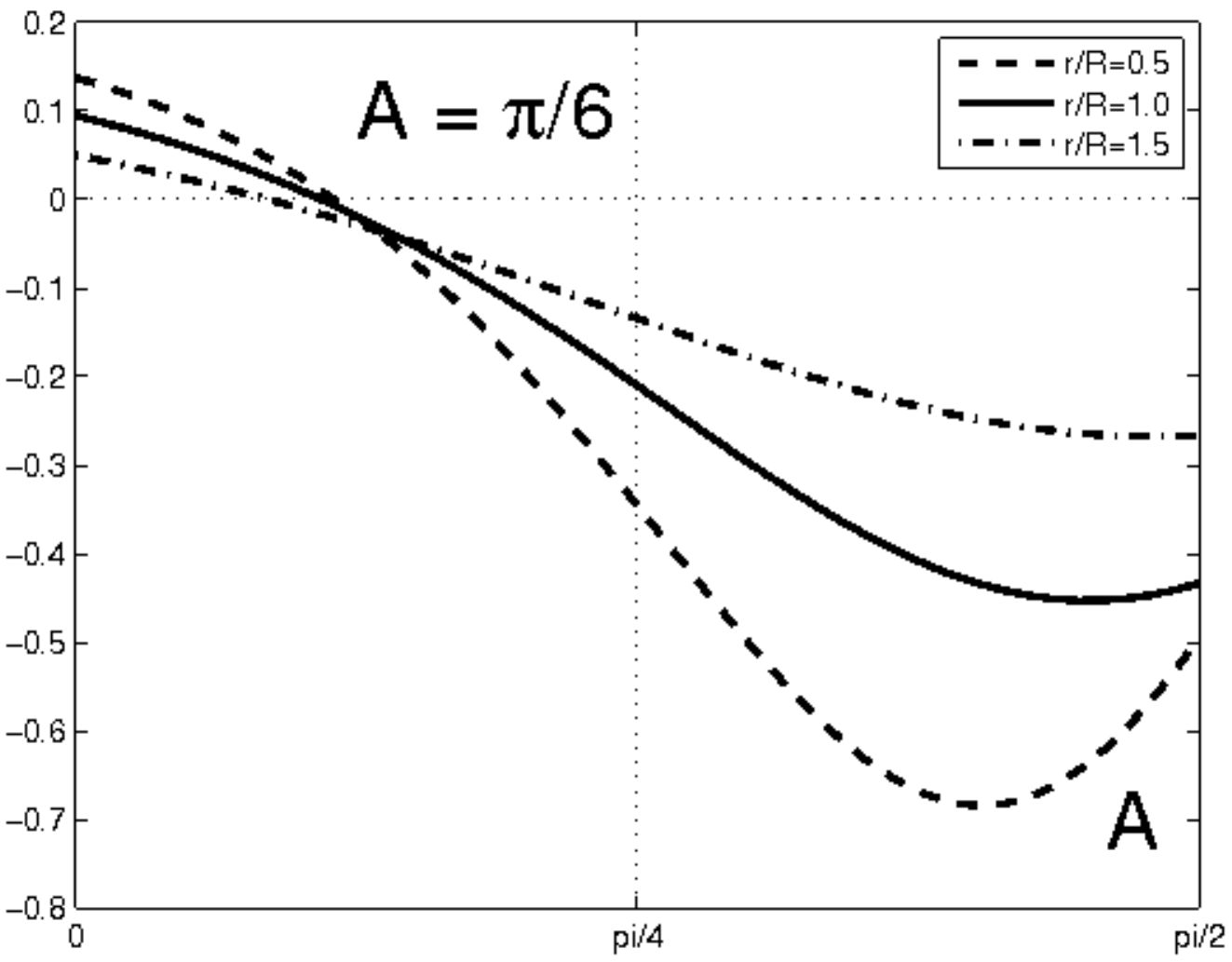}   \includegraphics[height=4cm, width=4cm] {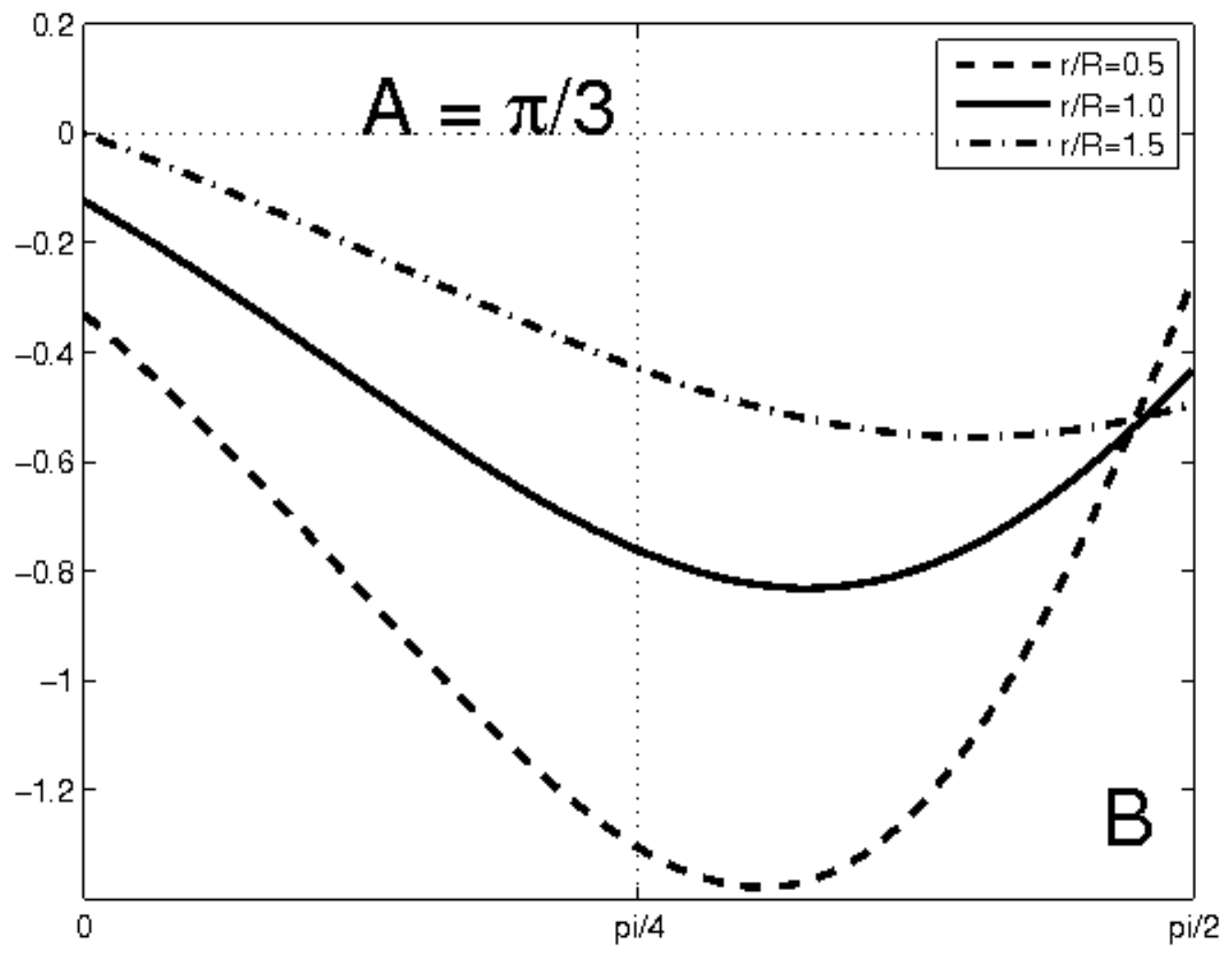}   \includegraphics[height=4cm, width=4cm] {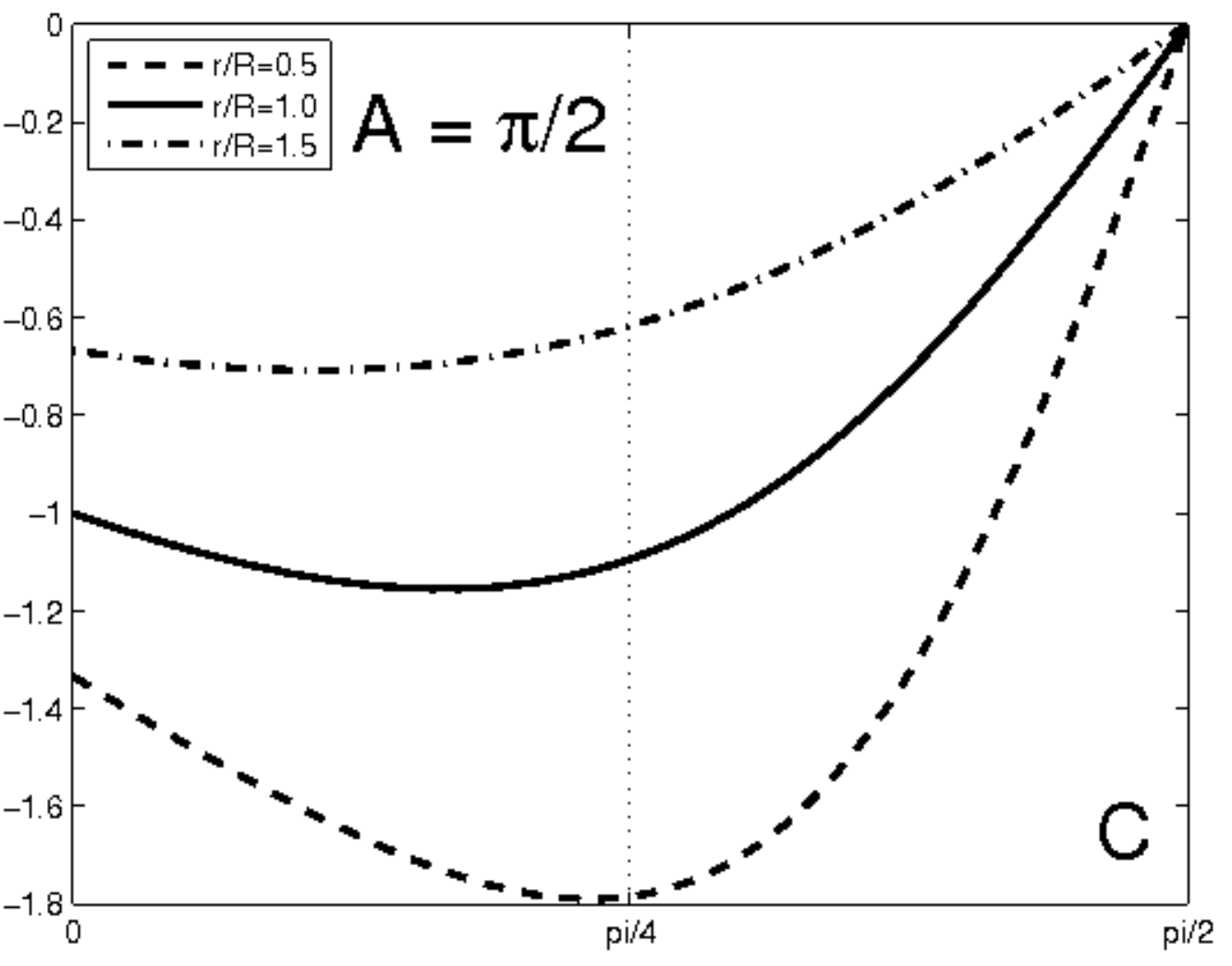} }
\caption{Relative acceleration for $\mathcal{A}=\pi/6$ (A), $\mathcal{A}=\pi/3$ (B), $\mathcal{A}=\pi/2$ (C) versus the longitudinal axis $\theta$ with $v^i=\partial v^i/\partial \lambda =1.0$ and $\varsigma^{\theta \theta} = \varsigma^{\phi \phi} = 4.0$. }
\label{fig:relpv1}
\end{figure}

\begin{figure}[ht]
\centering
\vbox{
 \includegraphics[height=4cm, width=4cm] {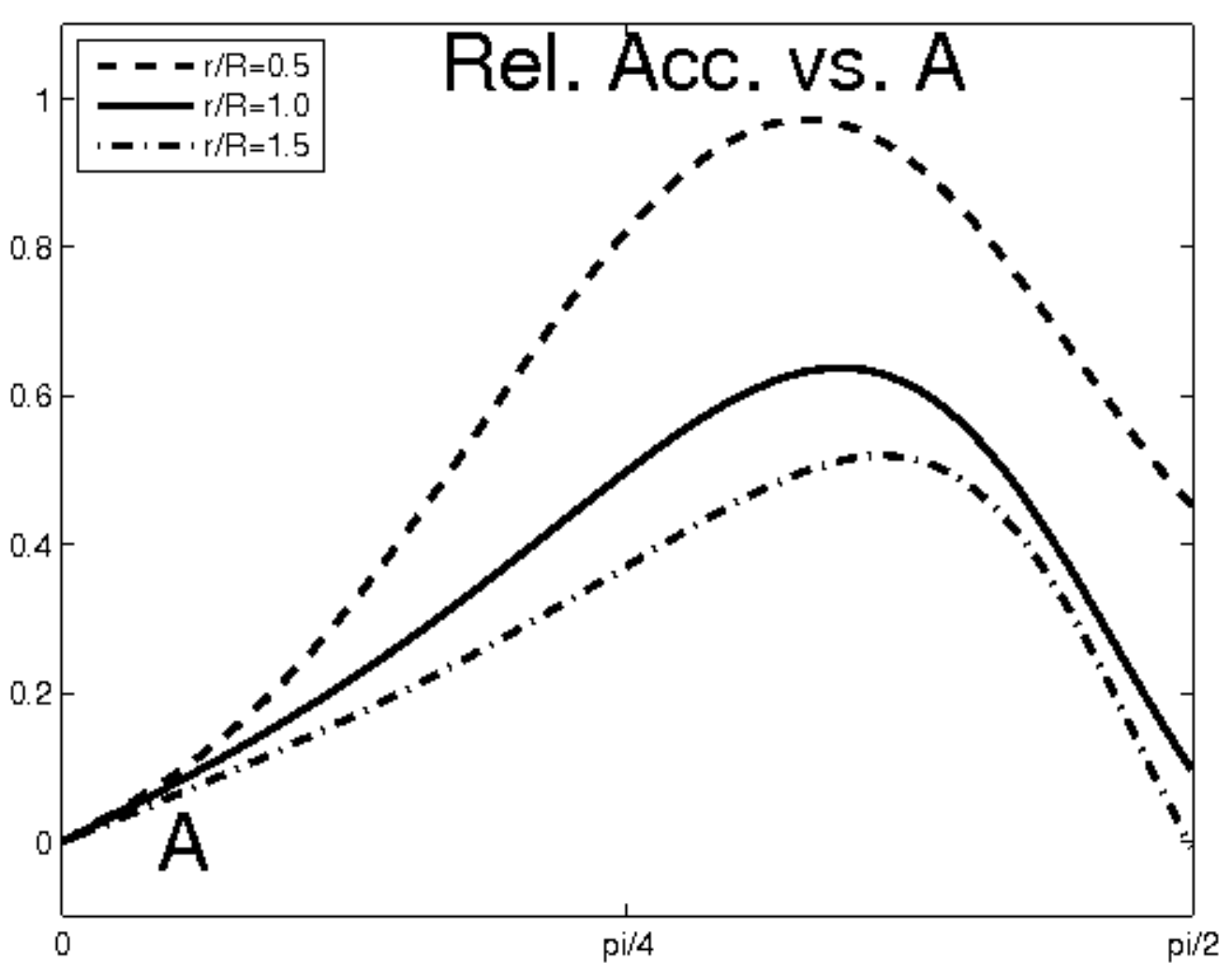}   \includegraphics[height=4cm, width=4cm] {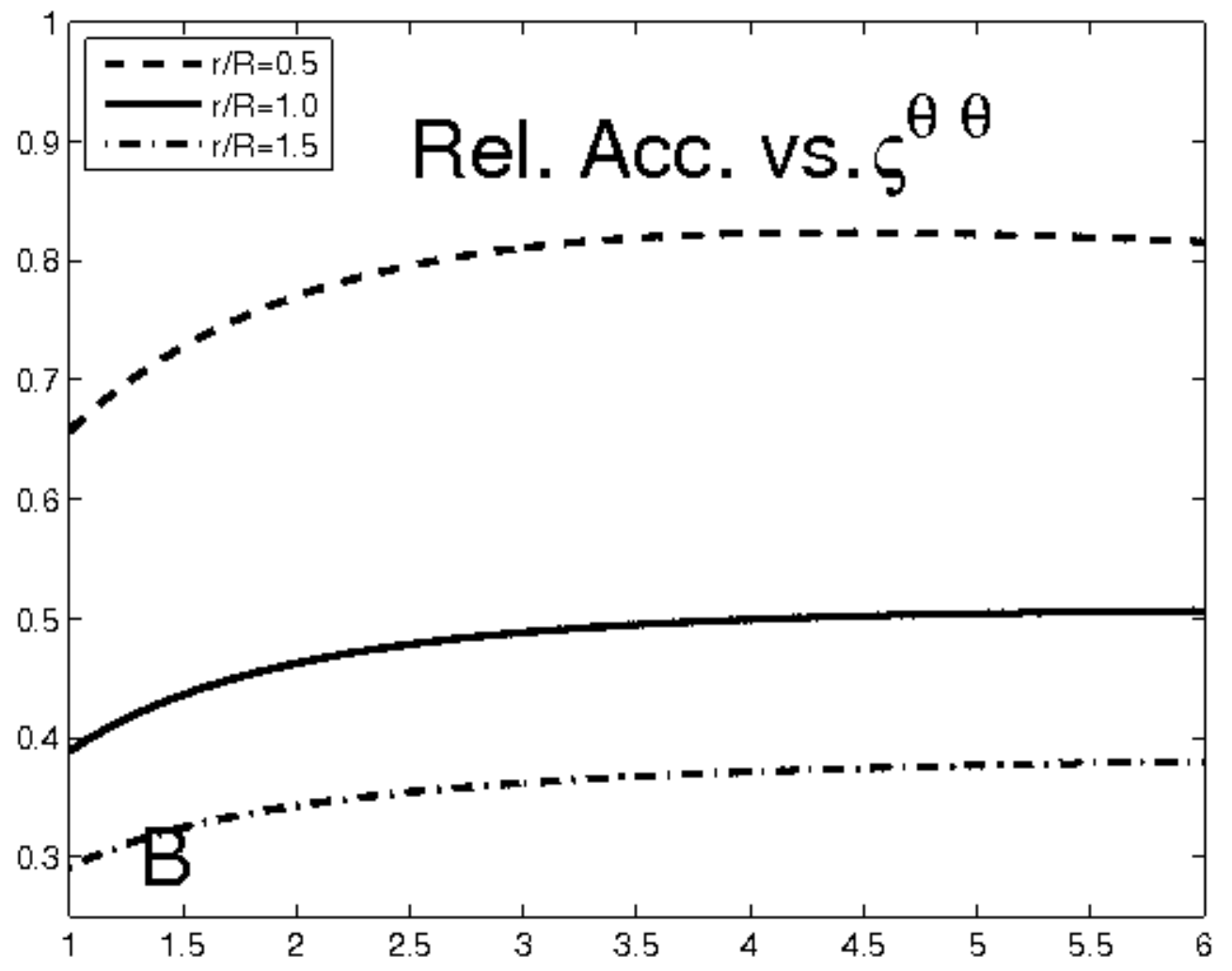}   \includegraphics[height=4cm, width=4cm] {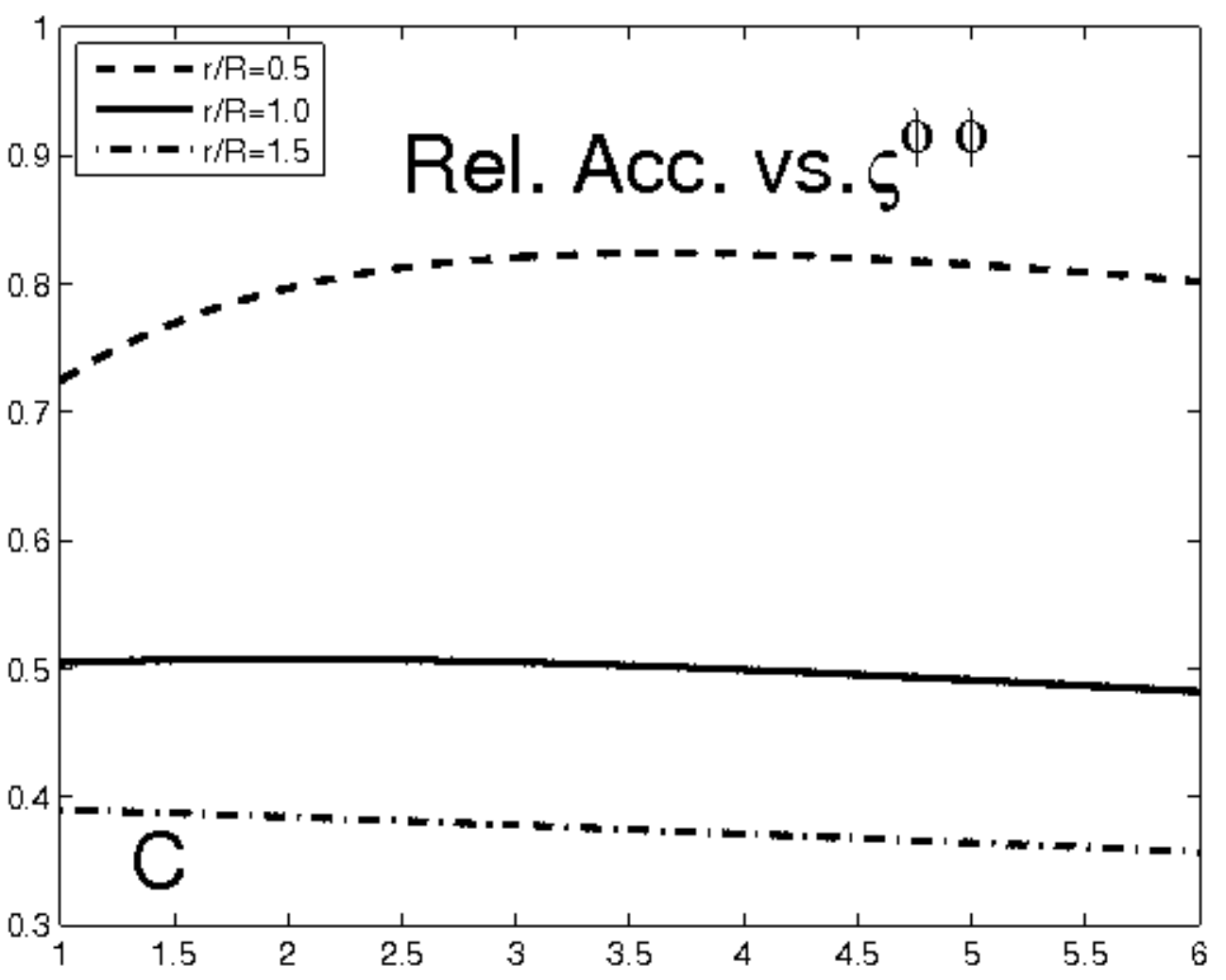} }
\caption{Relative acceleration difference between $\theta=0$ and $\theta=\pi/4$ versus the propagational angle $\mathcal{A} (A)$, the magnitude of $\theta$-anisotropy (B), and the magnitude of $\phi$-anisotropy (C). Default values are taken as $\mathcal{A} = \pi/4$ and $\varsigma^{\theta \theta} = \varsigma^{\phi \phi} = 4.0$.}
\label{fig:relpv2}
\end{figure}

Consider the changes of the relative acceleration with respect to the longitudinal axis $\theta$ for a constant angle $\mathcal{A}$. For a simple comparison, we let $v^i = \partial v^i / \partial n=1.0$ and use both anisotropies such that $\varsigma^{\theta \theta} = \varsigma^{\phi \phi} = 4.0$. With these conditions, we observe in Figure \ref{fig:relpv1} that the magnitude of the relative acceleration strongly depends on $\mathcal{A}$ and $\theta$ uniformly for almost all $r/R$ ratios. The distribution of the relative acceleration along $\mathcal{A}$ remains similar for $0 \le \theta \le \pi/4$, but it drastically changes for $\theta=\pi/2$. The structure of the PV is not symmetric at $\theta=0$ and we observe the different relative acceleration at $\theta=0$ and $\theta=\pi$.

Also, each plot in Figure \ref{fig:relpv1} indicates the conditions for possible conduction failure according to the previously mentioned hypothesis and propositions. Contrary to the test cases for simple surfaces in Appendix C where the relative acceleration has been compared to \textit{zero} acceleration, the relative acceleration at every points inside the PV is not zero, but has nontrivial magnitude possibly with different signs. Thus, the magnitude of relative acceleration for conduction failure can be measured more conveniently if it is compared to that of the normal propagation in the neighboring region. For example, let us pay our attention to the region of $0 \le \theta \le \pi/4$ which represents the region of the PV from the root to the middle. We only consider this region for conduction failure because the propagation comes from the root of the PV around $\theta=0$. For $A= \pi/6$, the relative acceleration ranges from $0.095$ to $-0.207$ for $r/R=1.0$ with a $0.302$ difference. For $A=\pi/3$, it ranges more widely from $-0.124$ to $-0.759$ with a $0.635$ difference. But this difference rapidly drops when $\mathcal{A}=\pi/2$, i.e., when the action potential propagates along the $\phi$-axis.  When $\mathcal{A}=\pi/2$, it ranges from $-1.0$ to $-1.096$ with a difference of $0.096$ at most. Therefore we may conclude that even if the action potential, which propagates with an angle close to $\pi/3$ or possibly up to $\pi/6$ depending on the magnitude of anisotropy, leads to conduction failure due to the PV, the action potential with $\mathcal{A}=\pi/2$ can propagate through the same PV of the same anisotropy without conduction failure.

\begin{figure}[ht]
\centering
\vbox{
 \includegraphics[height=4.0cm, width=4.0cm] {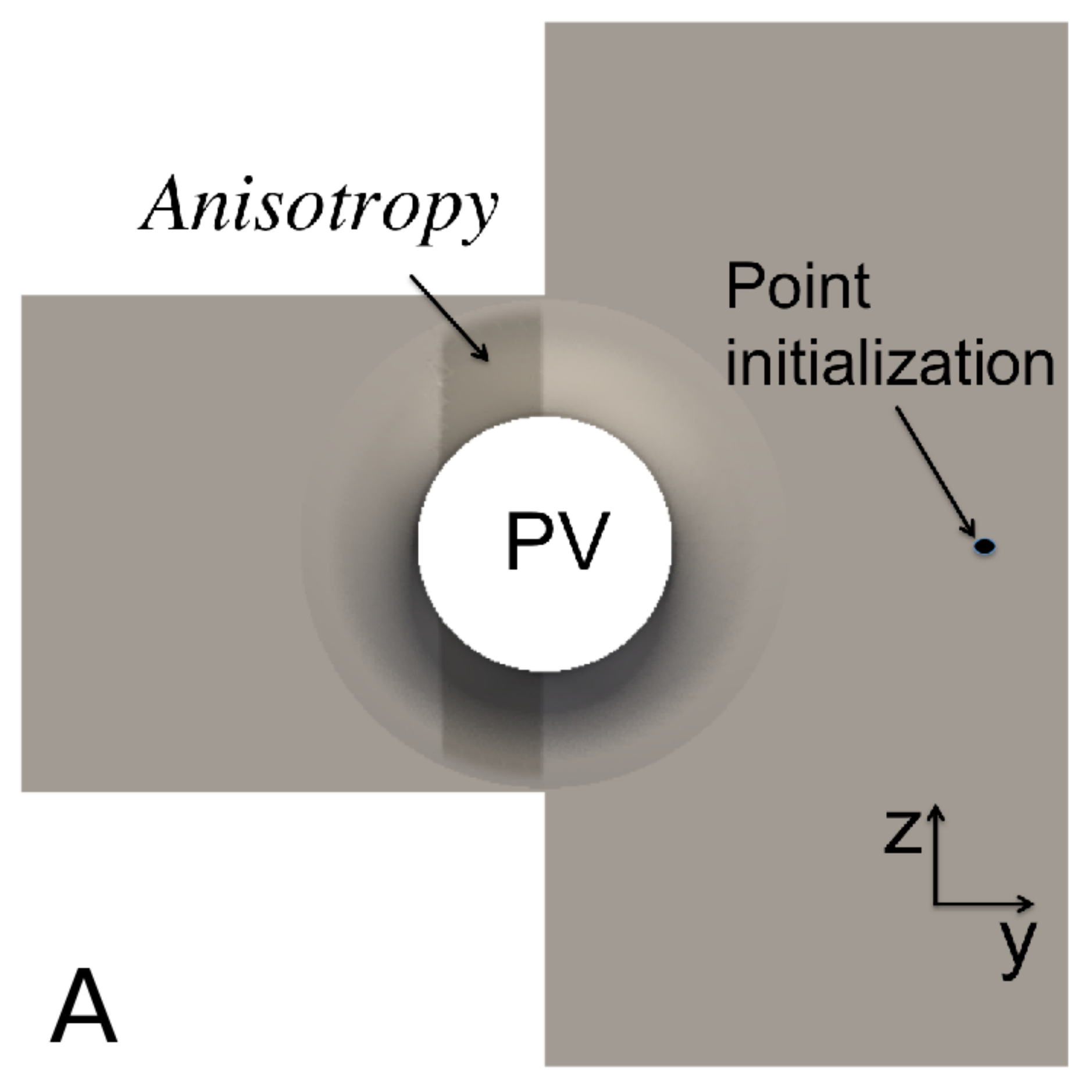}   \includegraphics[height=4.0cm, width=4.0cm] {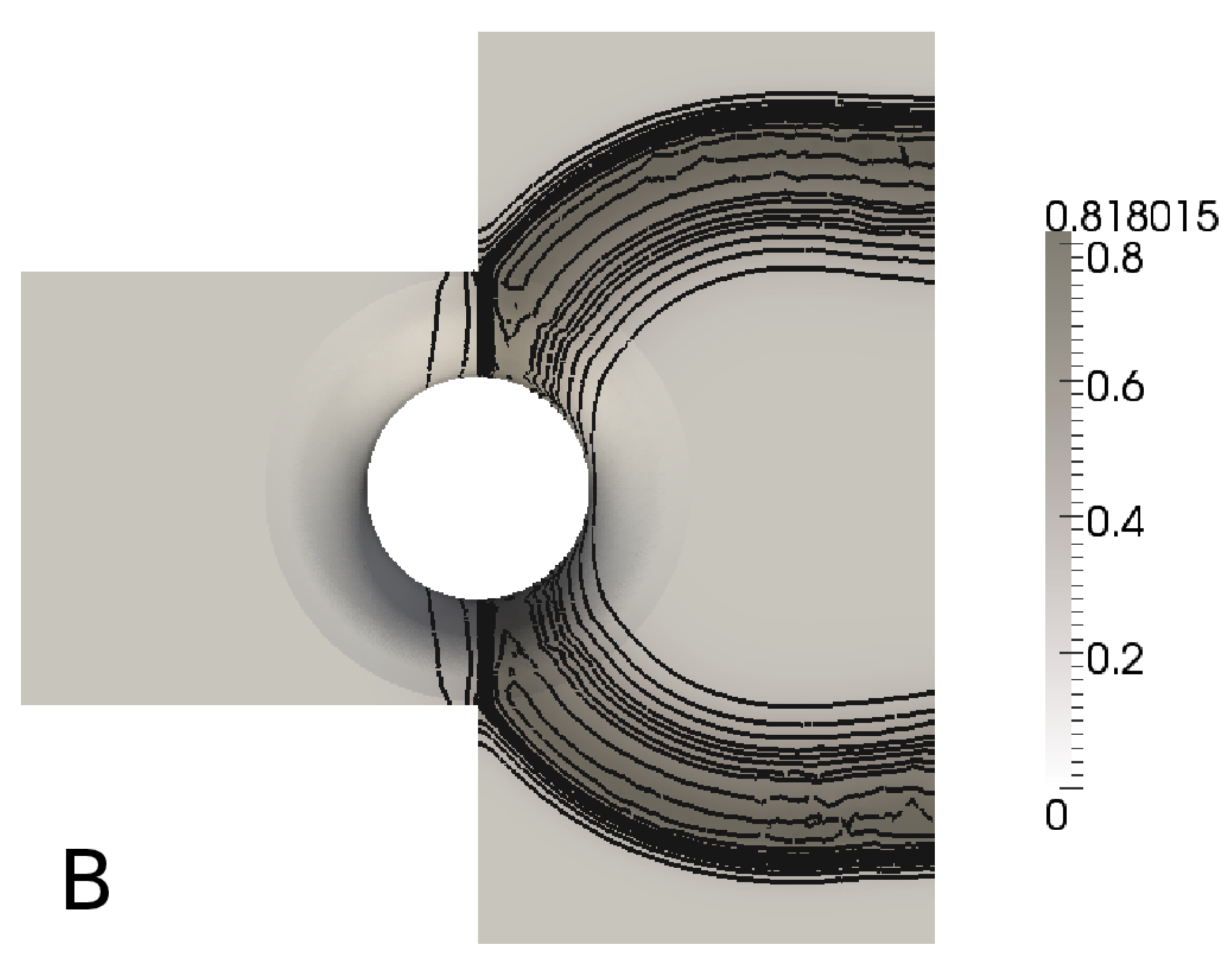}   \includegraphics[height=4.0cm, width=4.0cm] {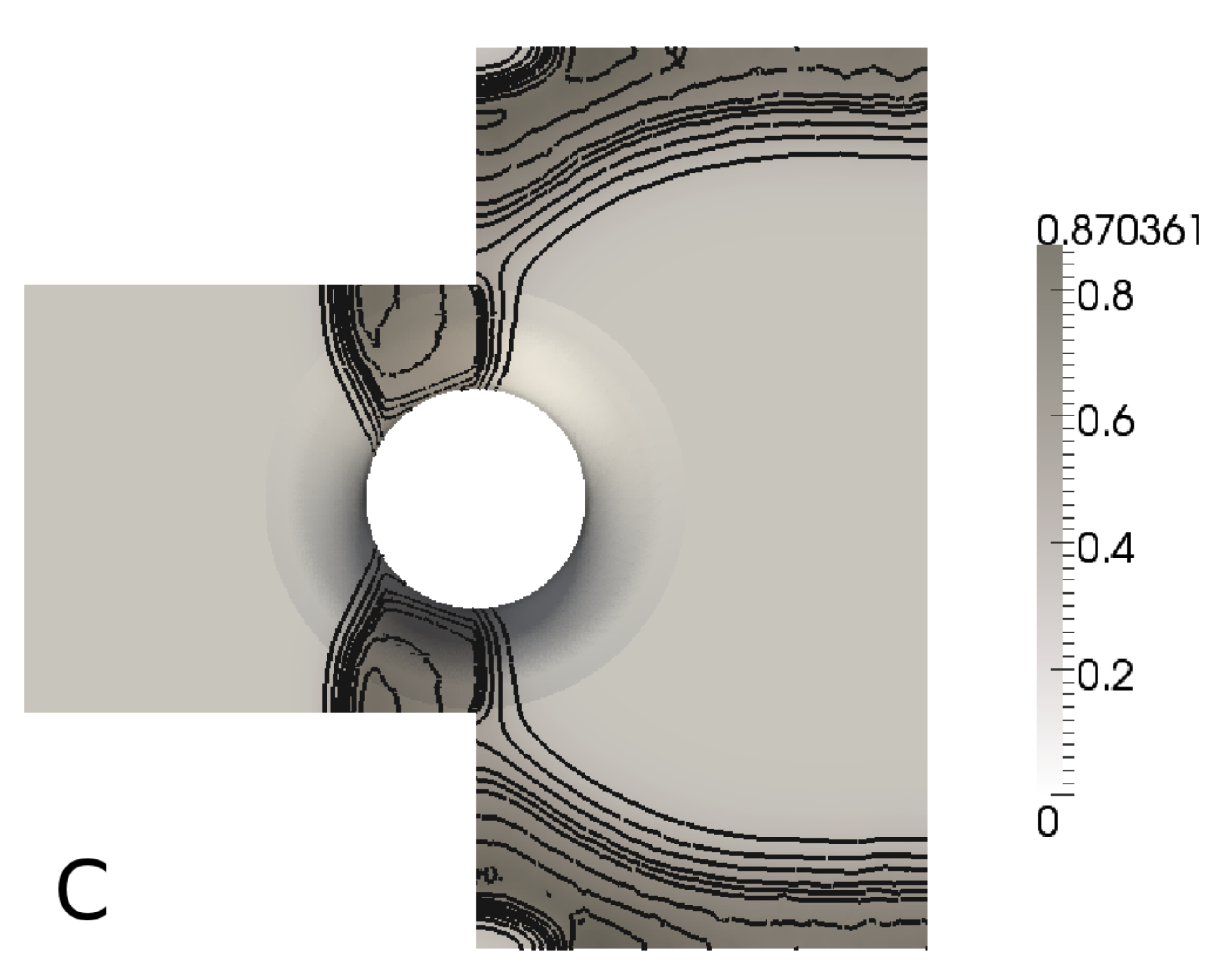} }
\caption{The PV with anisotropy of $\varsigma^{\theta \theta}=\varsigma^{\phi \phi}=8.0$ and the location of point initialization at $(0.0,~100.0,~0.0)$ (A). The action potential ($u$) at $T=800.0$ (B) and $T=1000.0$ (C) to show the propagation of the PV when $\mathcal{A} \approx \pi/2$.  }
\label{relpv2sim3}
\end{figure}

\begin{figure}[ht]
\centering
\vbox{
 \includegraphics[height=4.0cm, width=4.0cm] {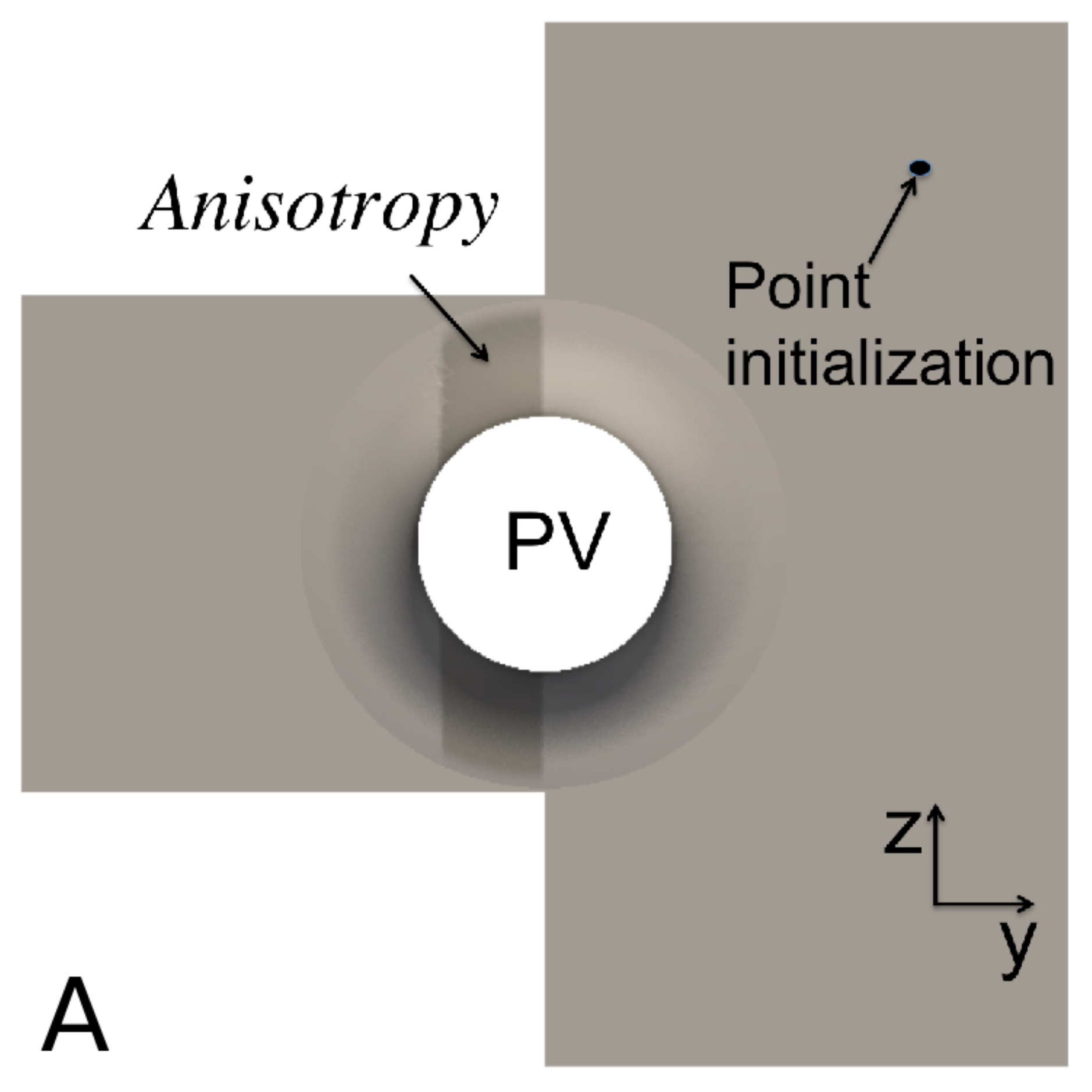}  \includegraphics[height=4.0cm, width=4.0cm] {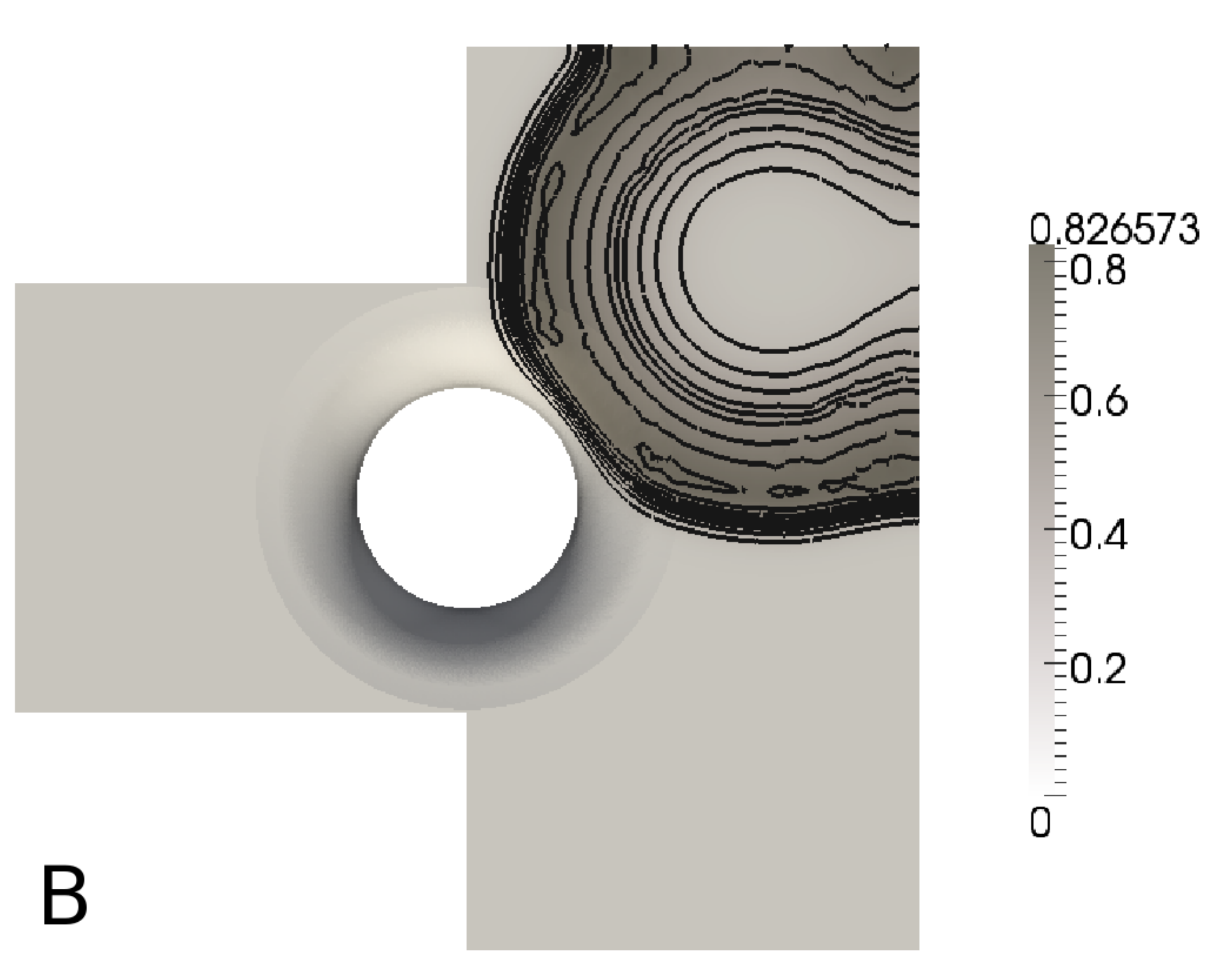}  \includegraphics[height=4.0cm, width=4.0cm] {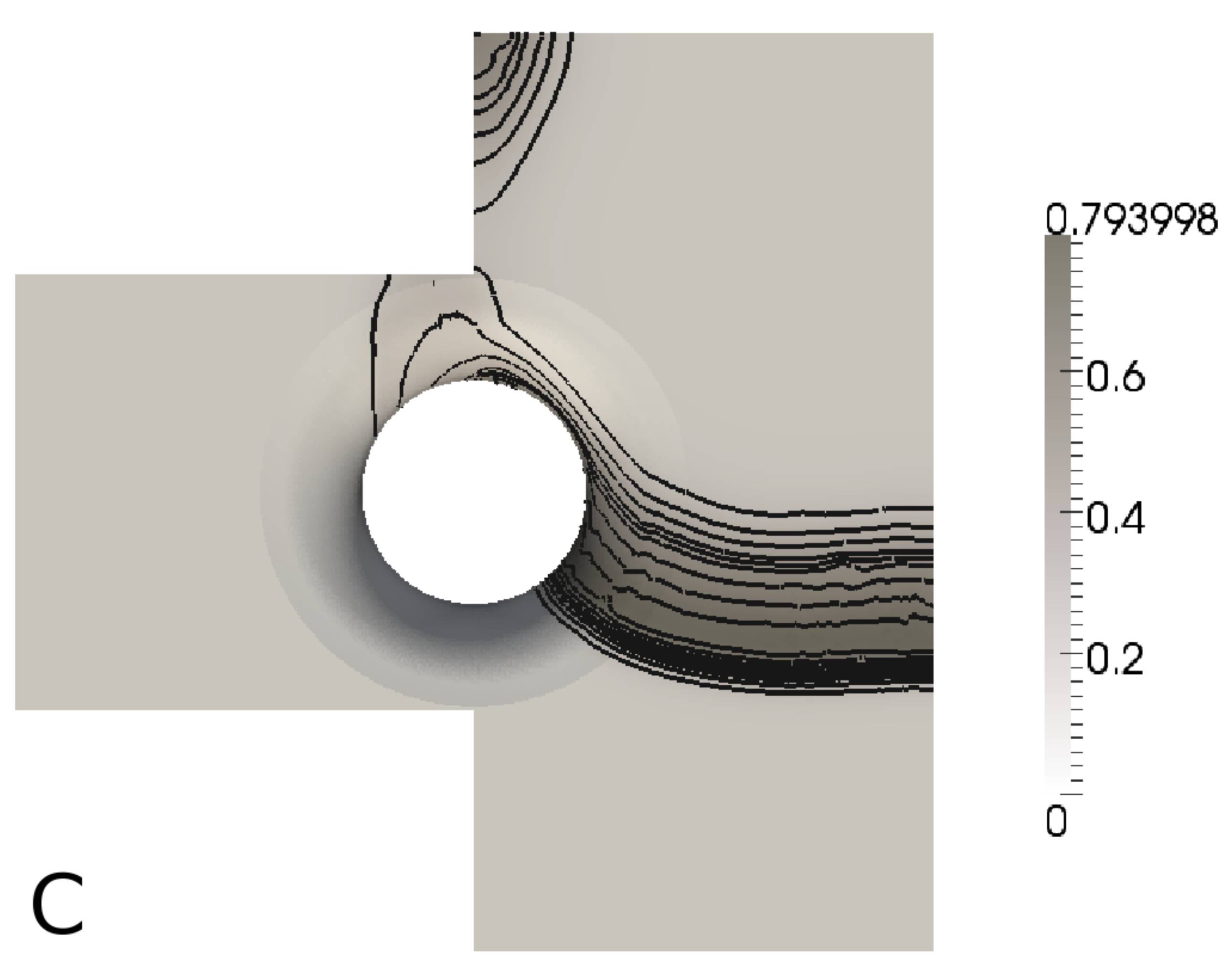} }
\caption{The PV with anisotropy of $\varsigma^{\theta \theta}=\varsigma^{\phi \phi}=8.0$ and the location of point initialization at $(0.0,~100.0,~ 80.0)$ (A). The action potential ($u$) at $T=600.0$ (B), $T=850.0$ (C) to show the conduction failure when $\mathcal{A} \approx \pi/4$. }
\label{relpv2sim4}
\end{figure}

Therefore, we confirm that the relative acceleration strongly depends on the propagational angle $\mathcal{A}$. Figure \ref{fig:relpv2}A displays the difference between the relative acceleration between $\theta=0$ and $\theta = \pi/4$. We observe that the maximum magnitude occurs around the angle of $\pi/3$ and that there is a drastic decrease as it gets closer to $\pi/2$. However, the propagational angle is not the only factor. As displayed in Figure \ref{fig:relpv2}B-\ref{fig:relpv2}C, not mentioning that the ratio between $r/R$ which indicates the three-dimensional shape of the PV changes relative acceleration, the magnitude of each anisotropy, i.e. $\varsigma^{\theta \theta}$ and $\varsigma^{\phi \phi}$, is also an important factor in relative acceleration. The impact of the changes of $\varsigma^{\theta \theta}$ or $\varsigma^{\phi \phi}$ seems to be restricted with a maximum reduction of approximately $10\%$, but as shown in the computational modeling in the next section, this small change can make a significant impact on the role of the PV as a unidirectional block. Moreover, in more general cases without our simplified assumptions on the wavefront, the role of $\varsigma^{\theta \theta}$ and $\varsigma^{\phi \phi}$ can be more significant. Thus, we have the following lemma:\\
\\
\textbf{Lemma 3}: If $\varsigma^{\theta \theta}$ and $\varsigma^{\phi \phi}$ are not sufficiently large, then the PV with both anisotropy can not block the action potential propagating along the $\phi$-axis even when the cardiac action potential propagation at an oblique angle with the $\theta$-axis causes conduction failure. \\
\\
\textbf{Computational modeling}: To demonstrate the validity of the above analysis, a T-shaped domain is designed by placing a PV in the middle of it. The left rectangle is $-150 \le y \le 0$ and $-70 \le z \le 70$, while the right rectangle is $-0 \le y \le 150$ and $-150 \le z \le 150$. At the center $(0,0,0)$, a PV is constructed with $R=70$ and $r=35$. Moreover, anisotropy is aligned with $\varsigma^{\phi \phi} = \varsigma^{\theta \theta} = 8.0$ in the area of $-30 \le y \le 0$ of the PV which is defined exclusively in the range of $0 < x \le 60$. With a point-initialization in the right rectangular domain, the only pathway to propagation to the left rectangular domain is through the PV with the $\theta$-anisotropy and the $\phi$-anisotropy. When the excitation is initiated at $(0.0,~100.0,~0.0)$ as shown in Figure \ref{relpv2sim3}A, the anisotropy of the PV does not generate a sufficient relative acceleration to block the propagation, since the angle $\mathcal{A}$ seems to be close to $\pi / 2$. As a consequence, the action potential propagates through the PV without conduction failure as shown in Figure \ref{relpv2sim3}B - \ref{relpv2sim3}C. On the other hand, when the excitation is initiated at $(0.0, ~100.0,~80.0)$ as shown in Figure \ref{relpv2sim4}A, the excitation propagates toward the PV with an oblique angle $\mathcal{A}$ which is more acute than the previous case and is possibly around $\pi/4$. As displayed in Figure \ref{relpv2sim4}C, this acute angle $\mathcal{A}$ induces a large relative acceleration which causes conduction failure.         Since the only difference between the two simulations is the location of point-initialization and subsequently the angle $\mathcal{A}$ toward the anisotropy in the PV, these models seem to confirm the previous analysis from the relative acceleration equation for the PV with anisotropy. By generalizing the lemma 3, we propose:\\
\\
\textbf{Proposition 3}: The PV is a unidirectional block of which unidirectionality is determined by the propagational direction and the magnitude of anisotropy.\\
\\
\textbf{Proof}: This argument is directly given from lemma 3 which means that the PV may or may not cause conduction failure depending on the propagational direction toward the axis of the PV and/or the magnitude of the anisotropic of both directions. $\square$.

\subsection{Comparison with the \textit{in silico} study \cite{Cherry}}

Observe that proposition 3 is strikingly similar to the conclusion of the ref. \cite{Cherry}, denoted [CHE] from now on, which has been known to the author long after the analysis and conclusion of this paper were drawn. The \textit{in silico} study [CHE] is based on microscopic (cellular-level) ion models with experimental data, while the analysis given in this paper is on macroscopic (organic-level) scale with theoretically-justified hypothesis and mathematical simplifications, but it is remarkable to notice the striking coincidence of the crucial factors of the PV for the reentry.

One noticeable difference is that the size of the PV is a deciding factor in [CHE], instead of the propagation direction in proposition 3, but a simple calculus can easily reveal that the former is just one component of the latter. In the \textit{in silico} study, the two parameters of the geometric shape of the PV were considered: one is the length of the PV and the other is the circumferential diameter of the PV. As seen in Figure \ref{figgeom2}, the former is roughly represented by $R$ and the latter by $r$, though the direct comparison may be not legitimate since the geometry of [CHE] is just a cylinder attached to the plane which has no geometrical curvature. The dependency of the parameter $r/R$ on relative acceleration and subsequently on the reentry is displayed in Figure \ref{fig:relpv1} and \ref{fig:relpv2}.  One conclusion of [CHE] is that the smaller and shorter PV is unlikely to generate the reentry from numerous simulations. By the relative acceleration approach, Figure \ref{fig:relpv1} and \ref{fig:relpv2} imply that the higher ratio $r/R$ generates higher relative acceleration, thus a unidirectional pathway is impossible because the PV prevents the propagation in most directions. Thus, only the PV with a lower ratio of $r/R$ can be more likely to be a unidirectional block leading to the reentry. Since lower $r/R$ means a larger diameter independent of length, the analysis of this paper is coincident with [CHE] which are also confirmed by clinical studies that arrhythmogenic PVs have larger diameters \cite{Yamane} \cite{Herweg}. Nevertheless, the proposition implies that the curvature of the PV is more important than the size of it, though the size of the PV also changes the curvature.

On the other hand, the conduction properties of the PV seem to mean the same both in [CHE] and proposition 3, but should be interpreted with different scales. The conductivity property of the PV was implemented in [CHE] by the \textit{randomization procedure} to generate heterogeneity by disconnecting the cell-to-cell connections. As mentioned in the section of \textit{potential limitation}, this randomization procedure seems to have a strong influence on the generation of the reentry, but as the name suggests, there is no unique model of the procedure, thus the conductivity cannot be parameterized for the reentry. By the relative acceleration analysis, the conductivity property of the PV is considered on macroscopic scale and can be decided uniquely in the formulation. For example, the anisotropic coefficients for both directions can be uniquely determined mathematically for the PV to generate the reentry. But, the biological meaning of anisotropy of \textit{both directions} should be interpreted microscopically, especially on the relation between macroscopic anisotropy of both directions and microscopic heterogeneity by randomization procedure, which may be discussed in the future publications.

\begin{figure}[ht]
\centering
\vbox{
  \includegraphics[height=4.0cm,width=4.0cm]{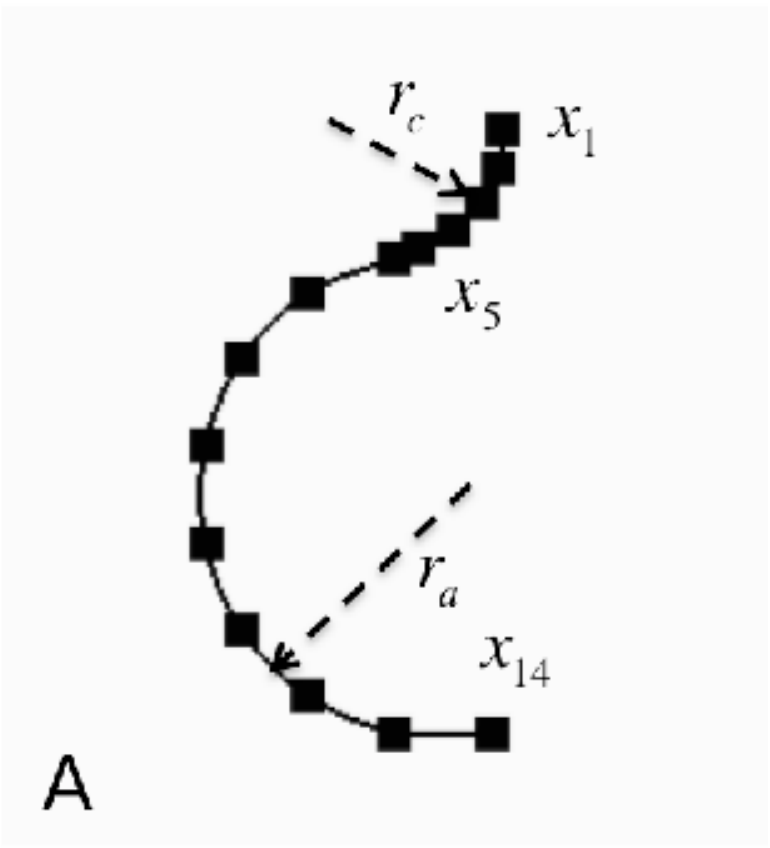}  \includegraphics[height=4.0cm,width=4.0cm]{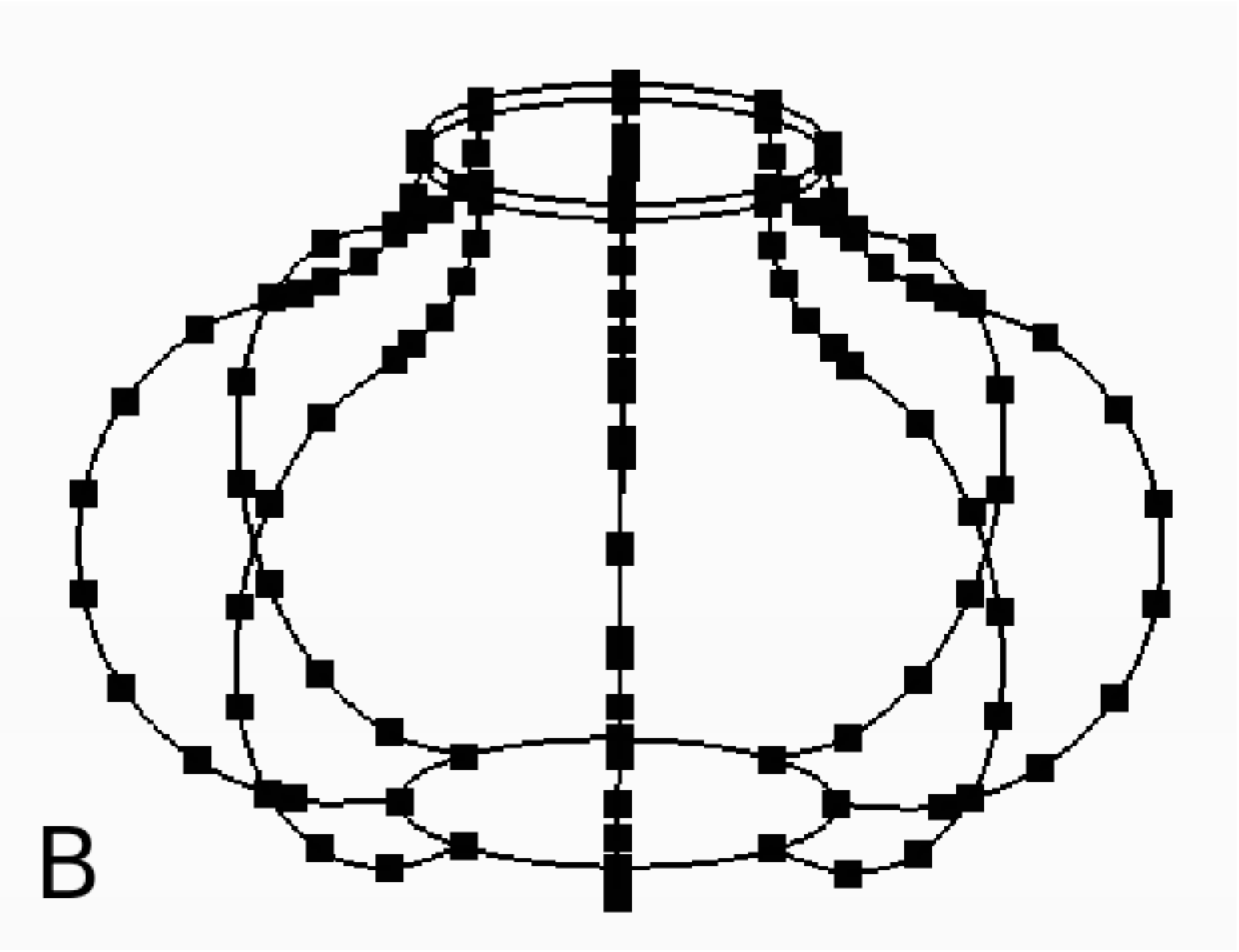}  \includegraphics[height=4.0cm,width=4.0cm]{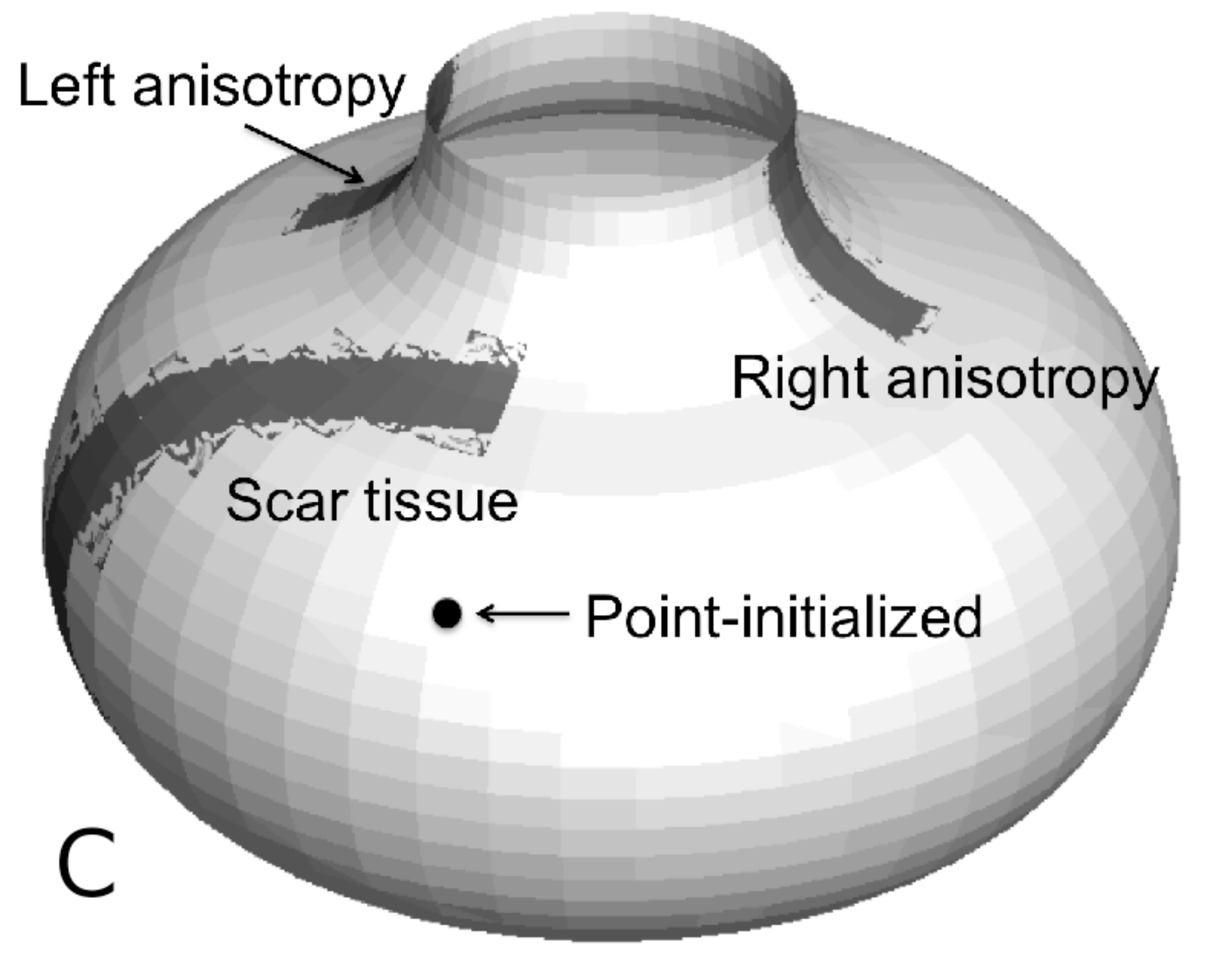}  }
\caption{Construction of a surface of revolution for the atrium with the PV. Adapted from \cite{MMF2}.}
\label {AFdomain}
\end{figure}

\section{Computational modeling of 3D atrial reentry}
In support of the geometric analysis expressed in proposition 3, two computational simulations will be provided in this section to illustrate how 3D atria reentry can be generated by a unidirectional block--the PV with anisotropy. The computational simulations are performed on a simplified surface with a PV-like column, but is sufficient to demonstrate the key features of the PV as a unidirectional block to generate the reentry. The use of the simplified smooth model is common in the study of the reentry. See the quasi three-dimensional cylinder attached to the plane of the LV for the \textit{in silico} study \cite{Cherry}. For a surface of revolution to model a smooth atrium with the PV, let us construct a curved line at the plane of constant $y$ with the following grid points $(x_k,~z_k)$ as shown in Figure \ref{AFdomain}A:
\begin{equation*}
\left \{
\begin{array}{l}
  \left ( R - r_c \sin (  k \pi / {10} ), r_c - \sqrt{ r_c^2 - (x_k - R)^2}  \right ),~~~ 1 \le k \le 5,    \\
  \left ( c_m + R + r_A \sin (  k \pi / {8}   ),  r_A +  \sqrt{ r_A^2 - (x_k - R - c_m)^2}  \right ),~~ 6 \le k \le 14 ,
\end{array}
\right .
\end{equation*}
where we used $R=80.0,~r_c = 0.7R,~r_A = R,~c_m = -20.0$. Revolving the curved line around the center line $(-20,0,z)$ yields the curved surface as shown in Figure \ref{AFdomain}B. Moreover, scar tissue and anisotropy are also placed: a line of scar tissue in the area of $ 55.0 \le x \le 75.0,~ -200.0 \le y \le 0.0$ and two strips of anisotropy in the area of $-30.0 \le x \le -10.0,~ 0.0 \le z \le 60.0 $ as shown in Figure \ref{AFdomain}C. In reality, large solid scars do not frequently appear adjacent to the PVs, but for simplicity of modeling, we placed the scar tissue in that region to provide an oblique angle of the propagation towards the PV when the SAN is located at $(126.52,0.0,-35.55)$. The role of scar tissue can be substituted by other factors changing the propagational direction towards the PV. Anisotropy is implemented by rescaling the moving frames by the factor of the diffusivity coefficient \cite{MMF2} and the scar tissues are implemented by letting all the reaction constants be zero.

The FHN mono-domain equations with the reaction functions \eqref{reactionF} and \eqref{reactionG} are solved to simulate the cardiac action potential propagation on this surface of revolution \cite {Rogers}. Bearing in mind that the average height of the atrium is $5~cm$, the domain is at least three times larger and five times wider than the actual size. This magnification of the size of the atrium is just for the convenience of adjusting the parameters for the desired phenomena. The consequence of this modeling should be the same for the real size of the atrium in similar circumstances.

\begin{figure}[ht]
\centering
\vbox{ \includegraphics[height=3cm,width=3cm]{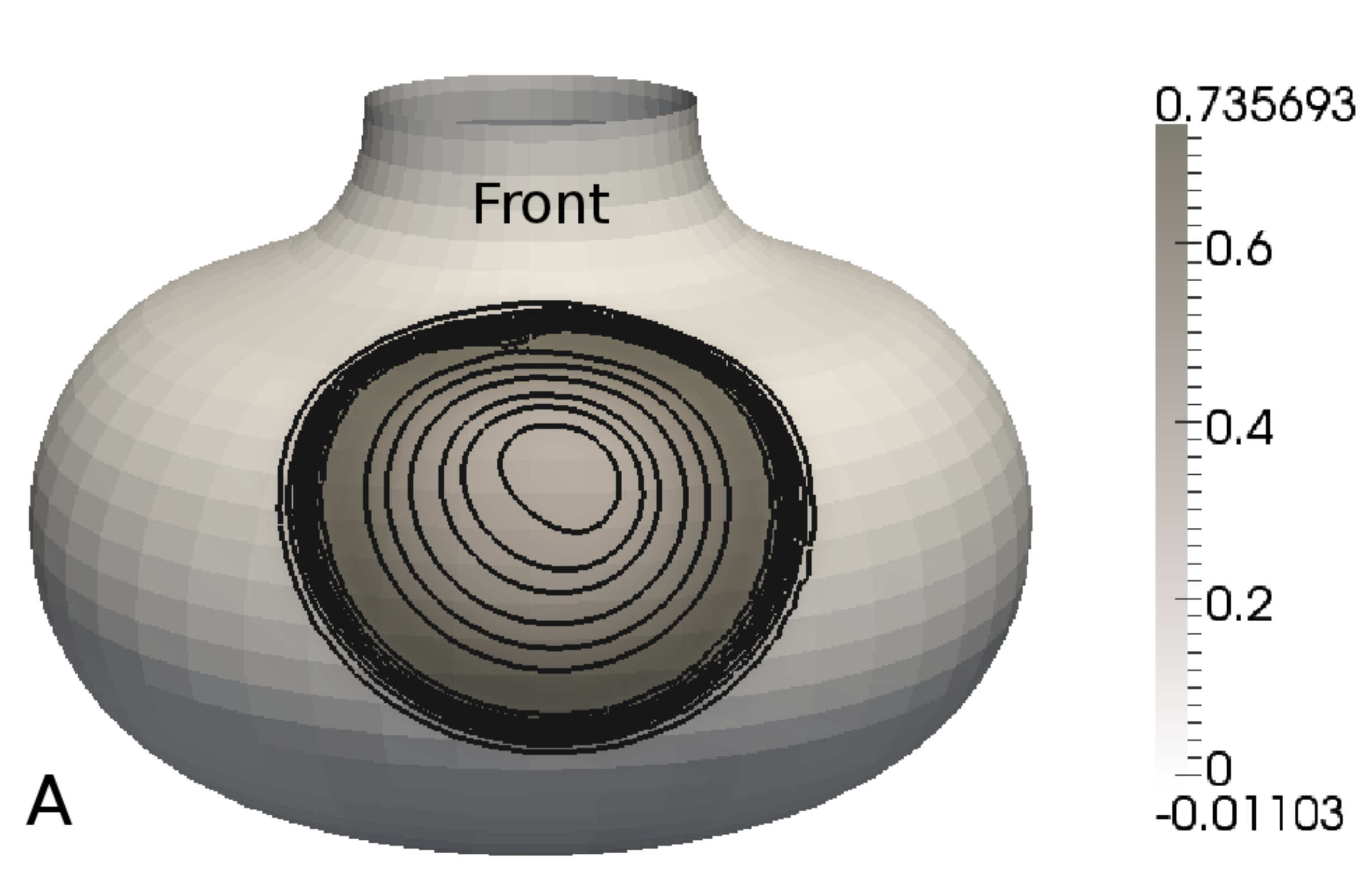}  \includegraphics[height=3cm,width=3cm]{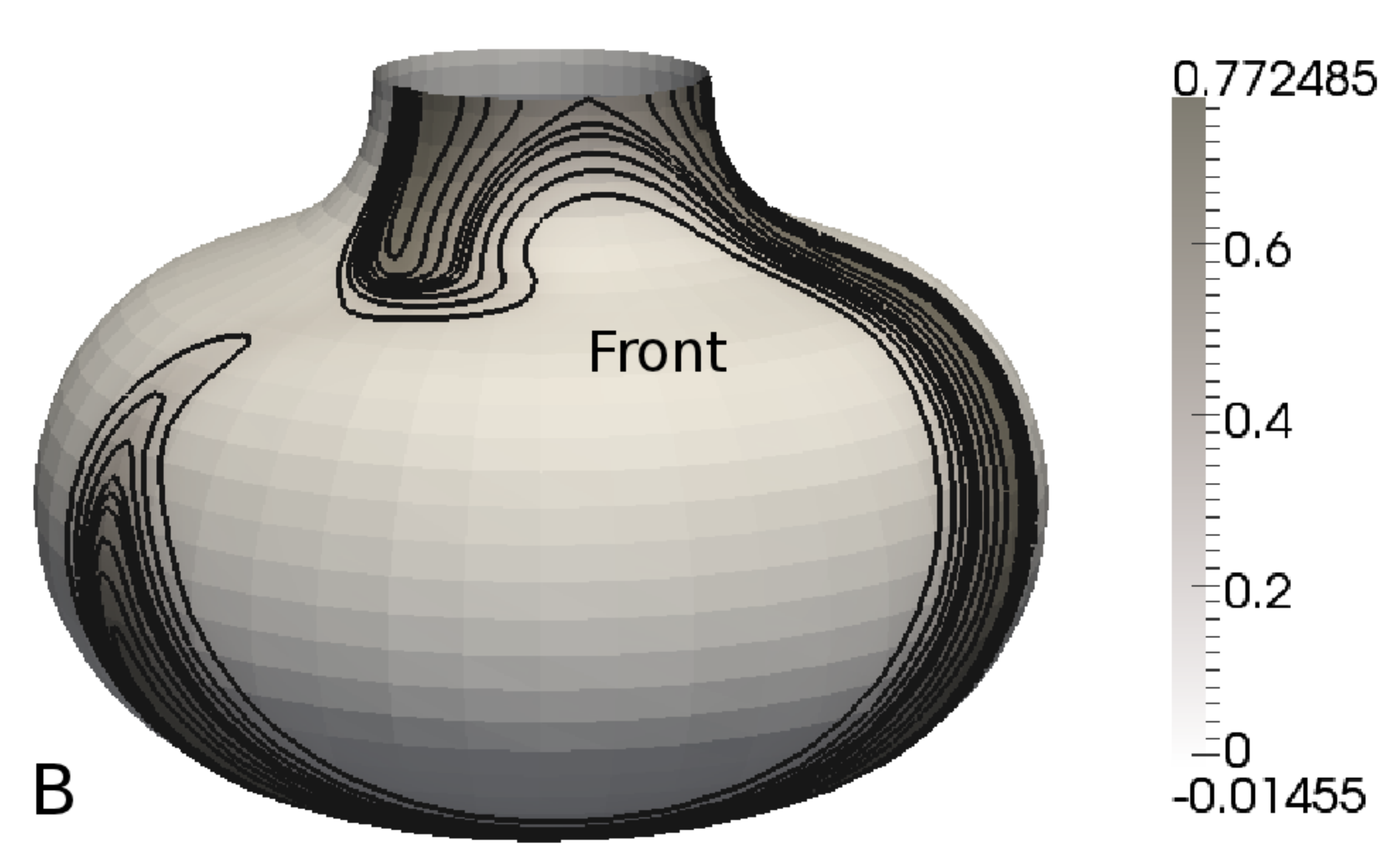}   \includegraphics[height=3cm,width=3cm]{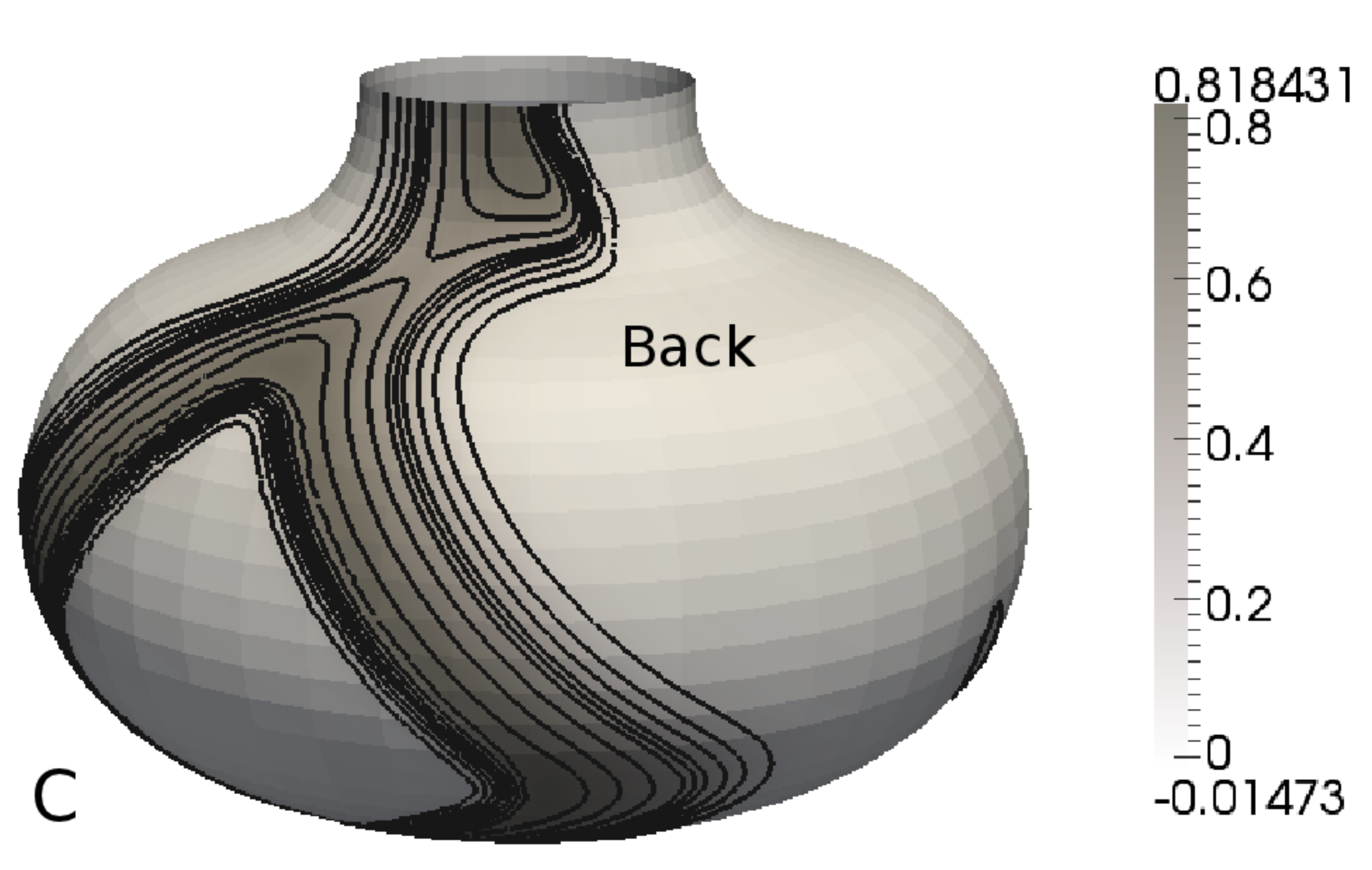} \includegraphics[height=3cm,width=3cm]{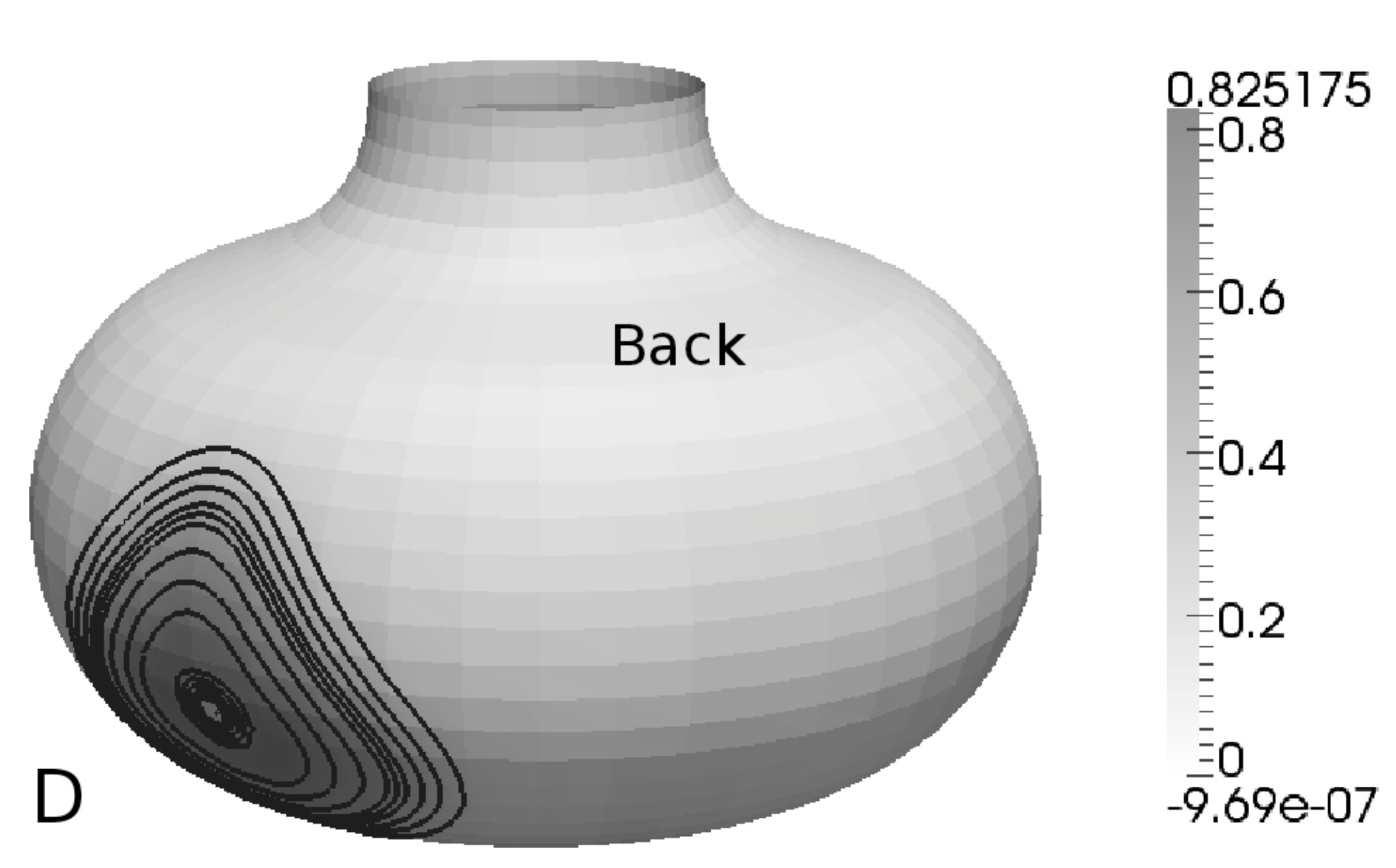} }
\caption{Normal propagation at $T$=$50.0$ (A), $T$=$100.0$ (B), $T$=$225.0$ (C), $T$=$250.0$ (D). $\varsigma^{\theta \theta} = \varsigma^{\phi \phi} =4.0$ for the anisotropic strips. Adapted from \cite{MMF2}.}
\label {AFnormal}
\end{figure}

First, let the magnitude of the anisotropy in the PV be $4.0$, i.e., $\varsigma^{\theta \theta}= \varsigma^{\phi \phi}=4.0$. Figure \ref{AFnormal} displays how the action potential propagation starts from the point of initialization and terminates at another point on the other side of the point of initialization. Because of scar tissue, the action potential propagates onto the PV with an oblique angle to yield two separate waves: the short wave on the left and the long wave on the right (Figure \ref{AFnormal}B). An oblique angle with anisotropy prevents the short wave from propagating on the PV, only leaving the propagation on the right. On the other hand, the right wave has sufficient distance to align its propagational direction in the $\phi$-direction for the right anisotropy: thus the propagation still remains on the PV. The two separate waves meet again at the back, but the collision of two waves creates an oblique angle to anisotropy again and, consequently, the action potential fails to propagate through the anisotropy on the PV (Figure \ref{AFnormal}C). Note that the left anisotropy blocks the propagation twice because the propagational angle of both cases is not close to $\pi/2$ or the magnitude of anisotropy is not sufficiently large; thus, it is not a unidirectional block. Without the propagation on the PV, the wave converges to a point (Figure \ref{AFnormal}D). Because the propagation starts from one point and terminates at one or two points, for example at the right and left atrial appendage, we describe this pattern of propagation as the \textit{normal propagation}.

\begin{figure}[ht]
\centering
\vbox{ \includegraphics[height=3cm,width=3cm]{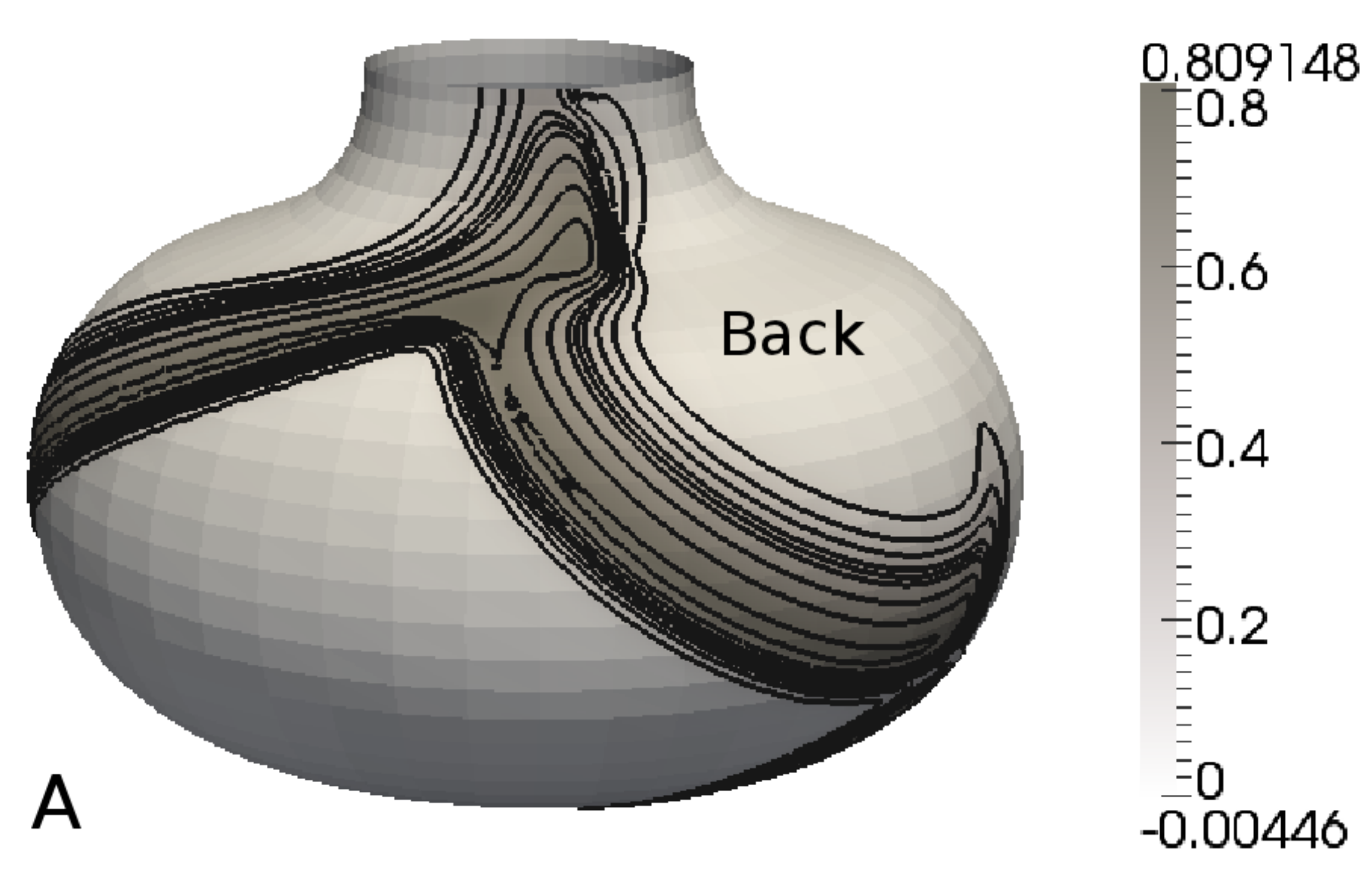}  \includegraphics[height=3cm,width=3cm]{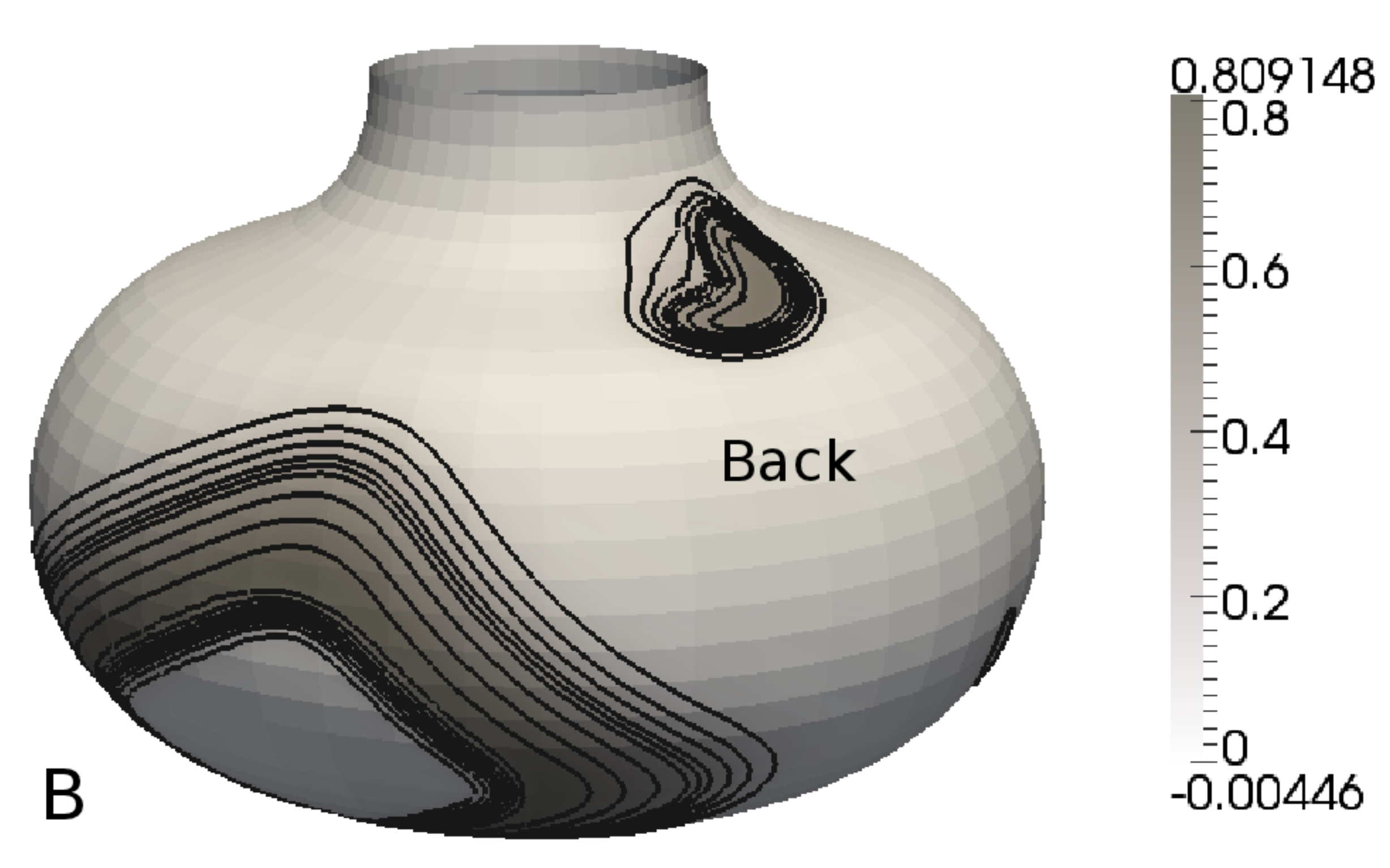}  \includegraphics[height=3cm,width=3cm]{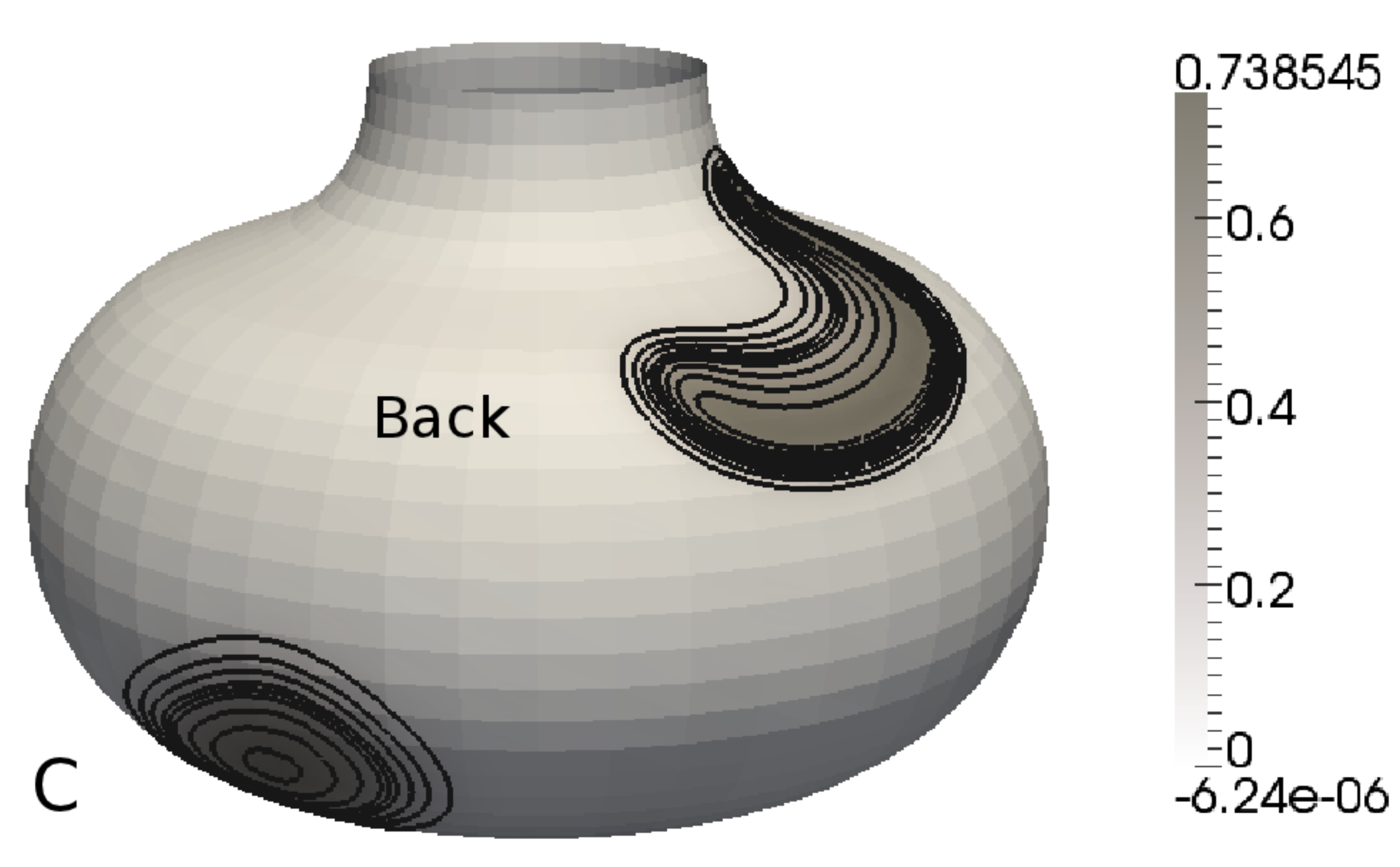} \includegraphics[height=3cm,width=3cm]{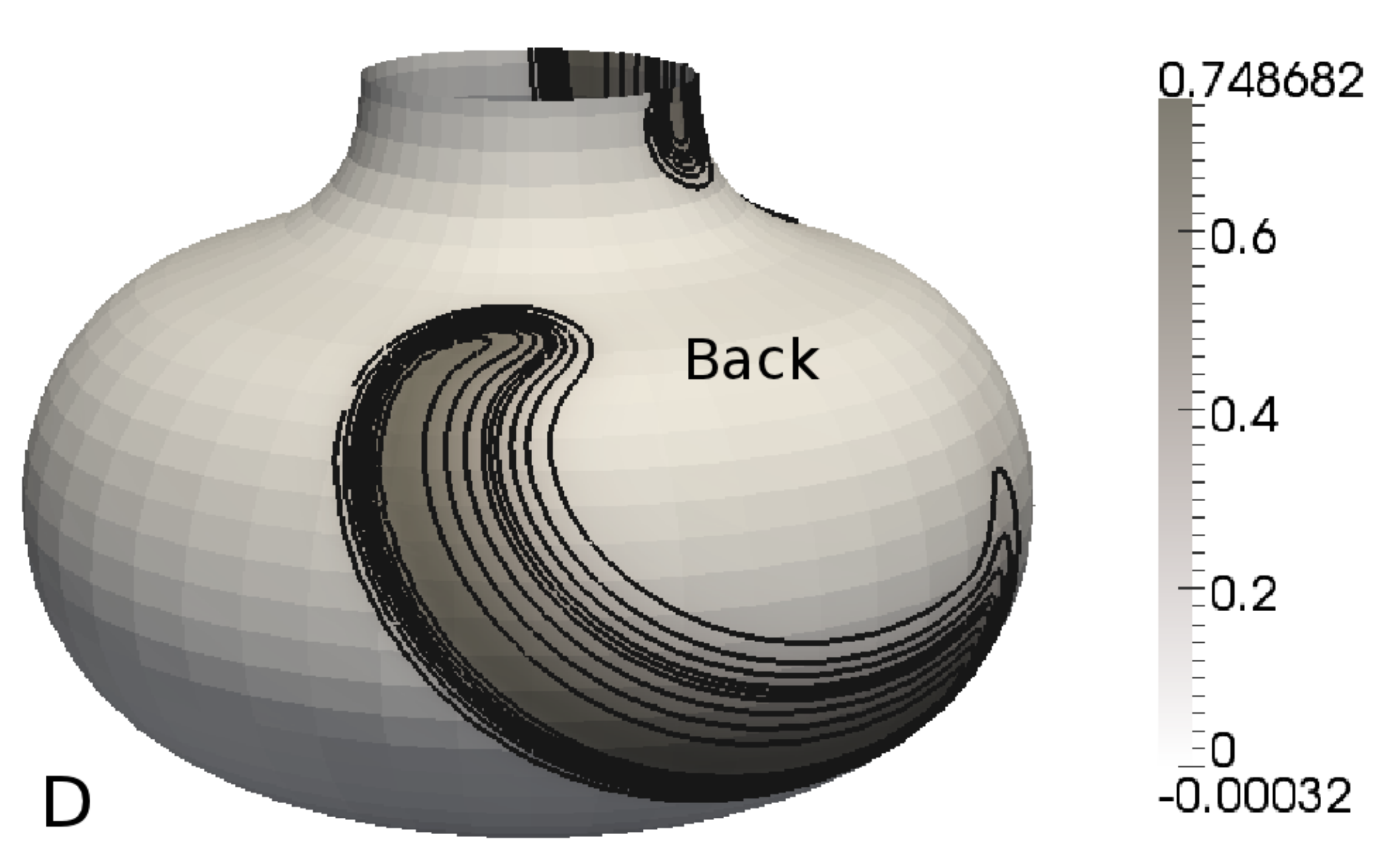} }
\caption{Modeling of atrial reentry at $T$=$195.0$ (A), $T$=$225.0$ (B), $T$=$250.0$ (C), $T$=$300.0$ (D). $\varsigma^{\theta \theta} = \varsigma^{\phi \phi} =3.0$ for the anisotropic strips. Adapted from \cite{MMF2}.}
\label {AFReentry1}
\end{figure}

\begin{figure}[ht]
\centering
\vbox{ \includegraphics[height=3cm,width=3cm]{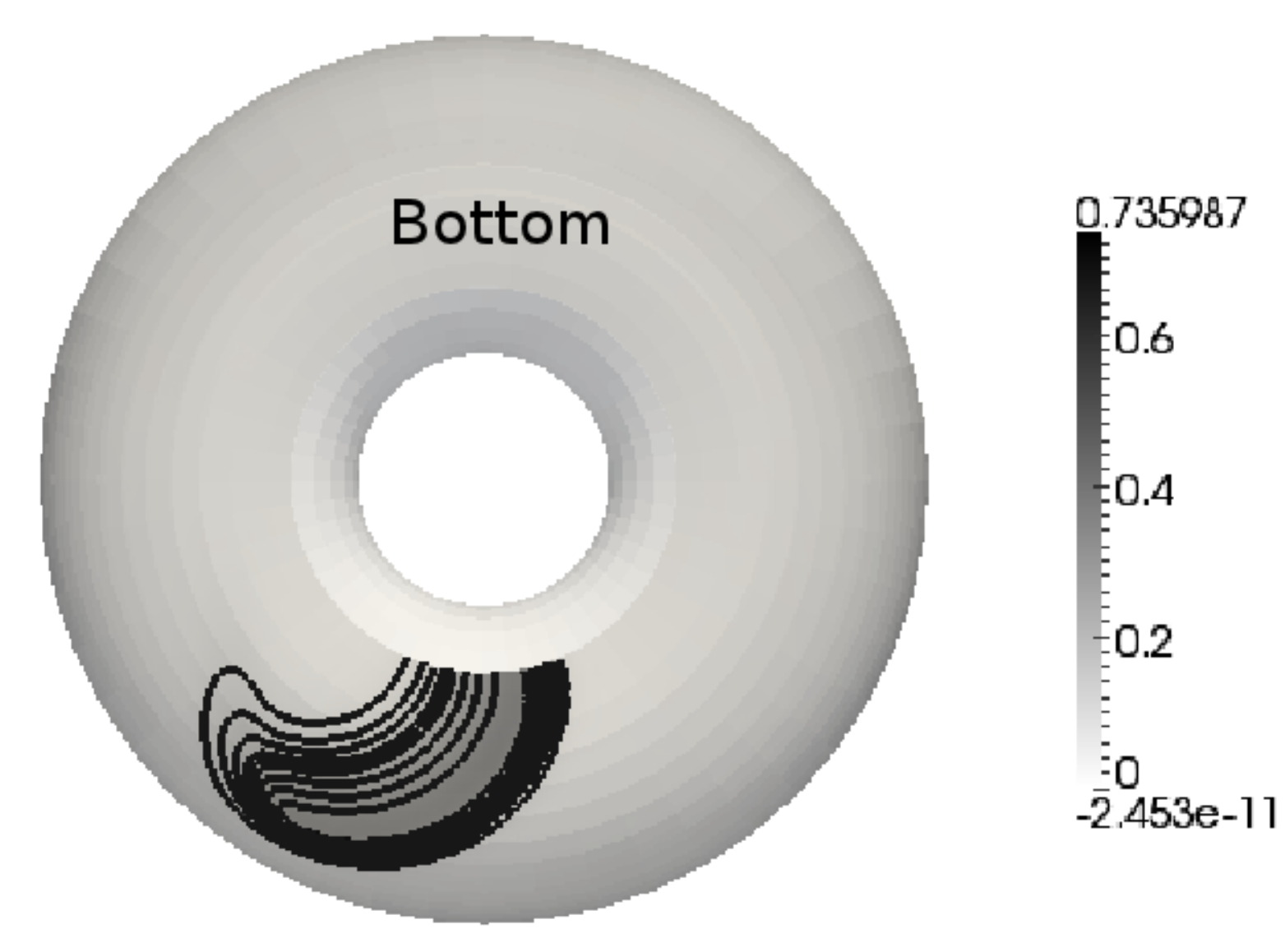}  \includegraphics[height=3cm,width=3cm]{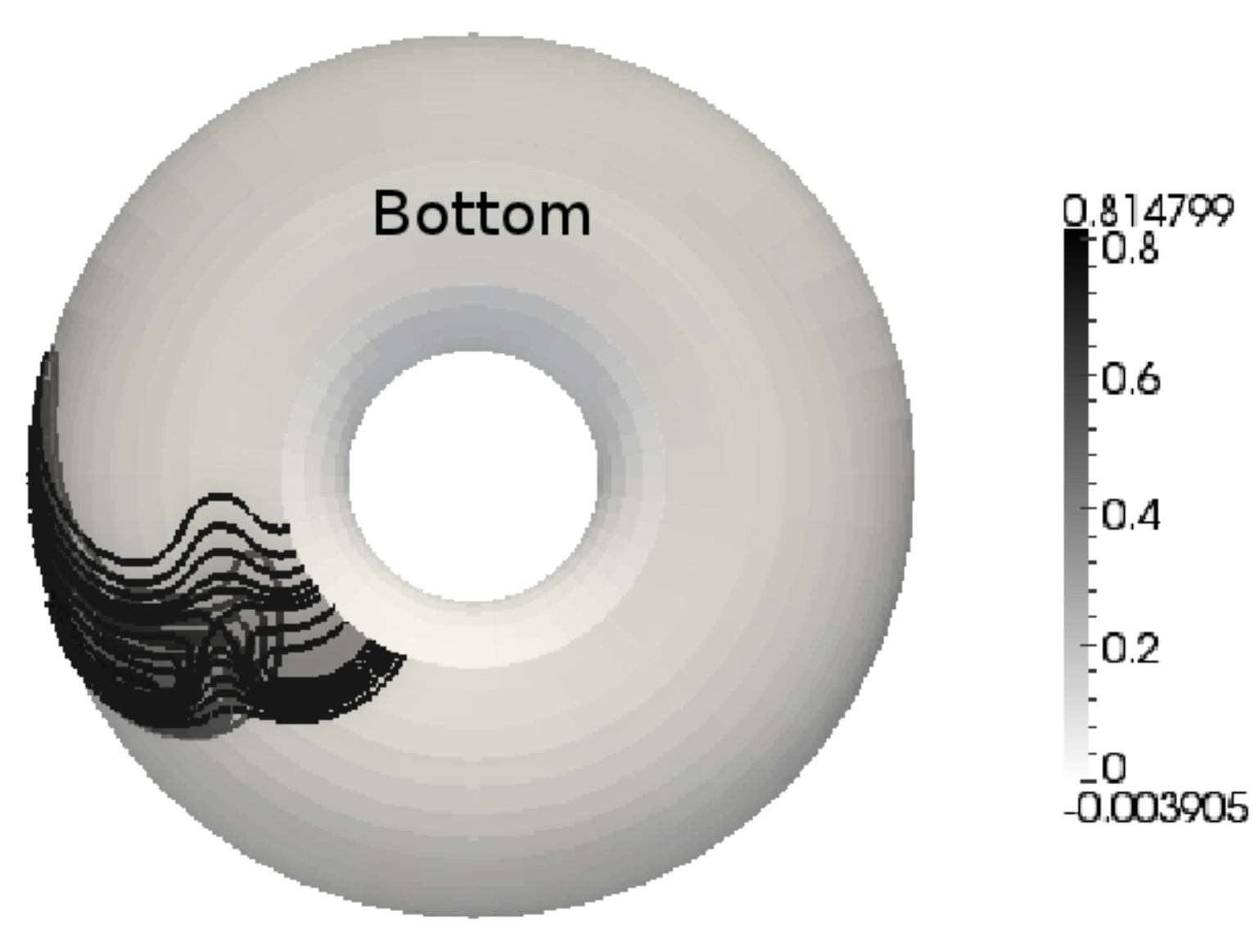}  \includegraphics[height=3cm,width=3cm]{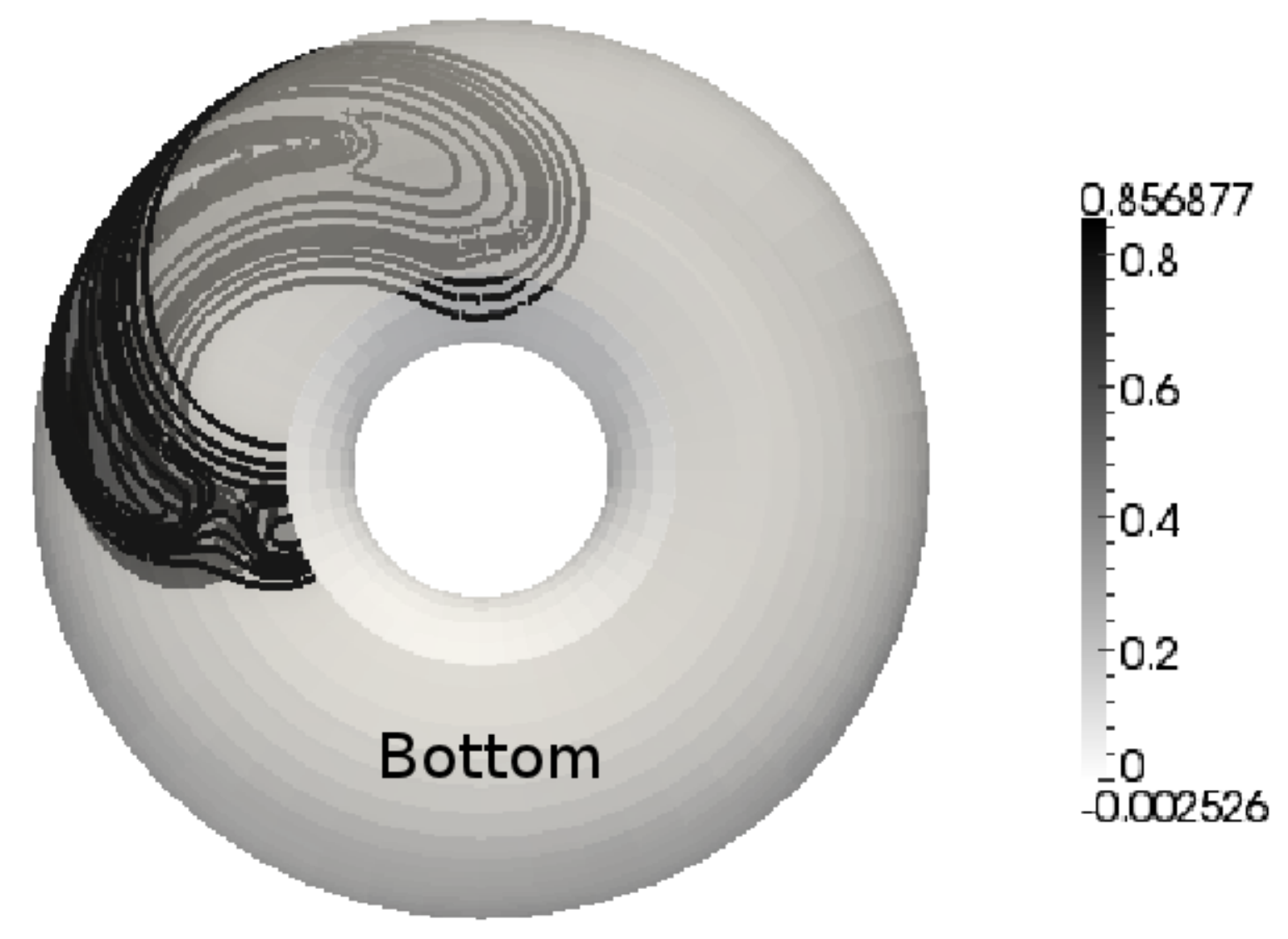} \includegraphics[height=3cm,width=3cm]{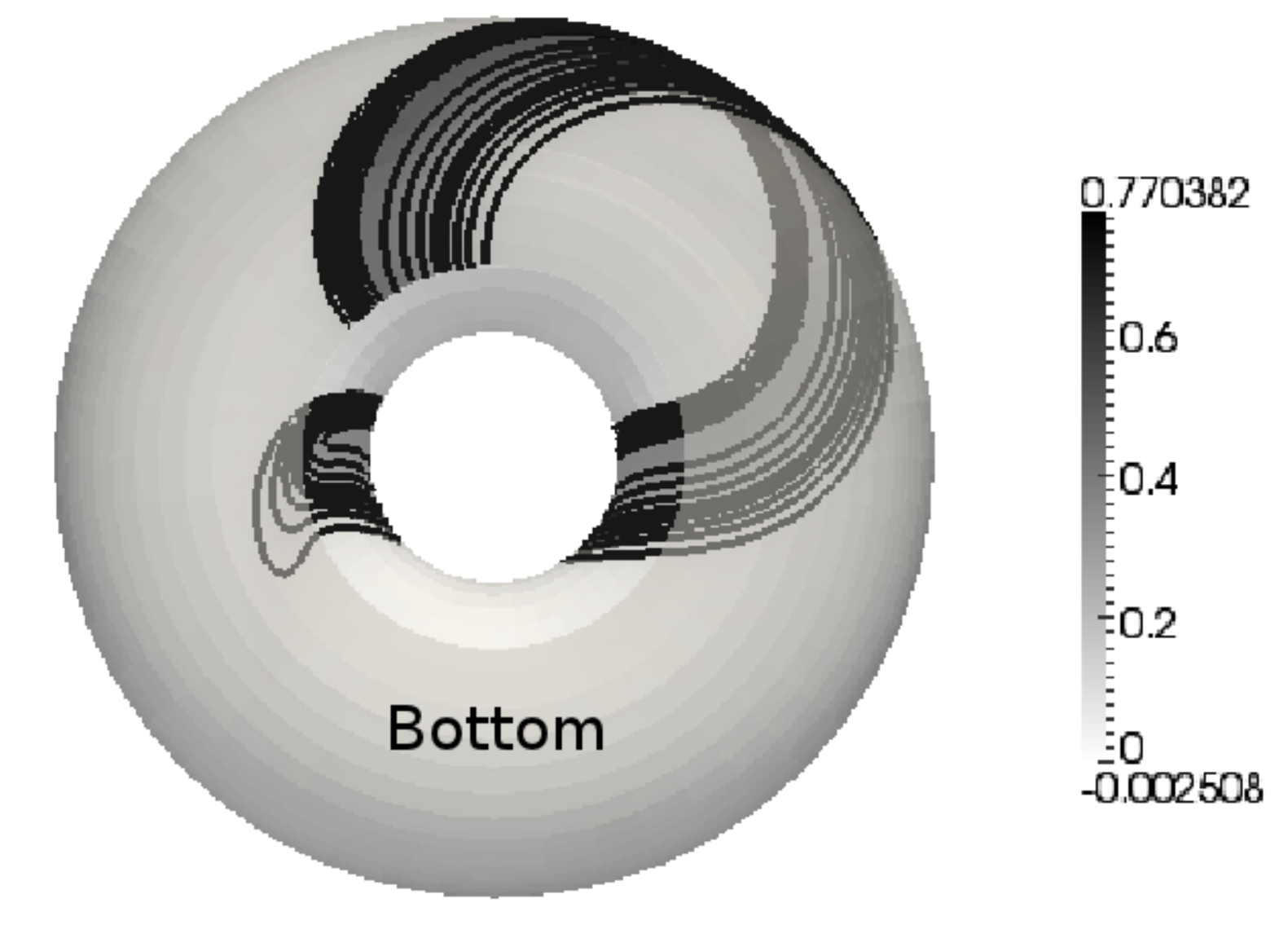} }
\caption{Perpetual propagation of the cardiac action potential. Transparent view from the bottom at $T$=$500.0$ (A), $T$=$1000.0$ (B), $T$=$1500.0$ (C), $T$=$2000.0$ (D). Same conditions as Figure \ref{AFReentry1}.}
\label {AFReentry2}
\end{figure}

If we change the magnitude of the anisotropy to 3.0, i.e., $\varsigma^{\theta \theta}= \varsigma^{\phi \phi}=3.0$, then the propagation appears almost the same until $T$=$200.0$ (roughly $0.126$ second after the initiation) as shown in Figure \ref{AFReentry1}A. Similar to the previous normal propagation, the short wave is blocked by anisotropy at the first time. However, the propagation of the collided wave significantly changes because of the weakened strength of the anisotropy. The collided wave at the back, with a less acute angle in the $\phi$-direction, passes through the anisotropy as shown in Figure \ref{AFReentry1}B. In other words, because the collided wave at the back approaches the anisotropy at a more acute angle than during the first propagation toward the anisotropy,  the changed magnitude of anisotropy fails to generate sufficient relative acceleration to stop the propagation, different from the normal propagation. Consequently, the propagation successfully proceeds throughout the anisotropy. Contrary to the normal propagation, the left anisotropy blocks the propagation the first time but allows the collided wave to pass through the second time. This is exactly the same role as a unidirectional block and seems to support proposition 3. However, the construction of the PV as unidirectional block can be numerous. For example, a similar phenomena can be observed if the width of anisotropy strip is moderately reduced even with the same magnitude of anisotropic coefficient. While most of the waves converge to a point, the remnant of the action potential on the PV keeps propagating along both the spherical shell and the PV to yield multiple wavelets (Figure \ref{AFReentry1}C-\ref{AFReentry1}D). Figure \ref{AFReentry2} also displays the \textit{perpetual} propagation of the cardiac action potential propagation in one direction to illustrate atrial reentry caused by this PV with the weakened anisotropy. This propagation is actually perpetual because it never terminates and self-initiates to override the SAN more than the approximate normal termination time of $T$=$250.0$ or equivalently $1.58$ second after the initiation.

\section{Discussion}
This paper proves the strong angular dependency of the cardiac action potential propagation in the PV to provide a model for the mechanism of unidirectional pathway which has been regarded as a necessity for atrial reentry resulting in atrial fibrillation.  In the perspective of the geometrical theory of wave trajectory which has been popularly used in space-time physics, the geometric analysis on the mechanism of unidirectional block is provided and some computation simulations are given to support this analysis. In this section, we mostly discuss the restrictions and the future work pertaining to the relative acceleration analysis.

In spite of several successful analyses, there are two main restrictions for applying the analytic technique directly to real cardiac electrophysiological phenomena. The first restriction is due to the assumption that the propagation is a traveling wave solution in the homogeneous media. The shape of the cardiac action potential should be roughly maintained in the propagation for valid analysis. For example, the dramatic upstrokes or downstrokes, the rapid slow-down or speed-up caused by the heterogeneity of the cardiac tissues cannot be dealt with when using the relative acceleration approach unless other approximations are introduced. In fact, anisotropy also induces changes in the shapes of the cardiac action potential, but the changes are relatively small, thus the analysis from the relative acceleration approach is not seriously flawed. The second restriction is regarding the singularities of anisotropy. The relative acceleration equation is derived in the neighborhood of a point in a curved surface, thus the discontinuity of anisotropy is not a problem, but the singularity of anisotropy is a problem. For example, when multiple anisotropies come across a point, it is equivalent to a singularity point in mathematics and this case cannot be considered with this analysis.

Nevertheless, the relative acceleration approach is an attractive option because it can be easily extended to more complex higher-dimensional anisotropic surfaces such as multiple layers of curved surfaces or three dimensional anisotropic space which are prevalent in the real heart. Also, revealing the relationship between the kinematics approach and the relative acceleration approach, or in other words the relationship between the shape of the wavefront and the distribution of the trajectories, may yield more practical tools applicable to clinical and surgical planning. In view of physiology, one of the most important consequences of this approach is the use of the trajectories and its validity in the cardiac electrophysiological phenomena. This result seems to implicitly suggest the existence of moving bodies in the propagation or at least something equivalent and it seems to open questions about what is in principle moving along the trajectories in the cardiac action potential propagation.

\section*{Acknowledgement}
Thanks are due to various readers to pointing out desirable modifications, but especially to Dr. Martins Bruveris (EPFL) and Dr. Emma Coutts (Durham University) for kind reading and suggestions. This paper is partially supported by the British Heart Foundation (BHF) to initiate the \textit{CardioMath} group lead by Professor Darryl D. Holm (Imperial College London) and Professor Nicholas S. Peters (Imperial College London). The use of Nektar++ for computational simulation is kindly advised and supported by Professor Spencer J. Sherwin (Imperial College London) and Robert M. Kirby (University of Utah) who also were the members of the group.

\section{Appendix A: Trajectory of the cardiac action potential on the surface $\mathcal{M}$}
To obtain the trajectory, we use the wavefront $S_j$ at time $t_j$. Let $n$ be the parameter of $S_j$ such that $S_j = S_j(n)$. Let's call $n$ the \textit{selector parameter}. Let $S_0=S_0(n)$ be the wavefront at time $t_0$. At $t_1 = t_0 + \Delta t$, we will have another wavefront $S_1=S_1(n)$ which is also parameterized by the same selector parameter $n$. See the left plot of Figure \ref{fig1}. Suppose the wavefronts $S_0$ and $S_1$ are both continuous and differentiable. For an arbitrary point $S_0(n)$ on the wavefront $S_0$, let the propagational velocity be $\mathbf{v}^0(n)$. Then, the trajectory will be defined similarly to that in classical mechanics \cite {Arnold}:\\
\\
\textbf{Definition}: For an interval $I \in R$, a \textit{trajectory} $P_0 (t)$ is a differentiable mapping $P_0: I \rightarrow \mathcal{M}^e$ to satisfy
\begin{equation*}
\left . \frac{\partial {P}_0}{\partial t} \right |_{S^0(n)} = \mathbf{v}^0 (n) = \lim_{\Delta t \rightarrow 0} \frac{ S_1 (n)  - S_0 (n) }{\Delta t}  .
\end{equation*}
\\
For example, let ${P}_0^h (t)$ be the trajectory for the selector parameter $0$ in $k$ discrete time steps. For the final time $T$, let $\Delta t$ be the time interval as $\Delta t = T/k$. In isotropic and homogeneous media, the propagational direction is normal to the wavefront, but in the presence of anisotropy, it may be aligned in the direction of anisotropy. For sufficiently small $\Delta t$, the trajectory at time $t_0 + \Delta t$ passing $S_0(n)$ is 
\begin{equation*}
{P}_0^h (t_0 + \Delta t) = S_0 (n) + \mathbf{v}^0 (n) \Delta t = S_1(n) .
\end{equation*}
Note that the trajectory ${P}_{n}^h$ meets the wavefront $S_j$ with the same parameter $n$, which is why $n$ is called the \textit{selector} of a trajectory. By repeating this procedure for $t_0 - \Delta t$, we can construct a continuous and differentiable curve $P_k$ in a neighborhood of $S^0(n)$ such that for a sufficiently small $\delta$, we have
\begin{equation*}
\lim_{\Delta t \rightarrow 0} \| {P}_k^h (t_0 + n \Delta t) \rightarrow P_k (t_0 + n \Delta t ) \|  < \delta ,~~~~ \mbox{for all } n .
\end{equation*}
With the affine parameter $\lambda = a t + b,~~ a,~b \in \mathbb{R}^+$, or the time that is measured by the moving body's clock, the trajectory is represented as $P_k = P_k (\lambda) \in \mathcal{M}$ as well as the wavefront $S_j = S_j (n) \in \mathcal{M}$. What remains is to prove the existence and uniqueness of a trajectory for each point on the curved surface $\mathcal{M}$.\\
\\
\textbf{Proposition A}: Consider a curved element $\mathcal{M}^e$ that is locally-Euclidean and two-dimensional manifold. For any point $p$ on the wavefront $S_j$ that is piece-wise continuous and differentiable in $\mathcal{M}^e$, there exists a unique and piece-wisely differentiable trajectory $P_k$ in the neighborhood of $p$. \\
\\
\textbf{Proof}: According to the above definition of the trajectory, it is sufficient to prove that, for any point $p$, there exists the unique vector $\mathbf{v}$ that is \textit{orthogonal} in the sense of the metric $g_{ij}$ to the tangent vector of the wavefront $S_j$. The consideration of the general orthogonality is because of anisotropy. Let $\mathbf{v}_{Sj}$ be the tangent vector of the wavefront $S_j$. The uniqueness of $\mathbf{v}_{Sj}$ is provided by the assumption that the wavefront is continuous and differentiable. Since $\mathcal{M}^e$ is a two-dimensional manifold, there is a function $\mathcal{C}$ which maps an Euclidean element $\Omega^2 \in \mathbb{R}^2$ into the curved element $\mathcal{M}^e$ such that $\mathcal{C}: \Omega^2 \rightarrow \mathcal{M}^e$. Without loss of generality, let $q$ be the point in $\Omega$ such that $\mathcal{C} (q) = p$. Let $s_k$ be the Euclidean axis of $\Omega^2$ and let $\partial / \partial s_k \equiv \mathbf{v}_{sk}$ be the tangent vector of the Euclidean axis for $k=1,~2$. Considering the tangent vector is mapped by $\mathcal{C}$ such as \cite{Weatherburn}
\begin{equation*}
\mathcal{C} \left (  \mathbf{v}_{s1}  \right )  \equiv \mathcal{C} \left (  \frac{\partial}{\partial s_k} \right ) = \frac{\partial \mathcal{C} }{\partial s_k} \equiv {\mathbf{v}}'_k,~~~ \mbox{for each } k,
\end{equation*}   
we see that each ${\partial \mathcal{C} }/{\partial s_k}$ becomes the tangent vector at $p \in \mathcal{M}^e$. Then, $({\mathbf{v}}'_1, {\mathbf{v}}'_2)$ constitute a linear basis of the tangent plane at $p$ where the tangent vector $\mathbf{v}_{Sj}$ lies, by definition. Thus, by an orthogonal procedure, for example by the Gram-Schmidt process in the tangent plane \cite{Strang}, there exists a unique $\mathbf{v}$ such that $\mathbf{v}$ is independent of $\mathbf{v}_{Sj}$ and
\begin{equation*}
( \mathbf{v} , \mathbf{v}_{Sj} ) = g_{ij}.~\square
\end{equation*}
It is clear to see that the uniqueness of the trajectory in the neighborhood of a point $p$ is the direct consequence of the uniqueness of the propagational vector and the smoothness of the wavefront. For example, for an isotropic sphere with a point initialization from the pole, the wavefront is aligned along the azimuthal angle and the trajectory along the polar angle for every point. It should be pointed out that the trajectory does not have to be orthogonal to the wavefront everywhere because of the feasible presence of anisotropy on the surface.

\section{Appendix B: Relative acceleration from a discrete model}

To qualitatively explain how a large relative acceleration can be interpreted as conduction failure for the justification of the following hypothesis, we first consider stopping conditions in a discrete model consisting of individual myocardial units that represent cells or tissues. Motivated by the widely known facts on myocardial cells or tissues, we give the following properties to myocardial units: \\
\\
(i) One unit faces multiple units such as hexagonal packing. \\
(ii) One unit can excite a limited number $(N_{\max})$ of neighboring units and when they try to excite more than $N_{\max}$, none of the neighboring units reach their threshold potential (TP) for excitation. \\
(iii) After a unit is excited, it undergoes the recovery process, called the refractory period (RP), and is not excited during RP.\\
\\
For the structural property (i) of myocardial tissues as reported in ref.  \cite{Hoyt} \cite{Peters}, the property (ii) provides the unique characteristics of excitation propagation to restrict the maximum number of excited cells that one cell can excite, known as \textit{impedance mismatch} or \textit{sink-source mismatch}. Property (iii) is also well known. Depending on ionic currents of $Ca^{2+}$ and $K^+$, the minimum time for the recovery of excitability is expressed as a function of RP that is also correlated with action potential duration (APD) \cite{Nattel}.

\begin{figure}
\centerline{\includegraphics[height = 8cm, width=8cm]{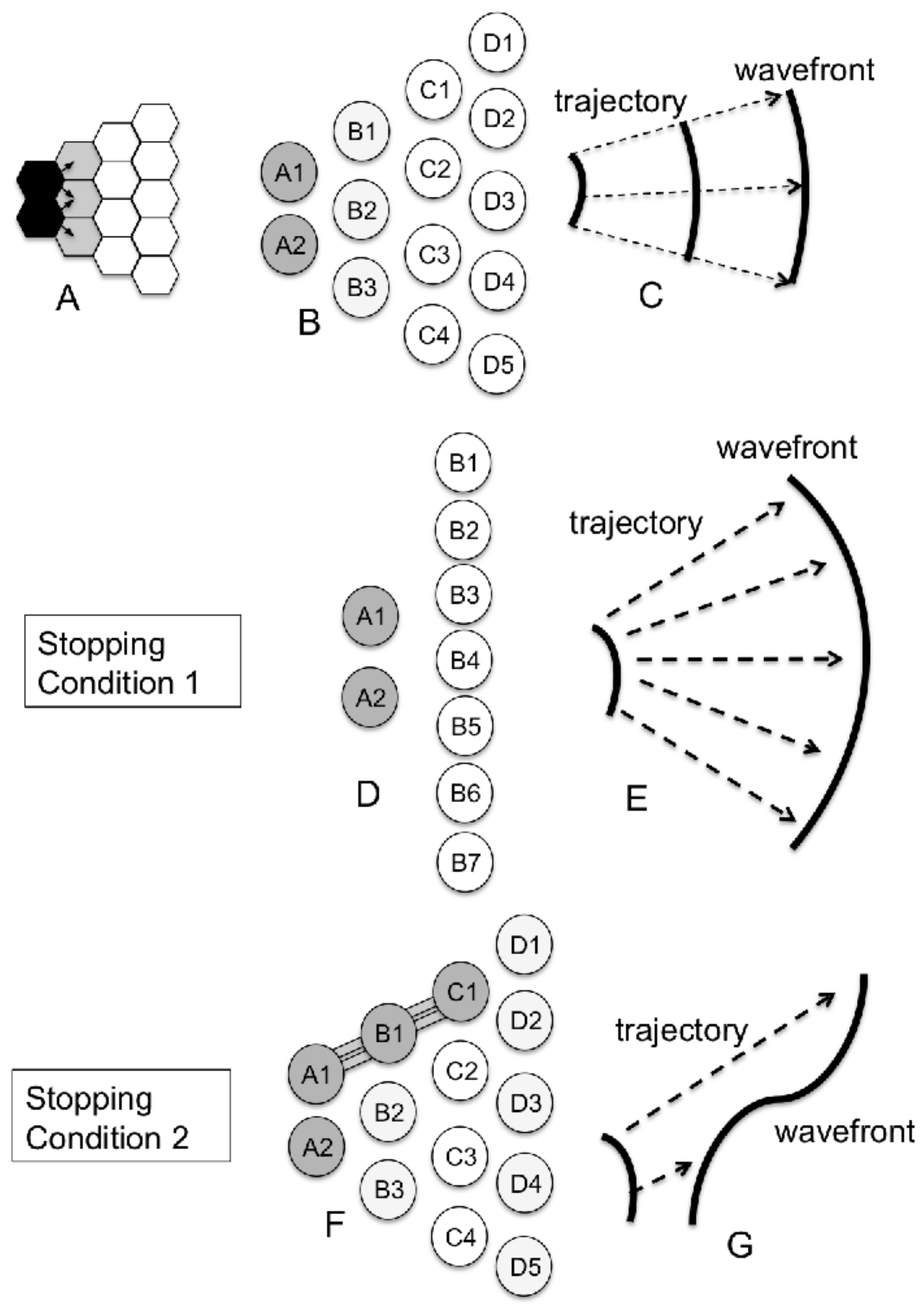} }
\caption{Representation of stopping conditions for conduction failure in individual myocardial cells and in continuous waves.}
\label{relaccel2}
\end{figure}

Figure \ref{relaccel2}A shows myocardial tissues modeled as hexagonal packing. The geometric-free discrete representation is shown in Figure \ref{relaccel2}B, where a dark color indicates the cell is excited and white colour indicates the cell is in excitable state. Figure \ref{relaccel2}C shows the wavefront to represent the line of excited tissues at each time.

For the hexagonal packing of myocardial units as shown in Figure \ref{relaccel2}A, we can also represent it with a geometric-free discrete model as shown in Figure \ref{relaccel2}B in which the number of units in each layer reflects the shape of the hexagonal packing. Starting from (A1,A2) units, the excitation propagates to (B1,B2,B3) units, (C1,C2,C3,C4) units and so forth. Note that the first letter of the unit indicates that the same layer must be excited in all of the units and the excitation occurs only in the alphabetical order, i.e., a C-unit cell cannot excite a B unit or a C unit. The goal of this discrete model is to expand the concept of the unit from cell to tissue, because when a cell satisfies the aforementioned properties, a tissue being a group of cells also satisfies the same properties if we neglect the time difference of excitation within a tissue.

No matter what this one unit in the discrete model represents, if we connect excited units at every time interval, we obtain the wavefront of cardiac excitation propagation. This wavefront represents a series of myocardial units that are in the process of depolarization. Consequently, the trajectory can be defined as if the cardiac wave is a physical wave, that is, a collection of moving particles. Upon these representations and notions, we describe the stopping conditions of the excitation propagation as the following: Without a loss of generality, \\
\\
\textit{Let three be the maximum number of units one can excite in the model of hexagonal packing}, i.e., $N_{\max}=3$. \\
\\
The first stopping condition is obtained directly from property (ii). In Figure \ref{relaccel2}D, if the A1 and A2 units attempt to excite the B-units, then each unit needs to excite an average of 3.5 units. Consequently, the electric potential of all B-units reaches below the TP and, consequently, fails to be excited. In brief, the excitation propagation stops or conduction failure occurs if: \\
\\
\textbf{[SG-i]} \textit{The average number of units to be excited exceeds the maximum number of excitable units.}    \\
\\
This stopping condition has often been cited when excitation propagation faces abrupt tissue expansion, such as a narrow cell strand connected to a large rectangular cell \cite{Rohr}. In the same context, we can explain why a small gap between ablation lesions, burned and dead myocardial tissues by catheter, does not allow excitation leakage. It is important to note that the excessive number of the B-units represents the geometric distribution of myocardial units, which can be interpreted as geometry in physics.

On the contrary, less attention has been given to the second condition. In Figure \ref{relaccel2}F, the geometry of hexagonal packing is the same as that of normal propagation in Figure \ref{relaccel2}B, although the A1 unit propagates rapidly only along the B1 unit, and finally to the C1 unit. This type of propagation is common in myocardial fibers or myocardial sheets, which we generally call anisotropy \cite{Panf}. Because the C1 unit is the only excited cell in C-units, the C1 unit attempts to excite all the five D-units. Let T1 be the time when the C1 unit uses all of the electric potential for the D-units. At the same time, the A2 unit proceeds to excite the (B2, B3) units and subsequently the (C2, C3, C4) units. Let T2 be the time when the (C2, C3, C4) units are all excited. There is a critical difference concerning the times T1 and T2: if T2 is earlier than T1, that is, if the (C2, C3, C4) units also can contribute to exciting D-units along with the C1 unit, then all the D-units can be excited and the propagation continues. However, if T1 is earlier than T2, that is, if when the (C2, C3, C4) units are excited, D-units are already in RP so that the (C2, C3, C4) units have no units to excite and propagation stops at the D-units. In other words, excitation propagation also stops, or conduction failure occurs if \\
\\
\textbf{[SC-ii]} \textit{The excitation time of other C-units exceeds the propagation time from the C1 unit to the D-units consisting of more than the maximum number of units one unit can excite}.  \\
\\
To simplify the above condition, we introduce another property of myocardial units as:\\
\\
(iv) The excitation time by a myocardial unit is roughly proportional to the number of cells that one cell has to excite. \\
\\
With the property (iv), the time T1 only depends on the propagational time from the A1 unit to the C1 unit, because the time required to propagate from the C1 unit to the D-units is fixed. Moreover, the ratio of velocity is at maximum four times higher on fibers than the normal excitable media \cite{Panf}. Thus, it is reasonable to say that the time T1 is relatively fixed. On the other hand, the excitation time of other C-units depends on the geometry of the B-units and the C-units other than the B1 unit and the C1 unit. Thus, the time T2 varies widely and is dependent upon the geometry of the C-units and the D-units.

These discrete models can be naturally interpreted in continuous wave forms. This procedure can be performed by drawing a line for excited units at each time which represents the front of the excitation propagation, called a wavefront in light of the waves. For example, Figure \ref{relaccel2}C displays the wavefront (solid line) and trajectories (dotted line) of the excitation propagation for the discrete model of Figure \ref{relaccel2}B. In waveforms, the shape of the wavefront and trajectories now reflect the geometry of myocardial tissues.

\section{Appendix C: Examples and validations of the relative acceleration analysis}

In the following examples on various curved surfaces, we will display how mathematical analysis from the relative acceleration equation \eqref{relacceqntensor} coincides with the computational results and how the curvature of the geometry changes the impact of anisotropy for conduction failure. For easier reading, each example is organized such that \textbf{Problem} means the setting of cardiac excitation propagation on a curved surface, \textbf{RA-analysis} means the analysis and predictions from equation \eqref{relacceqntensor}, \textbf{Computational modeling} and \textbf{validation} means the validation by computation simulations in the above computational scheme and by some references, respectively.

\subsection{Anisotropic plane}
\begin{figure}[ht]
\centering
\vbox{
\includegraphics[height=4.0cm, width=4.0cm] {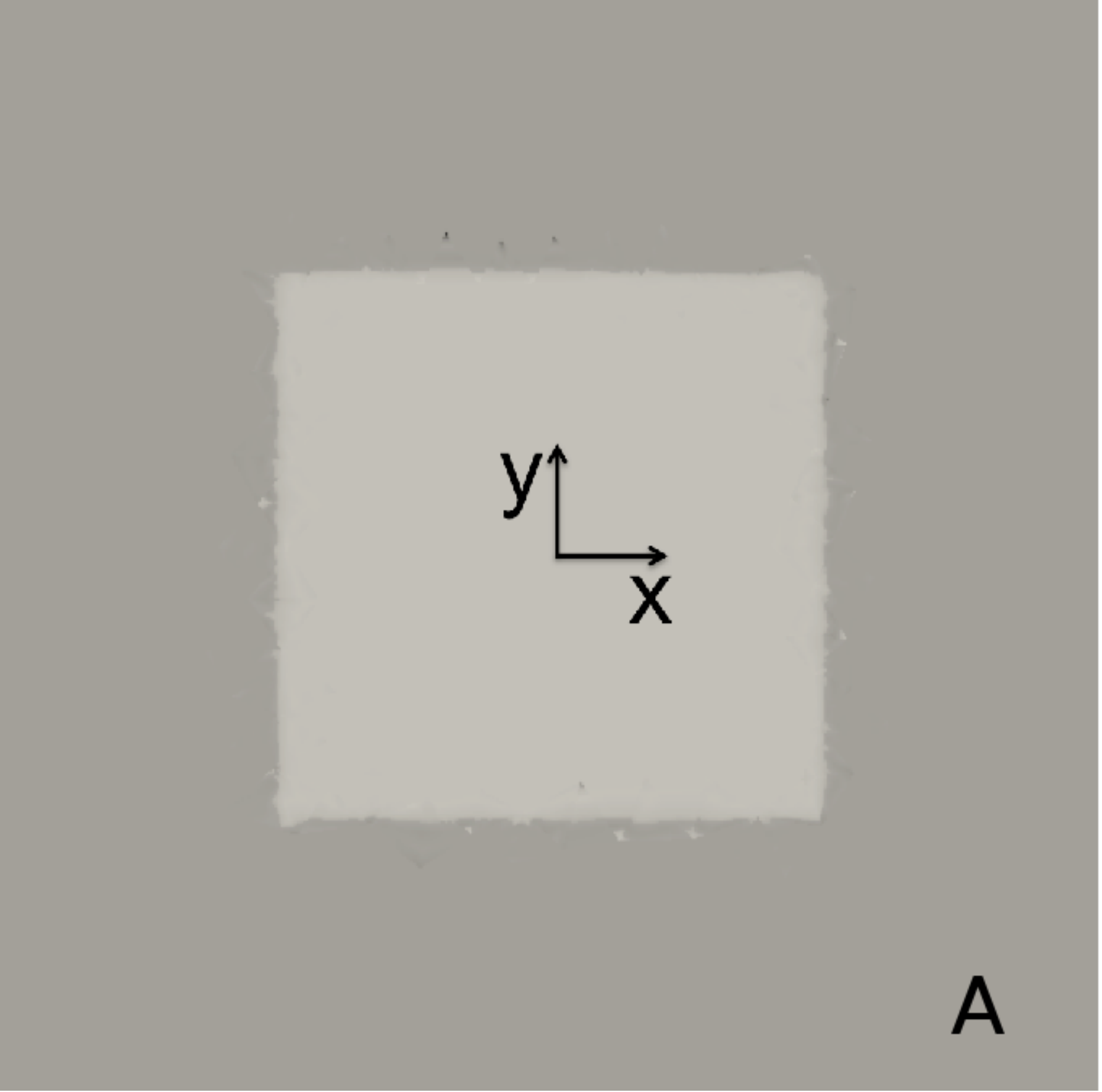} \includegraphics[height=4.0cm, width=4.0cm] {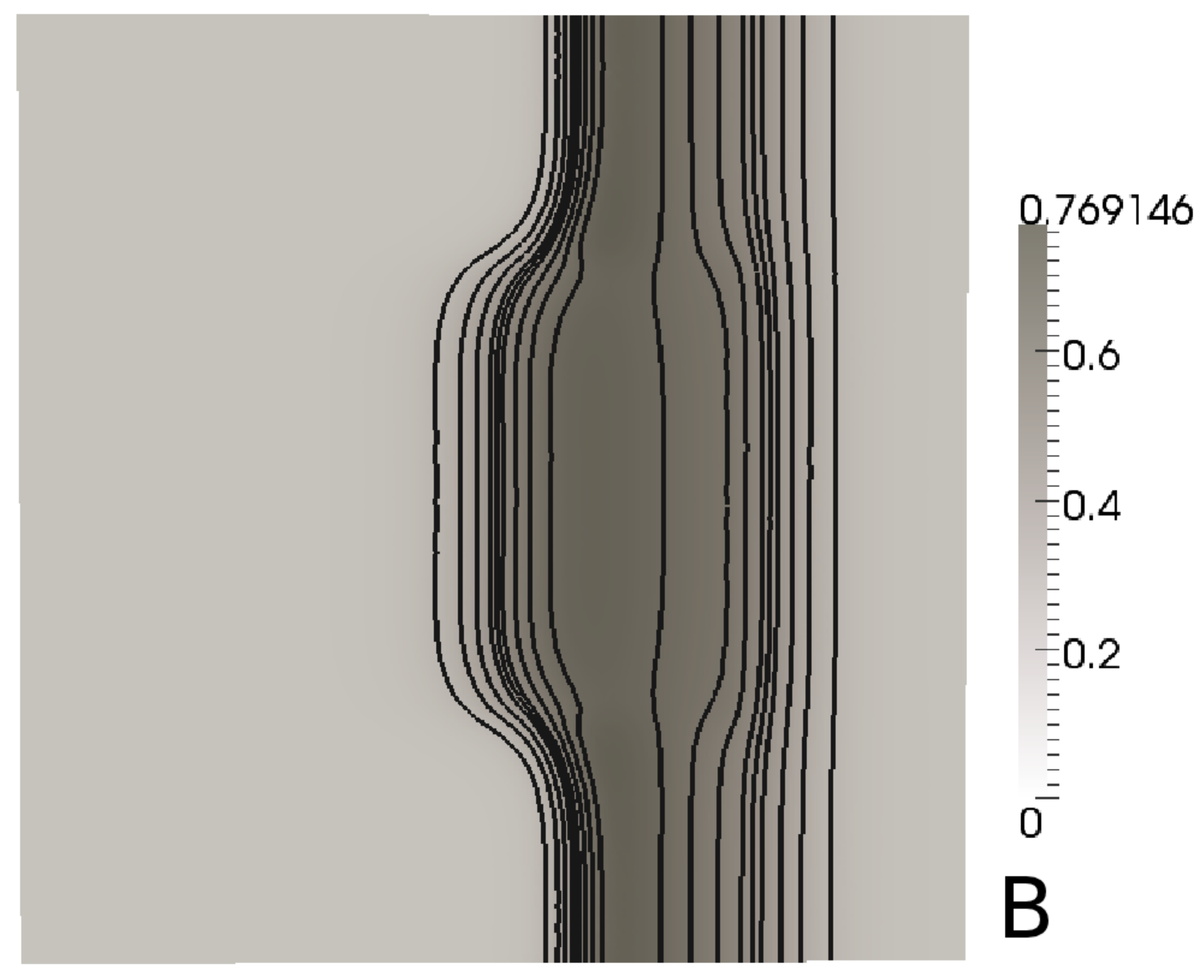}  \includegraphics[height=4.0cm, width=4.0cm] {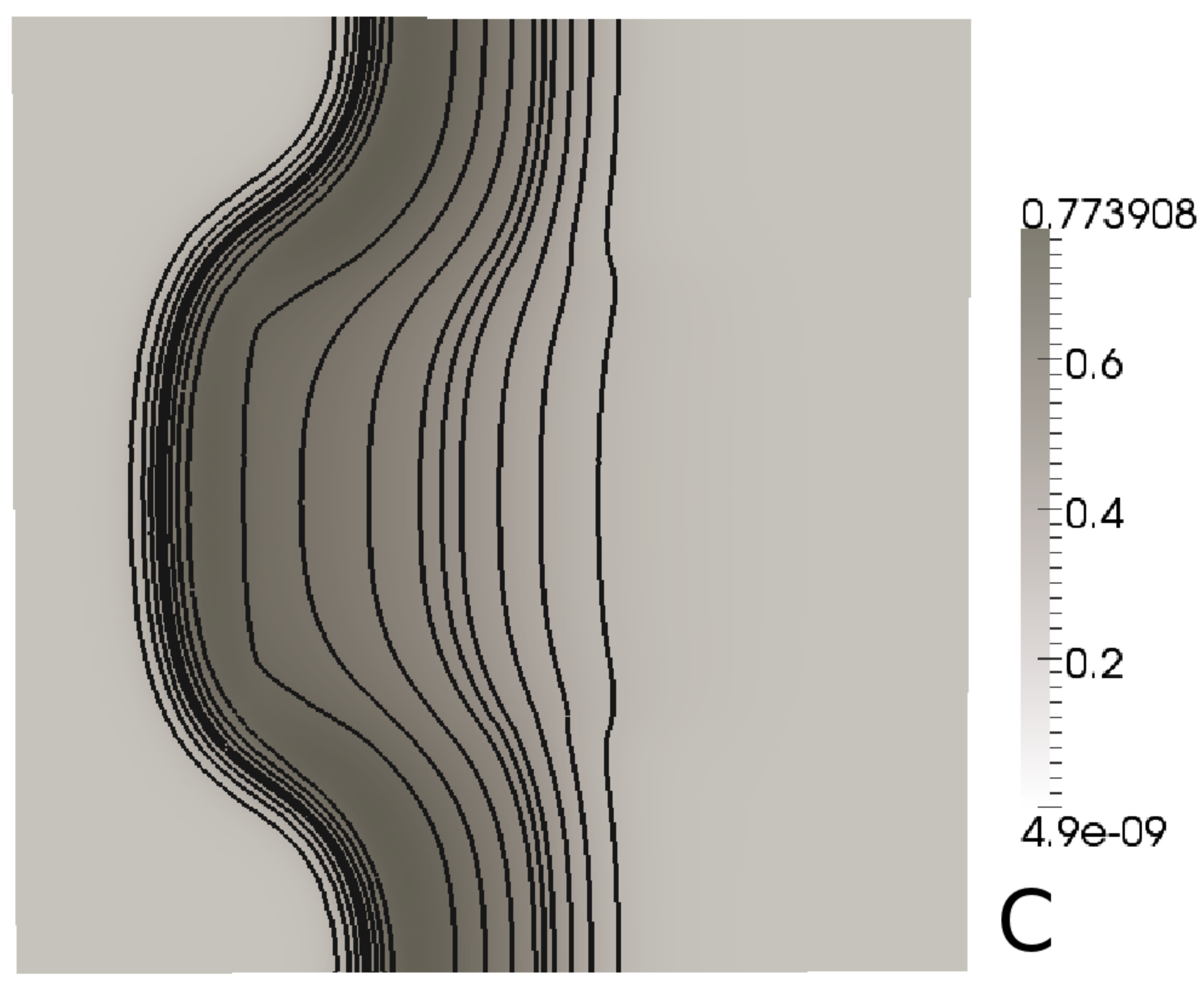} }
\caption{For $\varsigma^{xx} = 4.0$. Anisotropy block is located in the middle (A). After initiating from the right wall, the membrane potential ($u$) at $T=300.0$ (B), $T=500.0$ (C) from the right wall.}
\label{planeani1}
\end{figure}

\begin{figure}[ht]
\centering
\vbox{
\includegraphics[height=4.0cm, width=4.0cm] {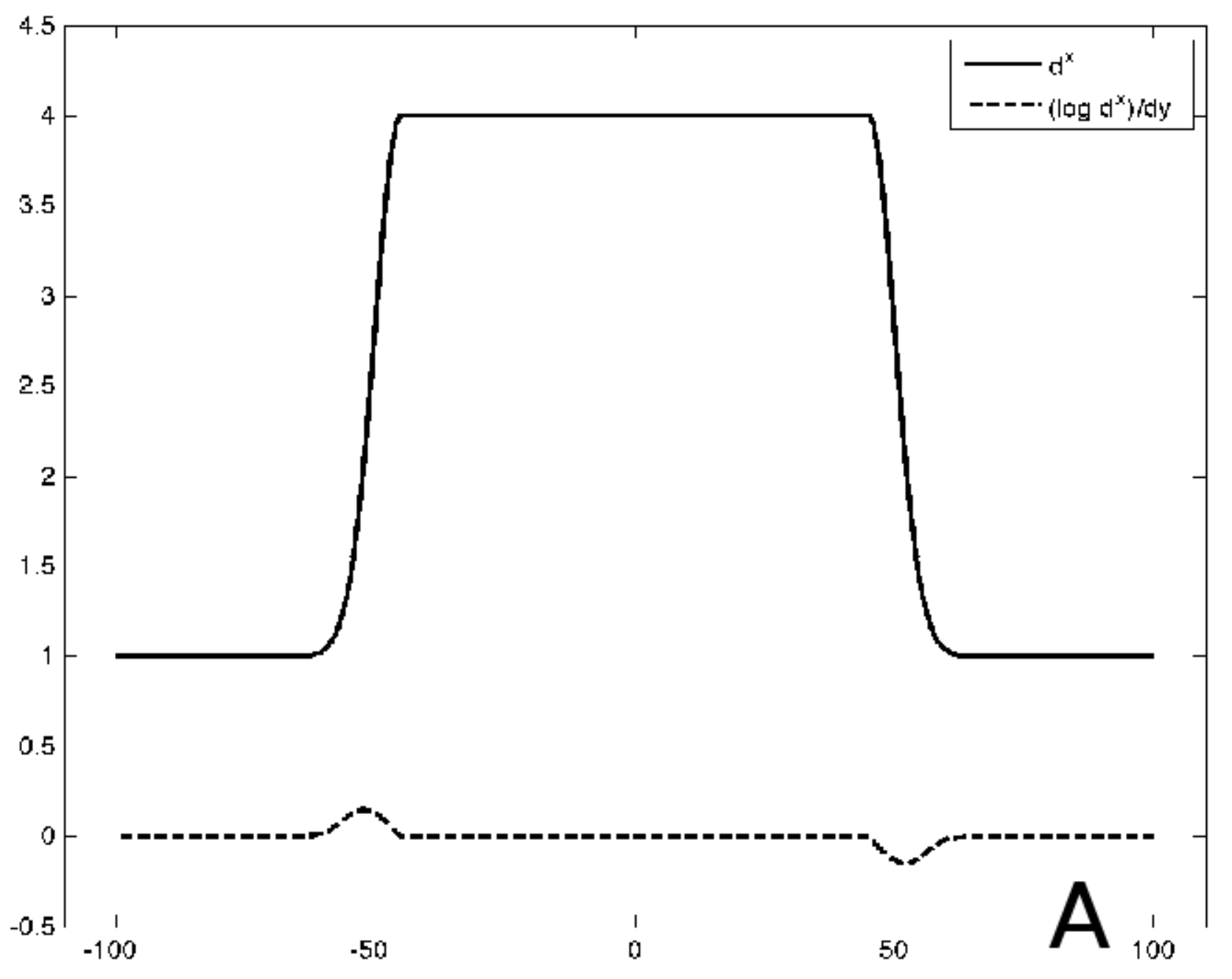} \includegraphics[height=4.0cm, width=4.0cm] {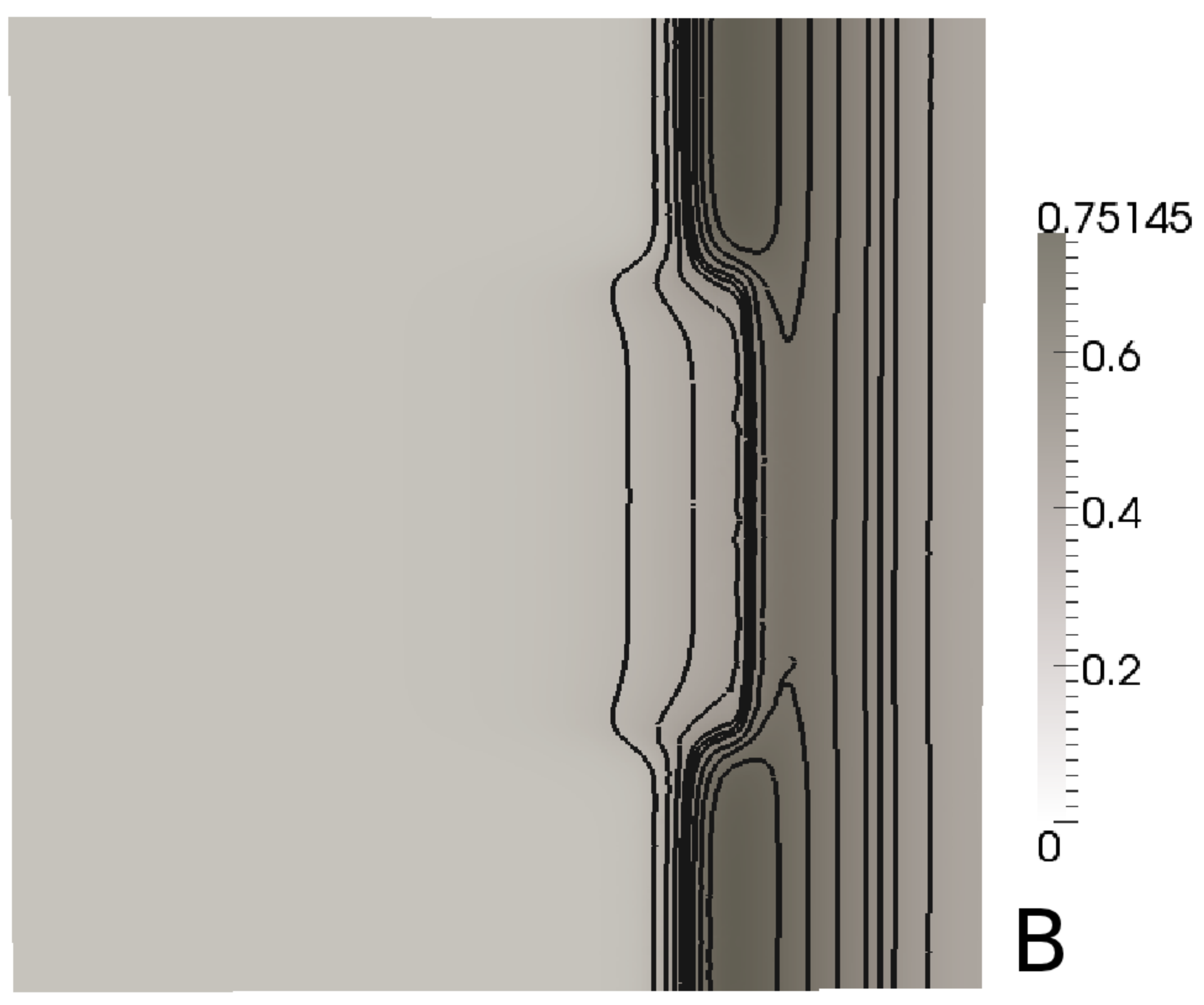}  \includegraphics[height=4.0cm, width=4.0cm] {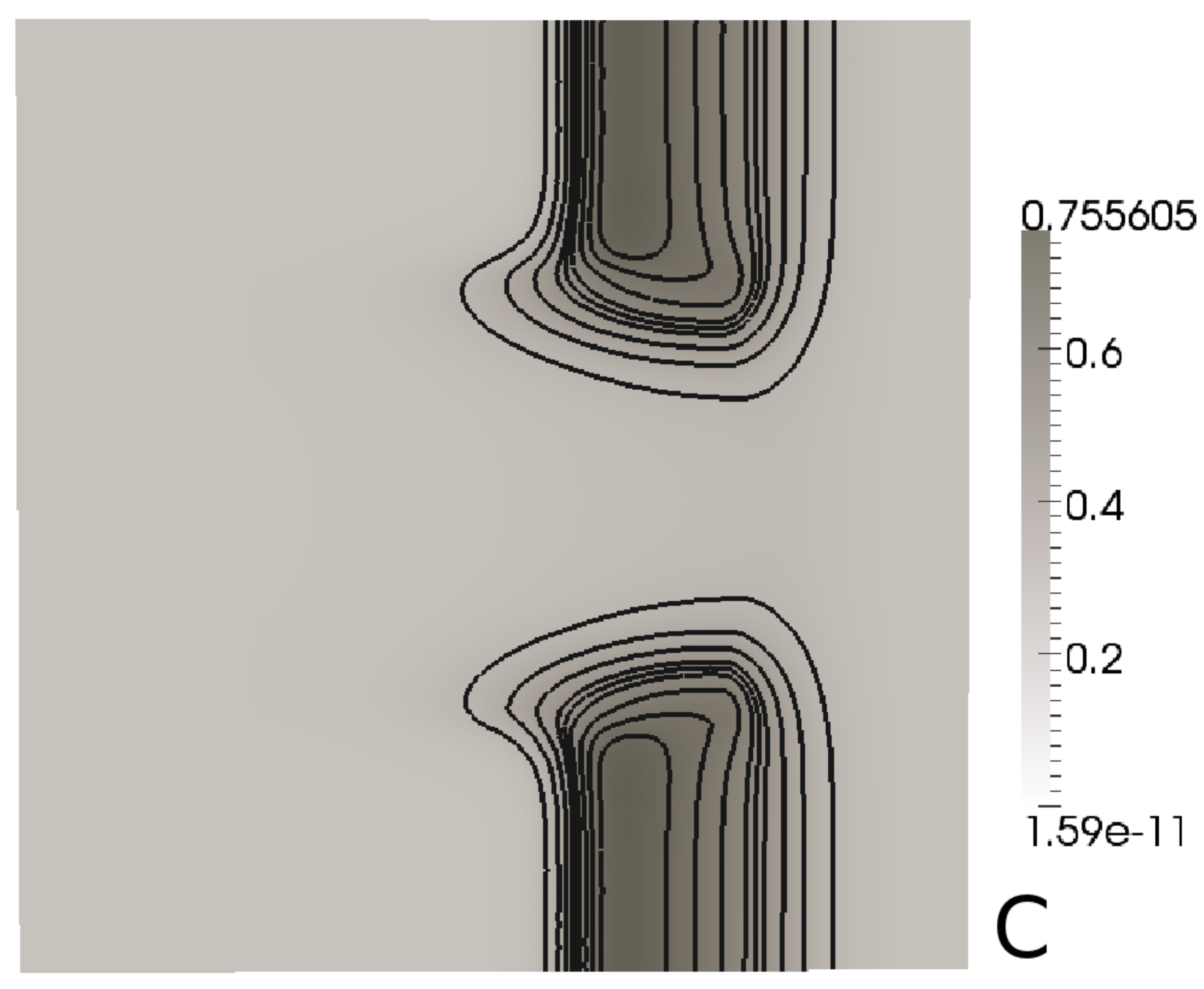} }
 \caption{For $\varsigma^{xx} = 10.0$. The distribution of $\varsigma^{xx}$ (solid line) and $\partial (\log \varsigma^{xx}) / \partial y$ (dashed line) along the $y$-axis (A). After initiating from the right wall, the membrane potential ($u$) at $T=200.0$ (B), $T=300.0$ (C).}
\label{planeani2}
\end{figure}

\textbf{Problem 1}: Consider a block of anisotropy in the middle of the plane. For the plane with $-100 \le x \le 100$ and $-100 \le y \le 100$, let the block be located in $-50 \le x \le 50$ and $-50 \le y \le 50$. Let the anisotropy be aligned with the direction of the $x$-axis, denoted as \textit{$x$-anisotropy} which expresses the diffusivity tensor $\mathbf{d}$ as $\mathbf{d} = \varsigma^{xx} \mathbf{x}$ in the block of anisotropy and $\varsigma^{yy}=1.0$ everywhere. See Figure \ref{planeani1}A. Also, let the excitation propagate in form of a plane wave in the $-x$ direction front the right wall and, consequently, the wavefront is in the direction of the $y$-axis. \\
\\
\textbf{RA-analysis}: Since these conditions mean that $g^{xx} = g^{yy} = 1$ and $\Lambda^x = 1,~\Lambda^y = 0$, the relative acceleration equation \eqref{relacceqntensor} boils down to
\begin{equation}
- \frac{\partial^2 n^i}{\partial \lambda^2} = \frac{\partial (\log \varsigma^{xx}) }{\partial y} \frac{\partial v^i}{\partial \lambda} + \frac{1}{\varsigma^{xx}} \frac{\partial^2 \varsigma^{xx}}{\partial x \partial y} v^i ,   \label{relplane1}
\end{equation}
where we used $n = y$ and $\partial v^i / \partial n = 0$. From the above equation, it is obvious to see that a large relative acceleration can be achieved \textit{only} by increasing the magnitude of anisotropy in the plane, especially at the interfaces of anisotropy where ${\partial^2 \varsigma^{xx}} / (\partial x \partial y)$ is large. This result is compatible with the break-up conditions drawn from the kinematics approach as shown by Morozov et. al. \cite {Morozov1999}. \\
\\
\textbf{Computational modeling}: In the computational simulation, we similarly observe that a large relative acceleration can only occur at the interfaces of the anisotropy block, i.e., in the line of $y=\pm50$ for the first component and two points $(50,50)$ and $(50,-50)$ for the second component. See Figure \ref{planeani2}A for the distribution of $\varsigma^{xx}$ and $\partial (\log \varsigma^{xx}) / \partial y$ along the $y$-axis. With $\varsigma^{xx}=4.0$, conduction failure does not happen as shown in Figure \ref{planeani1}B and \ref{planeani1}C. However, increasing the magnitude of anisotropy up to $\varsigma^{xx}=10.0$, though this large magnitude seems to be unrealistic in the biological system, leads to conduction failure in the block of anisotropy because the relative acceleration has increased significantly at the interface. Note that if the anisotropy changes slowly such that ${\partial^2 \varsigma^{xx}} / (\partial x \partial y)$ remains bounded, then no conduction failure would be induced by the anisotropy.\\
\\
\textbf{Problem 2, RA-analysis and validation}: Conduction failure can also occur even inside an anisotropic block. If the direction of anisotropy is oblique to the propagational direction, then conduction failure can happen inside anisotropy. Let $\varsigma$ be the magnitude of anisotropy and $\theta$ be the angle with respect to the direction of the excitation propagation. For the sake of simplicity, let the magnitude of anisotropy ($\varsigma$) be constant in $\Pi$. For the same planar propagation as before, the relative acceleration equation is expressed as
\begin{equation}
- \frac{\partial^2 n^i}{\partial \lambda^2} = - \varsigma \tan \theta \frac{\partial \theta}{\partial y}  \frac{\partial v^i}{\partial \lambda} + \left [ - \frac{\partial \theta}{\partial x} \frac{\partial \theta}{\partial y} - \tan \theta \frac{\partial^2 \theta}{\partial x \partial y} \right ] v^i.  \label{relplane2}
\end{equation}
Note that this equation implies that the action potential propagation can also be blocked by significant variable angle $\theta$ of the \textit{direction} of anisotropy as well as the magnitude. This result coincides with Davydov et. al. \cite {Davydov2004C} which displayed that the chiral anisotropy, which is aligned along the circumferential direction of circles, can break up the excitation propagation. But, equation \eqref{relplane2} suggests a host of possibilities for conduction failure by variations of the direction of anisotropy.

\subsection{Anisotropic sphere}
\begin{figure}[ht]
\centering
\vbox{
\includegraphics[height=4.0cm, width=4.0cm] {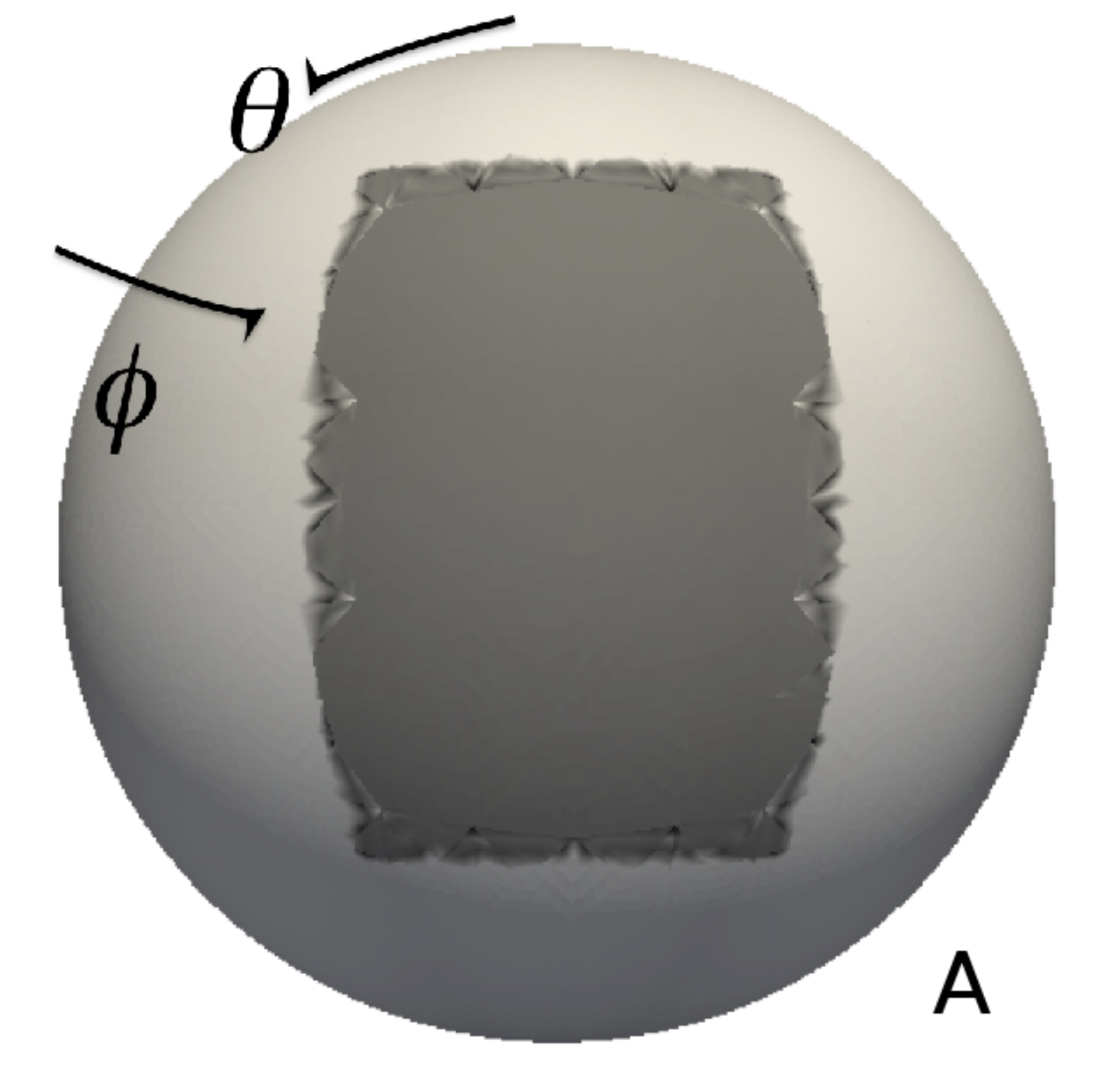} \includegraphics[height=4.0cm, width=4.0cm] {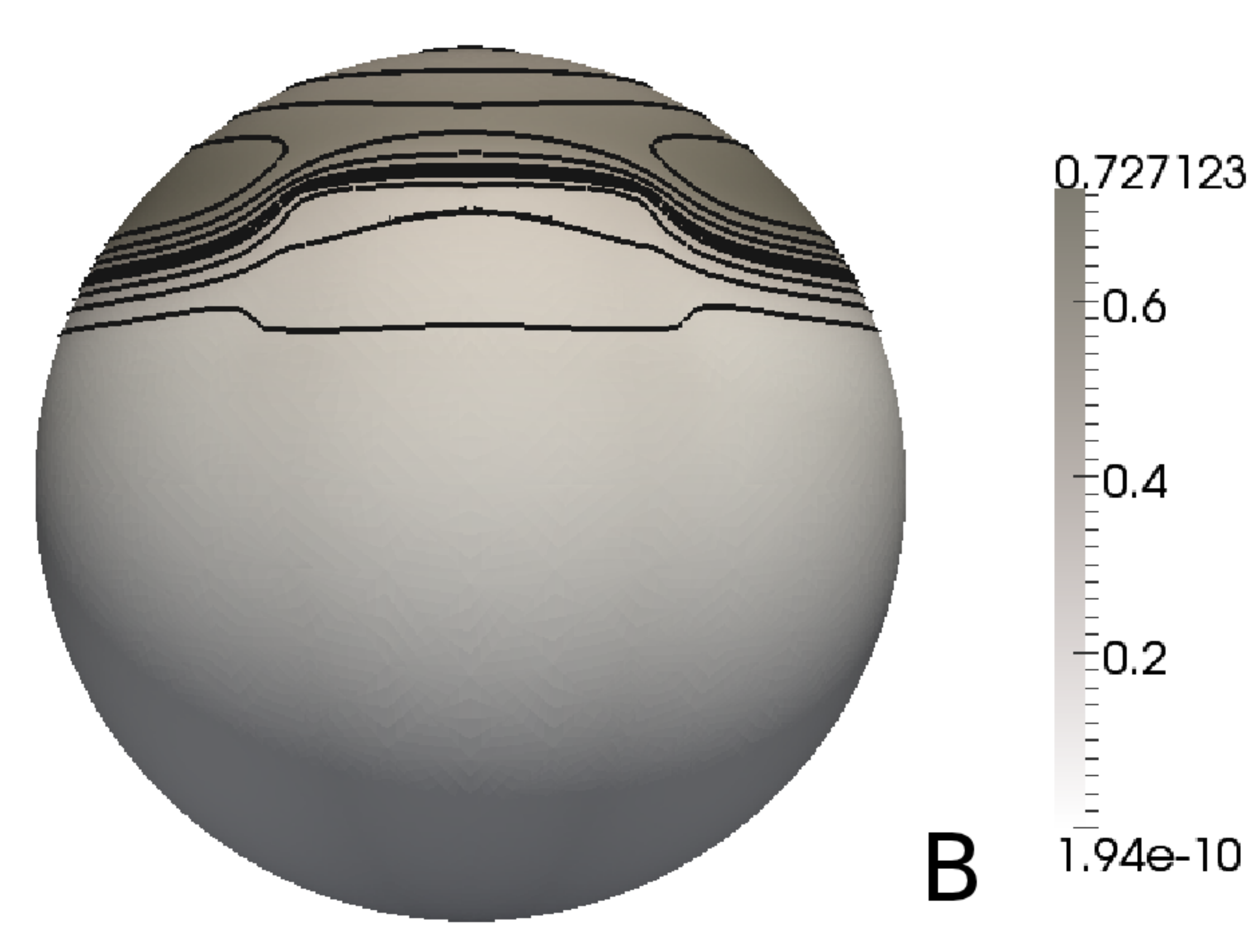}  \includegraphics[height=4.0cm, width=4.0cm] {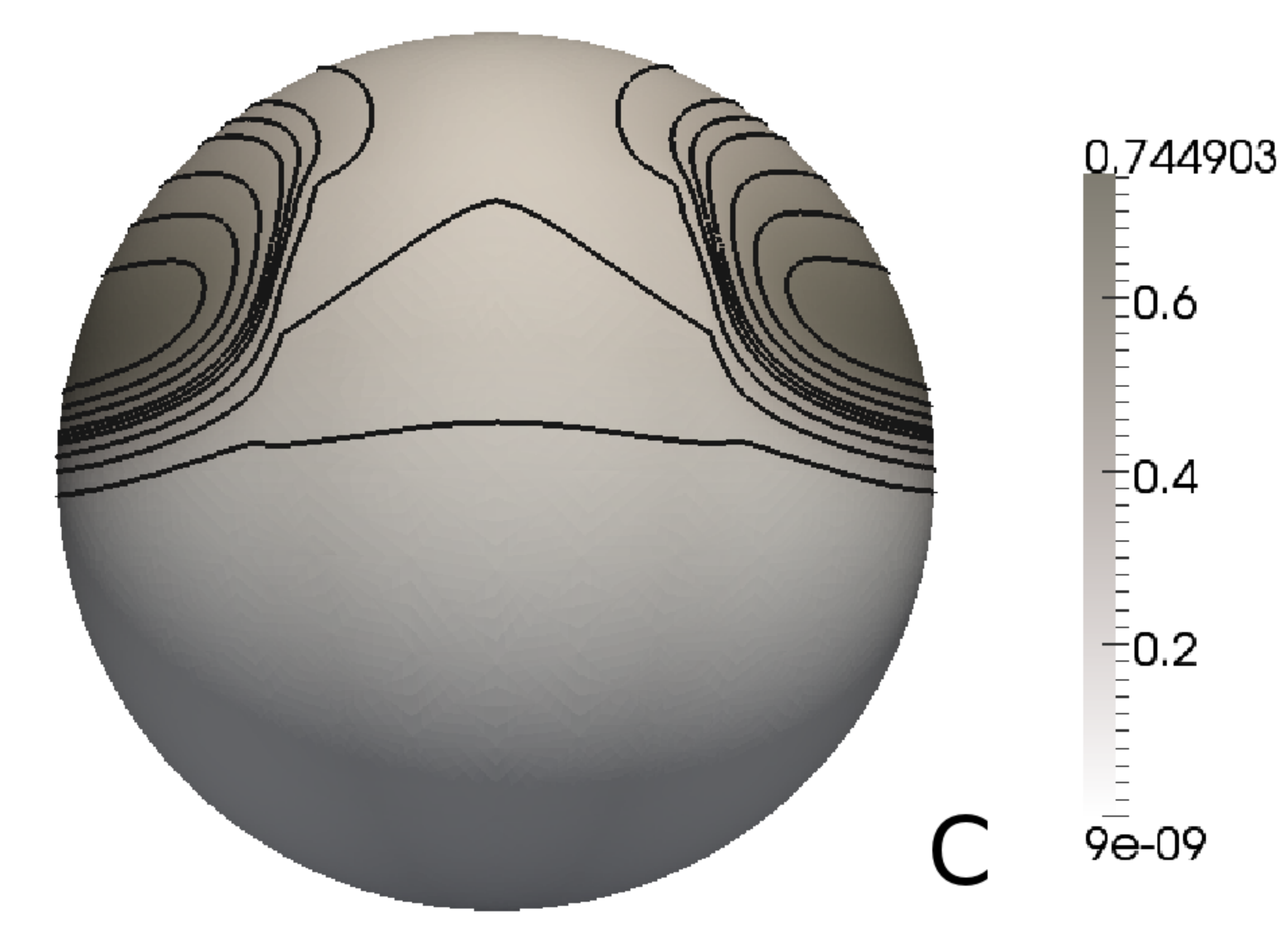} }
 \caption{For $\varsigma^{\theta \theta} = 10.0$. The block of anisotropy is located in the middle $\pi / 4 < \theta < 3 \pi / 4$ on a sphere (A). After point-initialized from the north pole, the membrane potential ($u$) at $T=300.0$ (B), $T=400.0$ (C).}
 \label{sphereani1}
\end{figure}

\textbf{Problem 3}: The relative acceleration analysis can be similarly applied to the propagation on anisotropic sphere. But, there is a difference. Contrary to the propagation in the plane, the planar propagation is impossible. Instead, the propagation should be point-wisely initiated, for example initiation from the north pole. Geometry tells us that the point-initialized excitation follows the line of the polar angle $(\theta)$ and the wavefront is in line with the azimuthal angle $(\phi)$. For sphere of radius $r=50.0$, consider a block of anisotropy located in $-\pi/4 \le \phi \le \pi/4$ and $-0.6 \le \theta \le 0.6$. See Figure \ref{sphereani1}A. Let the direction of anisotropy be both along the $\theta$- and $\phi$ axis in general, but because of the propagational direction, only $\theta$-anisotropy will contribute to the changes of the propagation.

\vspace{0.3cm}

\begin{figure}[ht]
\centering
\vbox{
\includegraphics[height=4.0cm, width=4.0cm] {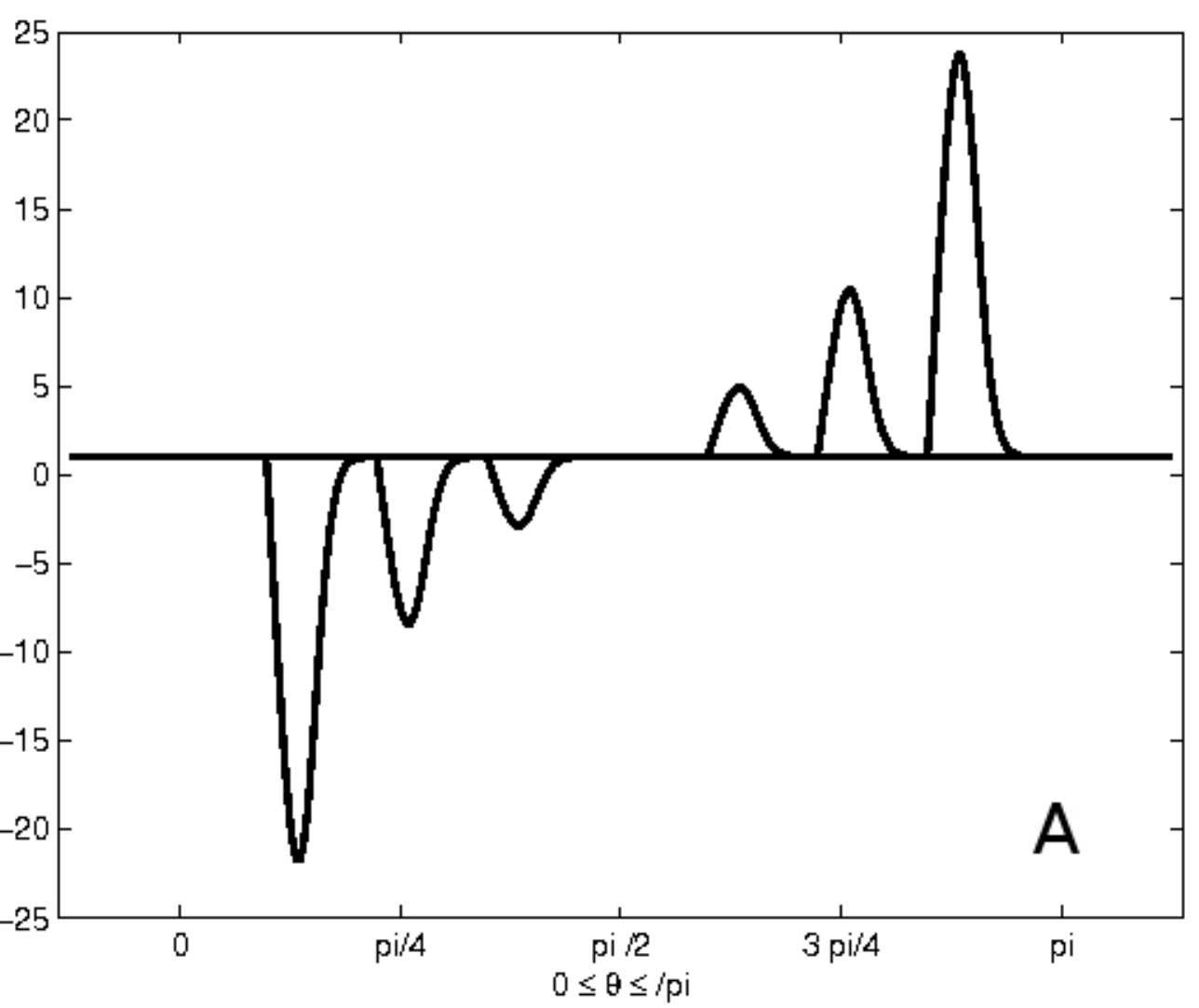} \includegraphics[height=4.0cm, width=4.0cm] {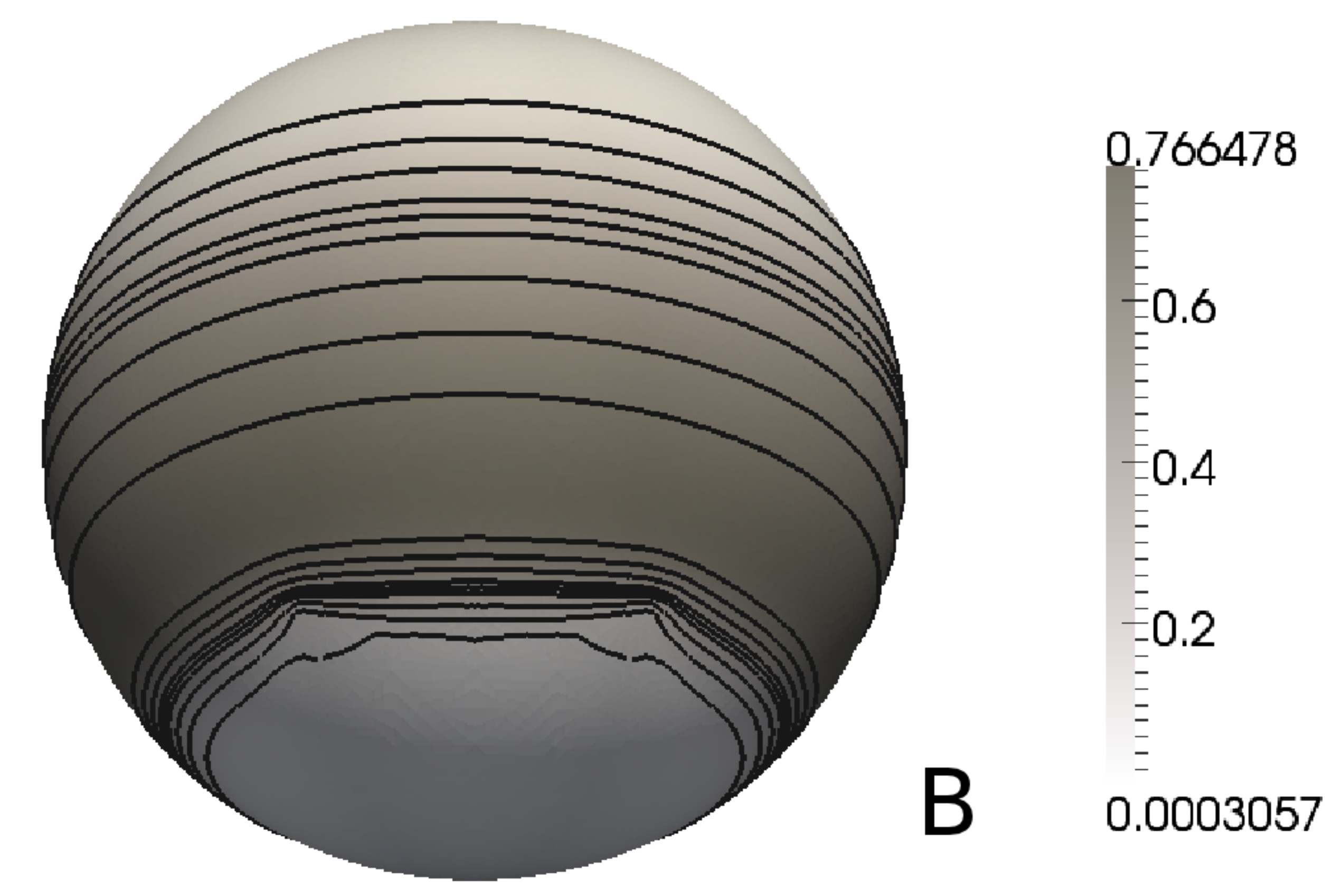} \includegraphics[height=4.0cm, width=4.0cm] {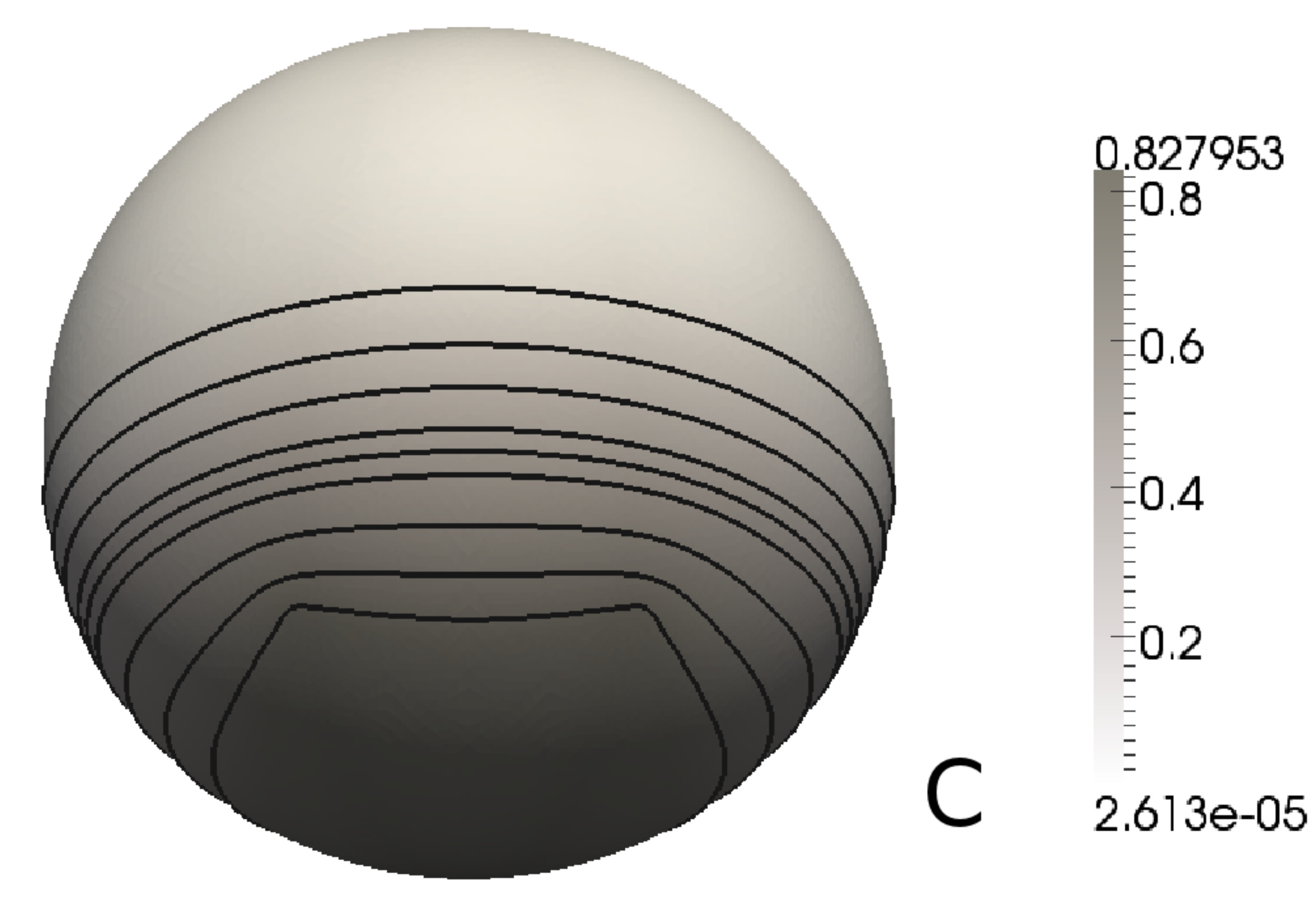} }
 \caption{For $\varsigma^{\theta \theta} = 10.0$. Magnitude of the last component in equation \eqref{relaccsphere} with respect to $\theta$ (A). The block of anisotropy is located in the bottom $ 3 \pi / 4 < \theta < \pi $. After point-initialized from the north pole, the membrane potential ($u$) at $T=600.0$ (B), $T=700.0$ (C).}
\label{sphereani2}
\end{figure}

\textbf{RA-analysis}: Geometrically, these conditions imply that $g^{\theta \theta} = {1}/{r^2}$ and $\Lambda^{\theta} = 1,~\Lambda^{\phi} = 0$, thus the relative acceleration equation \eqref{relacceqntensor} is expressed as
\begin{equation}
- \frac{\partial^2 n^i}{\partial \lambda^2} = \frac{\partial (\log \varsigma^{\theta \theta} ) }{\partial \phi} \frac{\partial v^i}{\partial \lambda} + \left ( \frac{1}{\varsigma^{\theta \theta}} \frac{\partial^2 \varsigma^{\theta \theta}}{\partial \phi \partial \theta} + \frac{\partial (\log \varsigma^{\theta \theta} ) }{\partial \phi} \cot \theta  \right ) v^i,   \label{relaccsphere}
\end{equation}
where we used $( {1} / {\sqrt{g}}) ({\partial (\sqrt{g} g^{kk} ) } / {\partial x_k } ) = g^{kk} \Gamma^m_{mk},~ m \neq k$ as  proved in Appendix D. Equation \eqref{relaccsphere} is similar to the equation \eqref{relplane1} for the plane, but the last additional component, i.e., ${\partial (\log \varsigma^{\theta \theta} ) } / {\partial \phi} \cot \theta $, appears. The above component indicates the dependency of the relative acceleration on the location of anisotropy with respect to the point of initialization. For example, for the point of initialization at the pole, let $\theta$ be the azimuthal angle of the sphere from the pole. Then, the \textit{magnitude} and \textit{sign} of the above component changes as the location of anisotropy changes. Figure \ref{sphereani2}A displays that the above component changes signs at $\theta=\pi/2$ and the magnitude of it increases as $\theta$ gets closer to $0$ or $\pi$. Consequently, since the change of signs of the above component can increase or decrease the total magnitude of the relative acceleration, equation \eqref{relaccsphere} implies that conduction failure also depends on the location of anisotropy with respect to the location of point initialization. In the perspective of the trajectory, this analysis is no surprise. With point initialization at the north pole, the trajectories diverge in the northern hemisphere up to the equator and from the equator the trajectories converge in the southern hemisphere. Thus, the use of anisotropy to increase the divergence of trajectories will add its relative acceleration only for the trajectories in the northern hemisphere and will decrease it for the trajectories in the southern hemisphere.\\
\\
\textbf{Computational modeling}: To confirm the above analysis, a block of anisotropy is placed on the sphere such that the first interface that the propagation meets is located at $\theta = \pi/4$ distance from the point of initialization, i.e., in the northern hemisphere. As shown in Figure \ref{sphereani1}B and \ref{sphereani1}C, the block of anisotropy with a sufficiently large magnitude causes conduction failure. On the other hand, move the point of initialization away from the block of anisotropy such that the first interface that the propagation meets is located at  $\theta = 3 \pi / 4$ distance from the point of initialization, i.e., in the southern hemisphere. Then, the block of the same anisotropy, with the same magnitude of anisotropy as before, does not cause conduction failure as shown in Figure \ref{sphereani2}B and \ref{sphereani2}C. This phenomenon seems to be well explained by the change of signs of the last component in equation \eqref{relaccsphere} representing the convergence and divergence of the trajectories.

\subsection{Anisotropic torus}

\begin{figure}[ht]
\centering
\vbox{
\includegraphics[height=4.0cm, width=4.0cm] {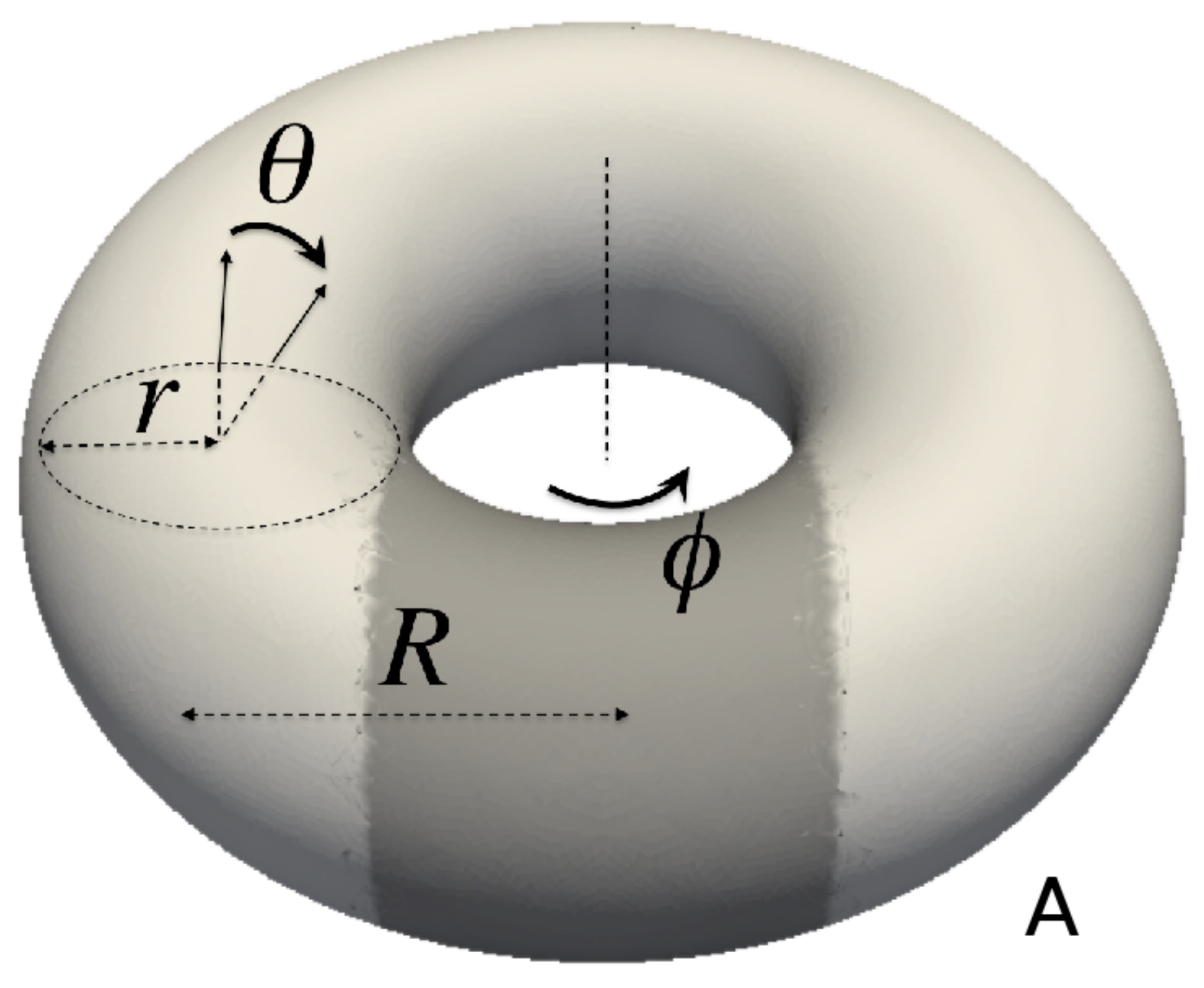}   \includegraphics[height=4.0cm, width=4.0cm] {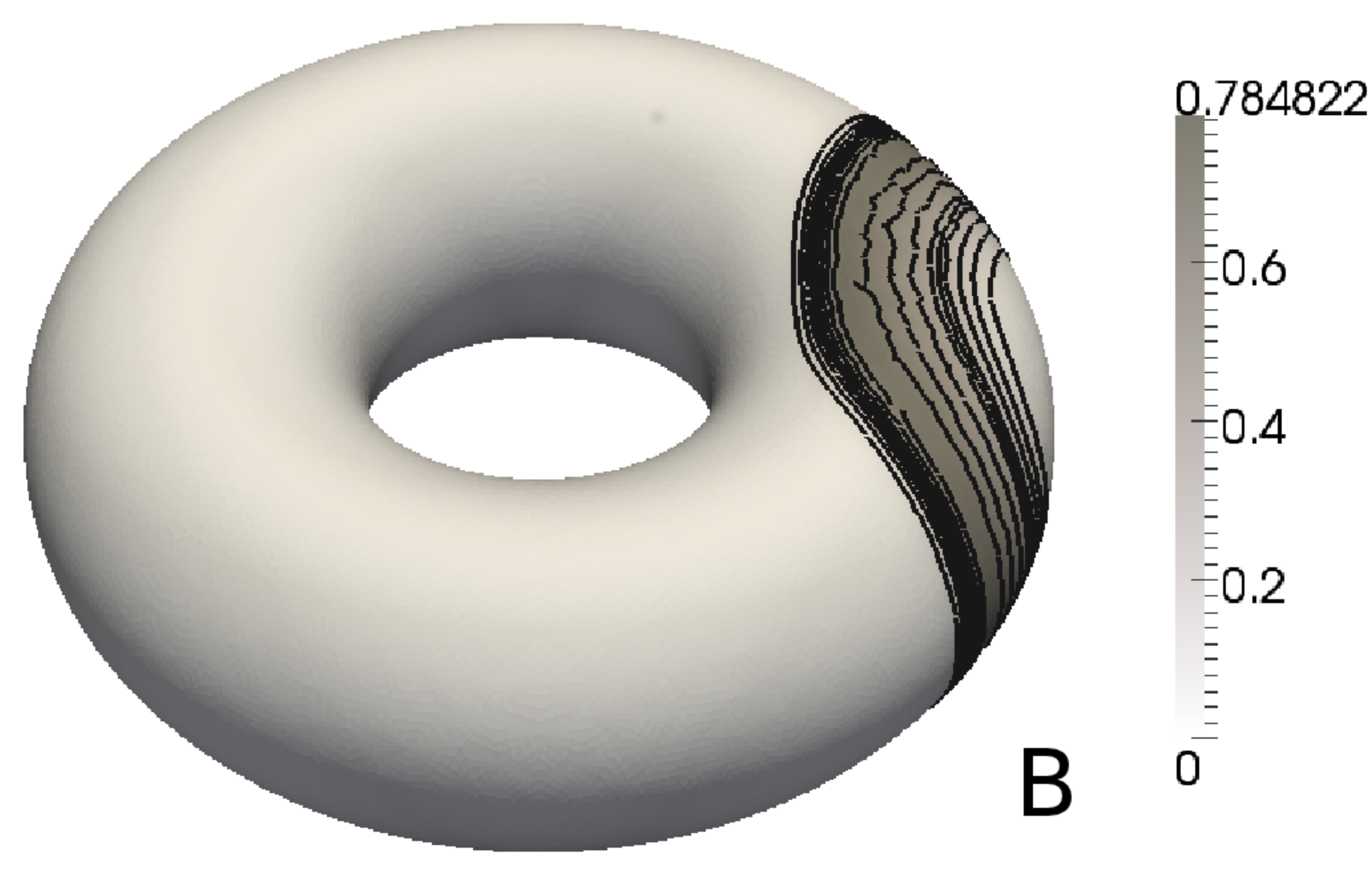}  \includegraphics[height=4.0cm, width=4.0cm] {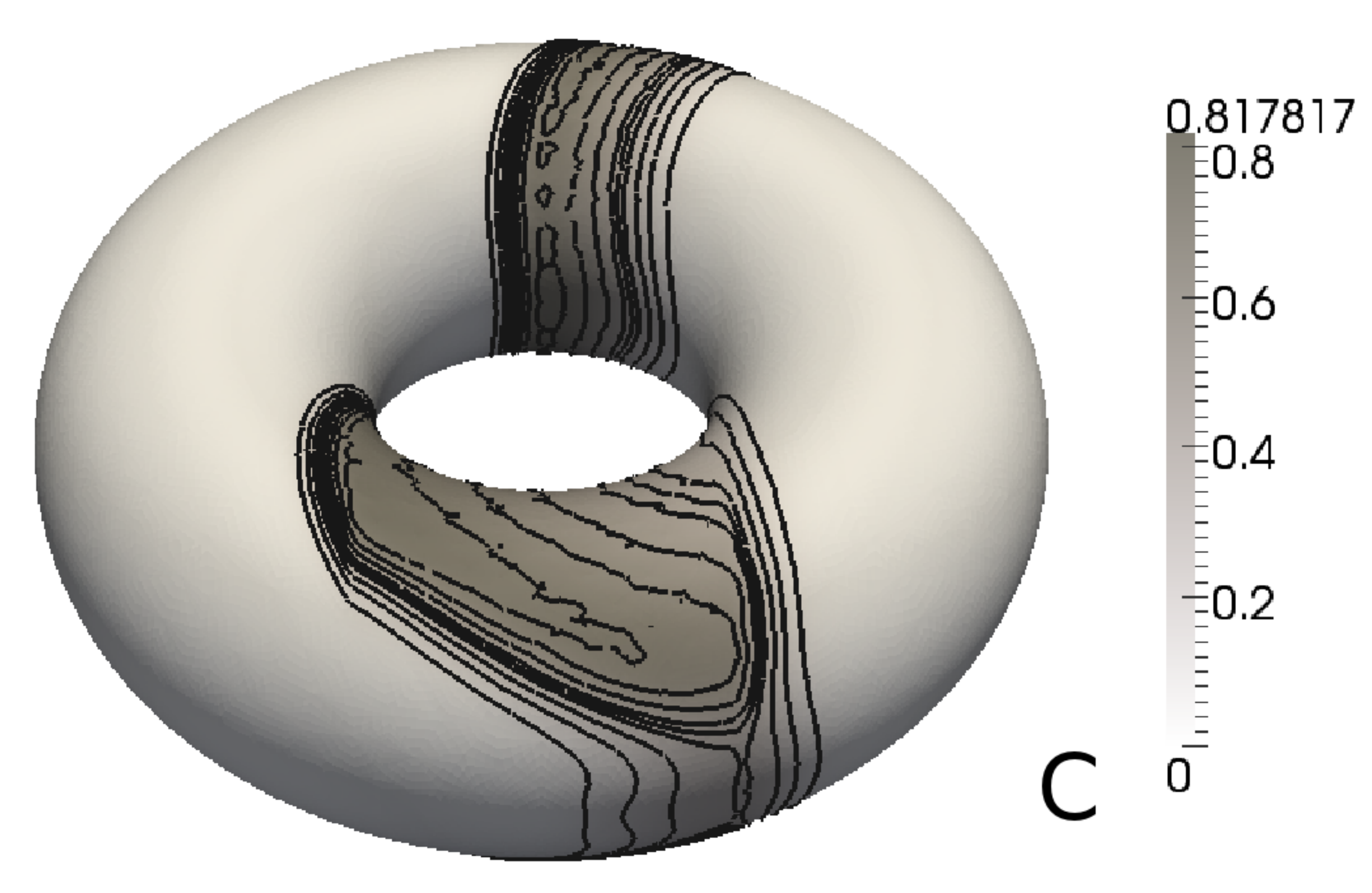} }
 \caption{Location of the anisotropy block (darkened) with $\varsigma^{\phi \phi}=8.0$ (A). After point initialized at the rightmost point of the above torus, the membrane potential ($u$) at $T=500.0$ (B) and at $T=1000.0$ (C).}
\label{torus2}
\end{figure}

\begin{figure}[ht]
\centering
\vbox{
\includegraphics[height=4.0cm, width=4.0cm] {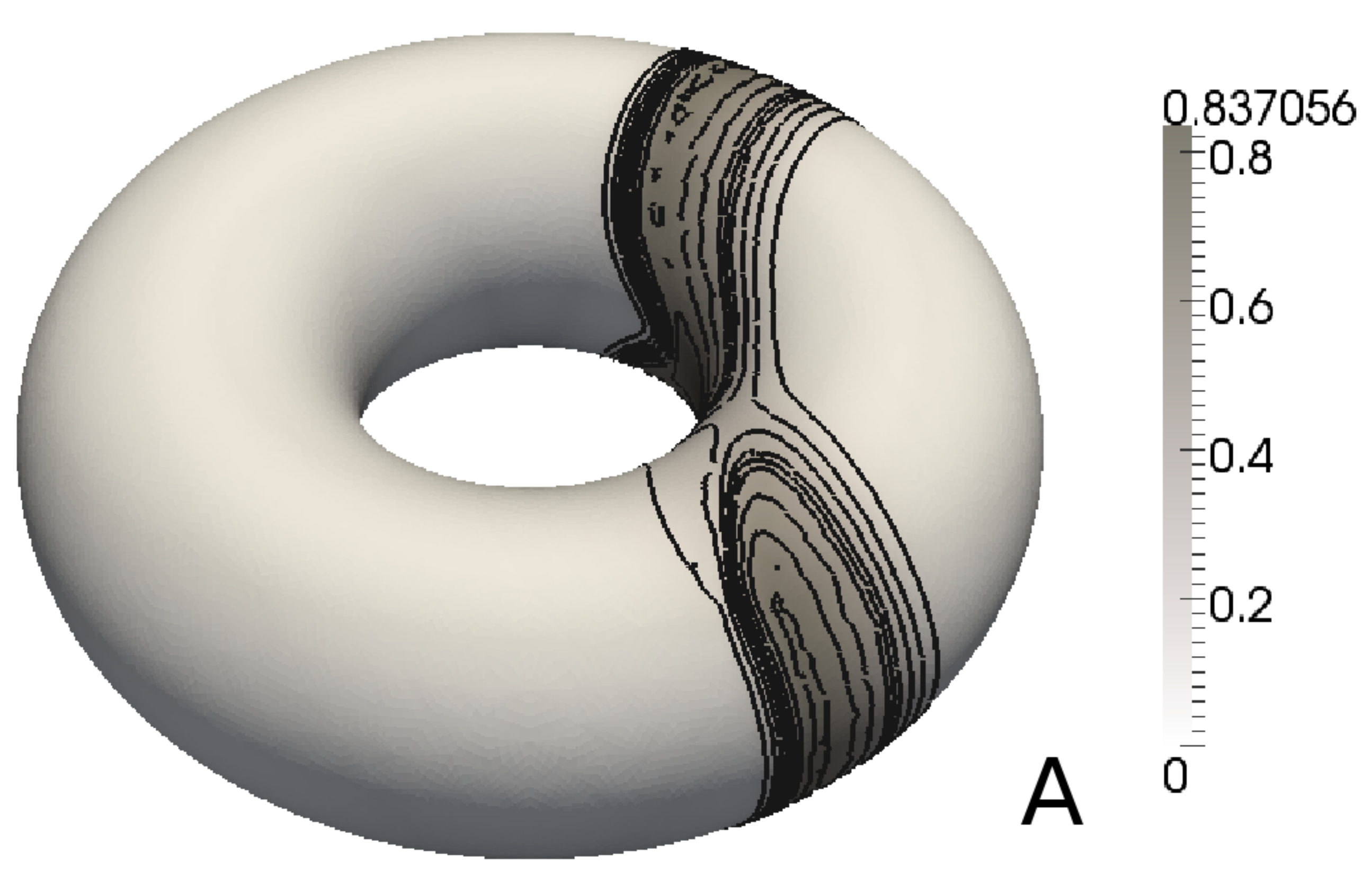}  \includegraphics[height=4.0cm, width=4.0cm] {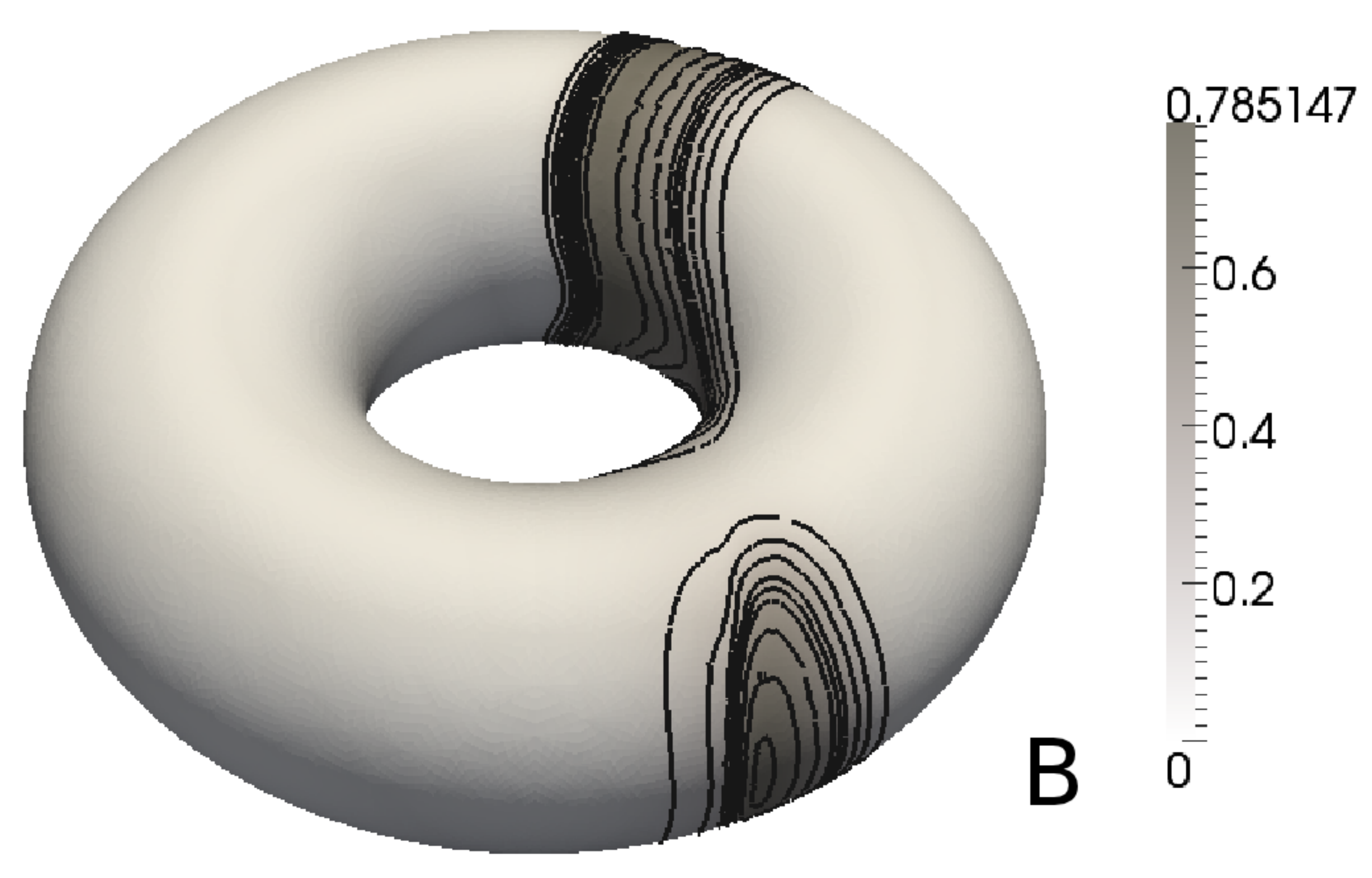}  \includegraphics[height=4.0cm, width=4.0cm] {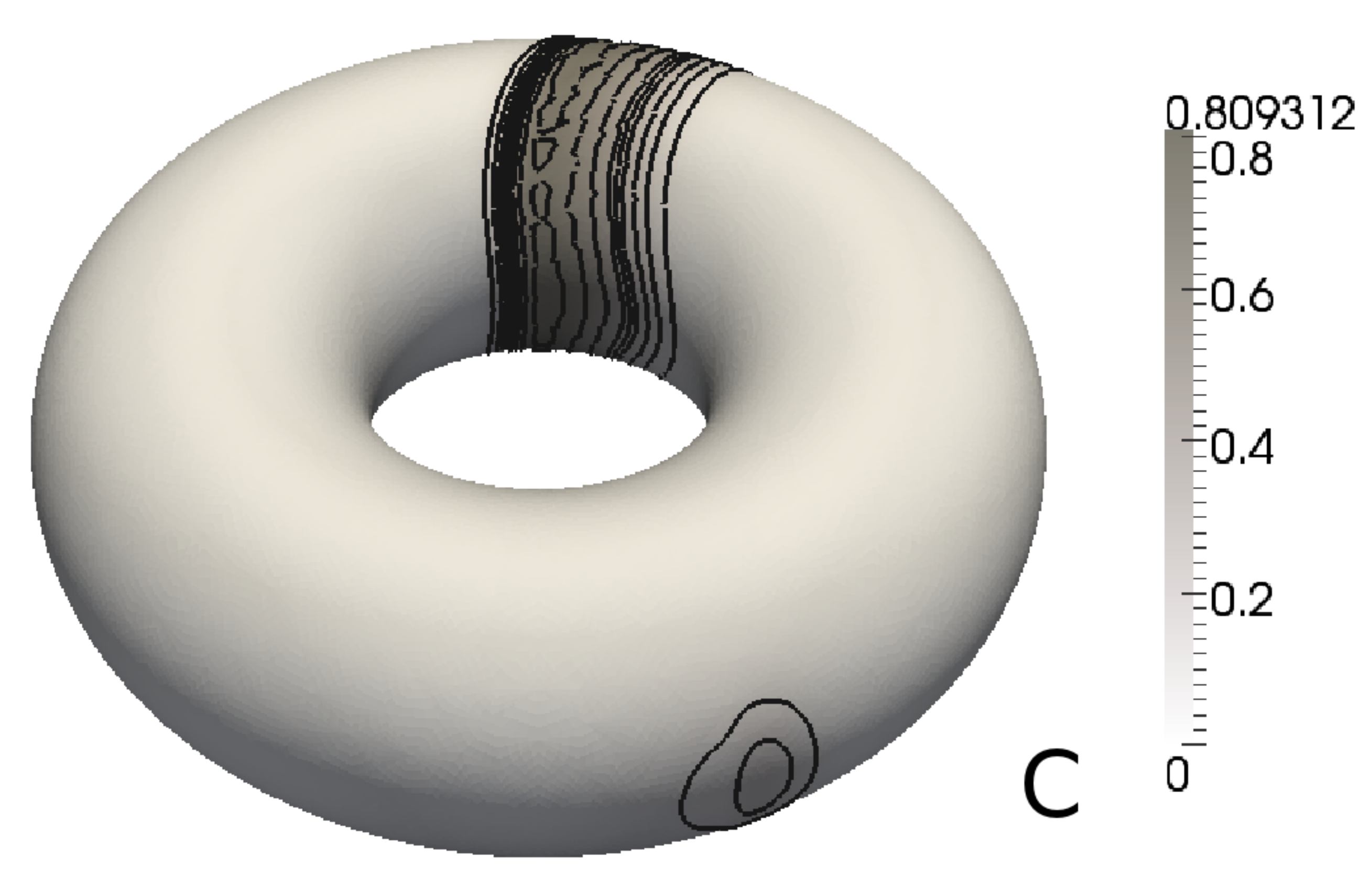} }
 \caption{Anisotropy block  with $\varsigma^{\phi \phi}=8.0$ and $\varsigma^{\theta \theta}=8.0$. After point initialized at the rightmost point of the above torus, the membrane potential ($u$) at $T=800.0$ (A), $T=900.0$ (B) and $T=1000.0$ (C).}
\label{torus3}
\end{figure}

\textbf{Problem 4}: The excitation propagation in torus is strikingly different from the propagation in the plane or sphere in the sense that we can not find one axis to represent the direction of propagation. For example, reminding that $\Lambda^k \equiv \partial y_1 / \partial x_k $, $\Lambda^x=1$ and $\Lambda^y=0$ for the planar propagation in the plane and $\Lambda^{\theta}=1$ and $\Lambda^{\phi}=0$ for sphere with point-initialization. However, $\Lambda^{\theta}$ and $\Lambda^{\phi}$ for torus is not always constant. With radius of the meridian $r$ and radius of the great circle $R$, let $\theta$ be the axis around the meridian and $\phi$ be the axis around the great circle. Then, a torus can be parameterized such as $
\mathbf{x} = ( (R+r \cos \theta ) \sin \phi, (R+ r \cos \theta) \cos \phi, r \cos \theta ),~0 \le \theta \le \pi,~ - \pi \le \phi \le \pi$. Then, no matter where or how the excitation starts, $\Lambda^{\phi}$ and $\Lambda^{\theta}$ is always neither constant or zero. This fact makes the analysis complicated, but to make the problem simpler, we first consider there is no anisotropy. \\ 
\\
\textbf{RA-analysis and validation}: First we consider an isotropic torus such as $\varsigma^{\phi \phi}=\varsigma^{\theta \theta}=1.0$. To make the problem even simpler, we suppose that the wavefront is at the angle of $\eta$ with respect to $\theta$ such that $\Lambda^{\theta} = \cos \eta$ and $\Lambda^{\phi}= \sin \eta$. Let $\Lambda^{\theta}$ and $\Lambda^{\phi}$ be constant along the wavefront. Since $g^{\theta \theta} = {1} / {r^2}$ and $g^{\phi \phi} = {1} / { (R+r \cos \theta)^2 }$, the relative acceleration equation \eqref{relacceqntensor} is expressed as
\begin{equation}
- \frac{\partial^2 n^i}{\partial \lambda^2} = \frac{ R + r \cos \theta}{ {\Lambda^{\theta}}^2 ( R + r \cos \theta )^2 + {\Lambda^{\phi}}^2 r^2 }   \left [ \cos a \frac{ R \cos \theta + r }{ R + r \cos  }  v^i  - \cos a  \sin \theta  \frac{\partial v^i}{\partial \lambda}   \right ] .   \label{relacctorus1}
\end{equation}
Note that each component shows strong dependency on the meridian $r$ for the fixed $R$. In other words, the above relative acceleration equation for isotropic torus implies that the relative acceleration can be significantly changed by resizing the meridian $r$. This result is coincident with the kinematic analysis by Davydov et. al. \cite{Davydov2000R} which demonstrated that the \textit{critical curvature} for breaking-up the wave can be generated by adjusting the ratio $R/r$ of torus. \\
\\
\textbf{Problem 5}: Consider anisotropy on torus. But, for the sake of simplicity, we suppose the propagation follows \textit{locally} the $\phi$ axis \textit{only} in a small region of torus. Let $R=100.0$ and $r=50.0$. Let the center of torus is located at $(0,0,0)$. The block of anisotropy is placed in the area of $-50.0 \le x \le 50.0$ and $0.0 \le y \le 150$. If the propagation is point-initialized at $(-150.0,0,0)$, then the propagation approximately follows the $\phi$-axis in the area of the anisotropy block. \\
\\
\textbf{RA-analysis}: With the anisotropy of both directions and $\Lambda^{\theta}=0$ and $\Lambda^{\phi}=1$, we obtain
\begin{equation}
 - \frac{\partial^2 n^i}{\partial \lambda^2} = \frac{\partial (\log \varsigma^{\phi \phi} )}{\partial n} \frac{\partial v^i}{\partial \lambda} + \frac{\partial^2 \varsigma^{\phi \phi}}{\partial n \partial \phi} v^i  + \frac{2r\sin\theta}{ R + r \cos \theta} \left ( \frac{\partial \theta}{\partial n}   \right )   \left [ \varsigma^{\phi \phi} \frac{\partial v^i}{\partial \lambda} + \frac{\partial \varsigma^{\phi \phi}}{\partial \phi} v^i  \right ] ,  \label{relacctorus2}
\end{equation}
where we used $\partial \phi / \partial n = 0$ according to the assumption that the propagation follows the $\phi$-axis in the area, thus $n = \theta$. The first two components are the same as the relative acceleration generated by anisotropy in the plane, but the last component is an additional component for anisotropic torus. Because of this last component, the behaviour of cardiac excitation propagation displays the following unique phenomena: the first phenomenon is that the relative acceleration depends on the $\theta$ angle. Considering $2 r \sin \theta / (R + r \cos \theta) $ increases as $\theta$ approaches $\pi/2$, it can be predicted that there is a larger relative acceleration in the area where $\theta$ is closer to $\pi/2$. The second phenomenon is as follows; because the last term adds additional relative acceleration, the critical magnitude of $\varsigma^{\phi \phi}$ for conduction failure is \textit{slightly} less than that for anisotropy in the plane. This is possible because all the components in equation \eqref{relacctorus2} are in the same sign when $\varsigma^{\phi \phi}$ is larger than $1.0$ which is the choice of our anisotropy. This result seems to support the conjecture that the curvature of geometry can increase the effects of anisotropy on curved surfaces.\\
\\
\textbf{Computational modeling}: The predictions of these unique phenomena in torus can be confirmed in computational simulations. Consider a torus of $R=100.0$ and $r=50.0$ centred at $(0,0,0)$. Let the block of anisotropy be located in $-50.0 \le x \le 50.0$ and $0.0 \le y \le 150.0$ as shown in Figure \ref{torus2}A. In Figure \ref{torus2}B and \ref{torus2}C, when the propagation is approximately in the direction of the $\phi$ axis, the anisotropy $\varsigma^{\phi \phi}=8.0$, which is less than $\varsigma^{\phi \phi}=10.0$ for the previous cases, generates the largest acceleration in the area where $\theta = \pi / 2$. But, the block of anisotropy fails to generate a sufficiently large relative acceleration for the area where $\theta$ is small. This agrees with the first predicted phenomenon from equation \eqref{relacctorus2}. However, the $\varsigma^{\phi \phi}$ anisotropy cannot block the propagation in result. This could be just a matter of direction of the propagation. Figure \ref{torus3}B and \ref{torus3}C displays that additional anisotropy in another direction $\varsigma^{\theta \theta}$ can add additional relative acceleration in the area where $\theta$ is small and consequently can stop the propagation. Note that in the plane, whatever the direction of the propagation is, anisotropy with the magnitude of $8.0$ cannot block the propagation for conduction failures. This confirms the second predicted phenomenon by the relative acceleration equation \eqref{relacctorus2}.

\section{Appendix D: Differentiation of $\sqrt{g} g^{kk}$}
\textbf{Lemma D}: Consider any index $k$ and the index $m$ different from $k$ and suppose that $x^m$ is in the Killing direction. Then, the following equation holds
\begin{eqnarray}
 \frac{1}{\sqrt{g}} \frac{\partial \left ( \sqrt{g} g^{kk} \right ) }{\partial x^{k}} =  g^{kk} \Gamma^{m}_{m k} . \label{AD3surfrev}
\end{eqnarray}
\textbf{Proof}: In the following calculations, the sum is taken over all the indices repeated above and below except $k$. Let's differentiate $\sqrt{g} g^{k k}$ with respect to the axis $x_{k}$. Using the chain rule and using ${\partial \sqrt{g}} / {\partial x_{k}} = \sqrt{g}\Gamma^{\mu}_{\mu k}$, we obtain
\begin{eqnarray}
\frac{\partial \left ( \sqrt{g} g^{k k} \right ) }{\partial x^{k}}  =  \sqrt{g}  \left ( g^{kk} \Gamma^{\mu}_{\mu k}  + \frac{\partial g^{kk}}{\partial x^{k}}   \right ) .
\label{AD30}
\end{eqnarray}
To replace the differentiation of $g^{kk}$ with the differentiation of the metric tensor $g_{kk}$, we use the identity of $g^{i \mu} g_{\mu j} = \delta^{i}_{j}$ to obtain
\begin{eqnarray}
\frac{\partial \left ( \sqrt{g} g^{kk} \right ) }{\partial x^{k}} =  \sqrt{g} g^{kk}  \Gamma^{\mu}_{\mu k}+ \sqrt{g} g^{k \nu} \left ( g_{\nu \mu} \frac{\partial g^{k \mu}}{\partial x_{k} } \right )  ,
\label{AD3first}
\end{eqnarray}
and the equation obtained by differentiating $g^{k k} g_{k \mu} = \delta^{k}_{\mu}$ as
\begin{eqnarray}
\frac{\partial g_{\nu \mu}} {\partial x^{k}} g^{k \mu} + g_{\nu \mu} \frac{\partial g^{k \mu}}{\partial x^{k}} = 0  . \nonumber 
\end{eqnarray}
Note that this reduce equation (\ref{AD3first}) to the following: 
\begin{eqnarray}
\frac{\partial \left ( \sqrt{g} g^{k k} \right ) }{\partial x^{k}} =  \sqrt{g} g^{kk} \Gamma^{\mu}_{\mu k}  -  \sqrt{g} g^{k \nu}  g^{k \mu}  \frac{\partial g_{\nu \mu}}{\partial x_{k}} .  \label{AD3two0}
\end{eqnarray}
Replacing the differentiation of $g^{k k}$ with the equalities in $\Gamma_{ijk}$ and $\Gamma^i_{jk}$, which are Christoffel symbols of the first kind and the second kind respectively, such that
\begin{eqnarray*}
\frac{\partial g_{\nu \mu}}{\partial x^{k}} = \Gamma_{\nu k \mu} + \Gamma_{\mu k \nu}, ~~\Gamma^{k}_{\mu k} = g^{k \nu } \Gamma_{ \nu \mu k}  .
\end{eqnarray*}
With these equalities and the condition of orthogonality of the curved axis $x_{k}$, equation (\ref{AD3two0}) becomes
\begin{eqnarray}
\frac{1}{\sqrt{g}} \frac{\partial \left ( \sqrt{g} g^{kk} \right ) }{\partial x^{k}} =  g^{kk} ( \Gamma^{m}_{m k} - \Gamma^k_{k k} ),~~~~~~~ m \neq k   .\label{AD3final}
\end{eqnarray}
For surfaces of revolution such as spherical shell and torus, the curved axis $x^{m}$ representing the rotational direction is in the direction of the Killing vector, or the \textit{Killing direction}, which satisfies the condition $\partial g_{\mu \nu} / \partial x^{m} =0$ \cite {Misner} and consequently equation \eqref{AD3surfrev}. $\square$

\section{Appendix E: Geometric factors of the surface of revolution to model the PV}

\begin{align*}
&R: \mbox{radius of the circular arc},  \\
&r: \mbox{radius of the circular hole}, \\
& \theta: \mbox{longitudinal angle from the root of the column}, \\
& \phi: \mbox{circumferential angle },
\end{align*}
\\
\textbf{(1) Parameterization}
\begin{align*}
& \mathbf{x} = ( R (1- \cos \theta), ( R (1- \sin \theta) + r ) \sin \phi, \\
& ~~~~~~~~~~~~~( R (1- \sin \theta) + r ) \cos \phi   ), \\
& \mathbf{x}_{\theta} = ( R \sin \theta, - R \cos \theta \sin \phi, - R \cos \theta \cos \phi    ), \\
& \mathbf{x}_{\phi} = ( 0, (R(1- \sin \theta) + r ) \cos \phi,  - (R(1- \sin \theta) + r ) \sin \phi  ),
\end{align*}
\\
\textbf{(2) The metric tensors}
\begin{align*}
& g_{\theta \theta} =  \mathbf{x}_{\theta} \cdot \mathbf{x}_{\theta}  =  R^2 ,~~~g_{\phi \phi} = \mathbf{x}_{\phi} \cdot \mathbf{x}_{\phi} =  ( R ( 1- \sin \theta ) + r )^2,\\
& g^{\phi \phi} =\frac{g_{\theta \theta}}{g} = \frac{1}{ ( R ( 1 - \sin \theta ) + r  )^2  },~~ g^{\theta \theta} =\frac{g_{\phi \phi}}{g} = \frac{1}{ R^2},
\end{align*}
\\
\textbf{(3) The Christoffel symbols}
\begin{align*}
& \Gamma_{\theta \phi \phi} = - \frac{1}{2} \frac{\partial g_{\phi \phi}}{\partial \theta} = R \cos \theta ( R ( 1 - \sin \theta ) + r ), \\
& \Gamma_{\phi \phi \theta} = \frac{1}{2} \frac{\partial g_{\phi \phi}}{\partial \theta} = - R \cos \theta ( R ( 1 - \sin \theta ) + r ), \\
& \Gamma^{\theta}_{\phi \phi} = g^{\theta \theta} \Gamma_{\theta \phi \phi} = \cos \theta ( ( 1 - \sin \theta ) + r/R ), \\
& \Gamma^{\phi}_{\theta \phi} = g^{\phi \phi} \Gamma_{\phi \phi \theta} =  \frac{ - \cos \theta }{ ( 1 - \sin \theta ) + r/R },  \\
& \frac{ \partial (g^{\theta \theta} \Gamma^{\phi}_{\theta \phi}) }{\partial \theta} = \frac{ 1 - \sin \theta (r/R)  }{ (R ( 1 - \sin \theta) + r )2 }, \\
& \frac{\partial g^{\phi \phi} } {\partial \theta} = \frac{ 2 R \cos \theta}{ ( R ( 1 - \sin \theta ) + r )^3 }.
\end{align*}

\bibliographystyle{plain}      
\bibliography{Article}

\end{document}